\documentclass[prd,twocolumn,superscriptaddress,floatfix,amsmath,amssymb,amsfonts,nofootinbib]{revtex4-2}
\usepackage{float} 
\usepackage{comment}
\usepackage[normalem]{ulem}
\usepackage[english]{babel}
\usepackage{graphicx}% Include figure files
\usepackage{dcolumn}% Align table columns on decimal point
\usepackage{bm}% bold math
\usepackage{blindtext}
\usepackage{verbatim}
\usepackage{mathrsfs}
\usepackage{musicography}
\usepackage{amsmath}
\usepackage{dsfont}
\usepackage{cancel}
\usepackage{physics}
\usepackage{epstopdf}
\usepackage{mathtools}
\usepackage{color}
\usepackage[usenames,dvipsnames]{pstricks}
\usepackage{epsfig}
\usepackage{pst-grad} % For gradients
\usepackage{pst-plot} % For axes
\usepackage{hyperref}
\usepackage{verbatim}
\usepackage{afterpage}
\usepackage{arydshln}
\hypersetup{
 colorlinks=true,
 linkcolor=blue,
 filecolor=magenta, 
 urlcolor=cyan,
}

\usepackage{tikz}
\usetikzlibrary{quantikz}

\newcommand{\be}{\begin{equation}}
\newcommand{\bel}[1]{\begin{equation}\label{#1}}
\newcommand{\ee}{\end{equation}}

\begin{document}

\title{The Pragmatic QFT Measurement Problem\\ and the need for a Heisenberg-like Cut in QFT}

\author{Daniel Grimmer}
\email{daniel.grimmer@philosophy.ox.ac.uk}
\affiliation{Faculty of Philosophy, University of Oxford, Oxford, OX2 6GG United Kingdom}
\affiliation{Reuben College, University of Oxford, Oxford, OX2 6HW United Kingdom}
\affiliation{Barrio RQI, Waterloo, Ontario N2L 3G1, Canada}

\begin{abstract}
Despite quantum theory's remarkable success at predicting the statistical results of experiments, many philosophers worry that it nonetheless lacks some crucial connection between theory and experiment. Such worries constitute the Quantum Measurement Problems. One can broadly identify two kinds of worries: 1) pragmatic: it is unclear how to model our measurement processes in order to extract experimental predictions, and 2) realist: we lack a satisfying metaphysical account of measurement processes. While both issues deserve attention, the pragmatic worries have worse consequences if left unanswered: If our pragmatic theory-to-experiment linkage is unsatisfactory, then quantum theory is at risk of losing both its evidential support and its physical salience. Avoiding these risks is at the core of what I will call the \textit{Pragmatic Measurement Problem}. 

Fortunately, the pragmatic measurement problem is not too difficult to solve. For non-relativistic quantum theory, the story goes roughly as follows: One can model each of quantum theory's key experimental successes on a case-by-case basis by using a measurement chain. In modeling this measurement chain, it is pragmatically necessary to switch from using a quantum model to a classical model at some point. That is, it is pragmatically necessary to invoke a Heisenberg cut at some point along the measurement chain. Past this case-by-case measurement framework, one can then strive for a wide-scoping measurement theory capable of modeling all (or nearly all) possible measurement processes. For non-relativistic quantum theory, this leads us to our usual projective measurement theory. As a bonus, proceeding this way also gives us an empirically meaningful characterization of the theory's observables as (positive) self-adjoint operators.

But how does this story have to change when we move into the context of quantum field theory (QFT)? It is well known that in QFT almost all localized projective measurements violate causality, allowing for faster-than-light signaling; These are Sorkin's impossible measurements. Thus, the story of measurement in QFT cannot end as it did before with a projective measurement theory. But does this then mean that we need to radically rethink the way we model measurement processes in QFT? Are our current experimental practices somehow misguided? Fortunately not. I will argue that (once properly understood) our old approach to modeling quantum measurements is still applicable in QFT contexts. We ought to first use measurement chains to build up a case-by-case measurement framework for QFT. Modeling these measurement chains will require us to invoke what I will call a QFT-cut. That is, at some point along the measurement chain we must switch from using a QFT model to a non-QFT model. Past this case-by-case measurement framework, we can then strive for both a new wide-scoping measurement theory for QFT and an empirically meaningful characterization of its observables. It is at this point that significantly more theoretical work is needed. This paper ends by briefly reviewing the state of the art in the physics literature regarding the modeling of measurement processes involving quantum fields.
\end{abstract}

\maketitle

\section{Introduction: Another Quantum Measurement Problem}\label{Introduction}
%\Out{Introduce the quantum MPs: dissatisfying theory-to-experiment disconnects.}
It is incontestable that quantum theory has been remarkably successful at predicting the statistical results of a wide range of experiments. However, despite its many predictive successes, many philosophers and physicists are nonetheless worried that quantum theory lacks some crucial connection between theory and experiment. Various dissatisfactions with various theory-to-experiment disconnects each deserve the title ``A Quantum Measurement Problem'': How should we understand/model measurement processes involving quantum systems? 

%\Out{Cite people and name drop ``pragmatic QFT MP'' before limiting attention to the non-relativistic context.}
Indeed, there is a wide literature aimed at identifying what the measurement problem is exactly. See, for instance Maudlin's ``Three Measurement Problems'' \cite{Maudlin1995ThreeMP} among many others \cite{muller2023measurement,sep-qt-issues,1987Saui,Wallace2020,DICKSON2007275}. The quantum measurement problems have also been much-discussed in the context of quantum field theory (QFT)~\cite{BARRETT2014168,barrett_2005,Barrett2002,1681666,Halvorson:2006wj,OAOQFTChap10,OAOQFTChap11,Redhead1995,Malament1996-MALIDO, papageorgiou2023eliminating}. Adding to these discussions, this paper will introduce a new set of worries which I will call the \textit{pragmatic measurement problem}. These worries relate to how we model our quantum measurement processes in order to extract experimental predictions. Since these worries have been more-or-less solved in non-relativistic quantum theory, the focus of this paper will be on the pragmatic measurement problem in the context of QFT. Before discussing this, however, allow me to first introduce the pragmatic measurement problem in a non-relativistic context.

%\Out{As a foil, introduce the sophomore's quantum theory and how they model measurements.}
In order to differentiate the various quantum measurement problems from each other, it is perhaps best to start from a version of quantum theory which (hopefully nearly) every physicist and philosopher is dissatisfied with. I have in mind the parts of non-relativistic quantum theory which students are urged to focus on after they are told to ``Shut up, and calculate!''. Let us call this \textit{the sophomore's quantum theory}. Students are here taught to model quantum experiments as follows.\footnote{A sophisticated sophomore may also learn about selective and non-selective measurements as well as post-measurement state updates via L\"uders rule. Moreover, they may also learn about density matrices, $\hat\rho$, and non-ideal measurements, i.e, Positive Operator-Valued Measures (POVMs)  $\hat{E}_\text{out}\geq0$ with $\sum_\text{out} \hat{E}_\text{out} = \openone$. In terms of POVMs, Born's rule is $p(\text{out}\vert\,\hat{U},\,\text{in})=\text{Tr}(\hat{E}_\text{out}\hat{U}\hat{\rho}_\text{in}\,\hat{U}^\dagger)$. The pragmatic worries discussed below apply equally well to this sophisticated sophomore.\label{FNSophSoph}} The sophomore is first told the following two tautologies: All measurements are of some observable and, moreover, all observables are measureable. 

The sophomore is then told that quantum theory's observables are exactly the self-adjoint operators. In order to model a measurement of a given self-adjoint operator, $\hat{Q}$, one begins by computing its eigensystem, $\hat{Q}=\sum_\text{out}q_\text{out} \ket{\text{out}}\bra{\text{out}}$. The projectors, $\hat\pi_\text{out}\coloneqq\ket{\text{out}}\bra{\text{out}}$, appearing in this decomposition define a projection-valued measure (PVM). Next one takes the given initial conditions, $\ket{\text{in}}$, and applies the given unitary evolution, $\hat{U}$. The sophomore is told that putting these computations together unambiguously yield a statistical prediction of the experiment's outcome via the Born rule, $p(\text{out}\vert\,\hat{U},\text{in})=\vert\!\bra{\text{out}}\hat{U}\ket{\text{in}}\!\vert^2$.

%\Out{Broadly distinguish pragmatic from realist worries: model vs understand}
Many physicists and philosophers are dissatisfied with the sophomore's quantum theory, claiming that it lacks the right kind of connection between theory and experiment, and rightly so. One can broadly distinguish two types of worries surrounding quantum measurement: pragmatic worries and realist worries. Pragmatic worries are methodological in nature and aim at clarifying how exactly these statistical predictions are to be pulled out of the theory. Specifically, they ask: How should one \textit{model} real-life measurement processes as a matter of experimental practice? By contrast, realist worries are aimed at establishing a metaphysical account of the measurement process. Realist worries ask: How should the measurement process be \textit{understood}, metaphysically speaking? See, for instance, Maudlin's ``Three Measurement Problems'' \cite{Maudlin1995ThreeMP} all of which I classify as realist/metaphysical worries.

%\Out{Highlight the sophomore's pragmatic failings: wrong observables, no connection with experiments, watch them falter.} 
The sophomore's quantum theory fails on both the pragmatic and realist fronts. Its realist failures are well known, but its pragmatic failures deserve some further comment. Firstly, the sophomore has misidentified the observables of quantum theory (see Secs.~\ref{SecObs} and~\ref{ImpossibleMeasurements} below). The deeper issue, however, is that the sophomore's quantum theory does not, in fact, give us a way of making unambiguous (or even approximate) statistical predictions for real-life experiments. While it is true that statistical predictions are unambiguously associated with initial states, unitaries, and projectors, (recall, $p(\text{out}\vert\,\hat{U},\text{in})=\vert\!\bra{\text{out}}\hat{U}\ket{\text{in}}\!\vert^2$) these themselves have not yet been suitably connected with our real-life experimental setups.

Specifically, the sophomore has no good answers to the following questions:\footnote{The same complaint holds for the sophisticated sophomore discussed in footnote \ref{FNSophSoph} if we here replace PVMs with POVMs.} How can one go about determining (even approximately) which observable \textit{this} apparatus measures? Under what conditions is it appropriate to model this piece of lab equipment using a PVM? If it is appropriate, then exactly which PVMs am I allowed to use? At what time point in the experiment am I allowed to model this PVM as occurring? There may be ready answers to these questions pre-written on the sophomore's problem sheets, but show them a piece of real-life lab equipment and watch them falter.

%\Out{The sophomore may correctly guess which PVM to use and when, ... but the issue is that they are guessing.}
Often the sophomore may intuitively guess which PVM to use and when. It is highly intuitive that in modeling a double-slit experiment the right PVMs are (at least approximately) the position projectors, $\hat\pi_\text{out}=\ket{x}\bra{x}$. Moreover, it is highly intuitive that the right time to apply them is (at least approximately) when the electron hits the detection screen. For most practical purposes this is effectively what happens. Indeed, it may often be the case that the sophomore's guesses consistently give accurate-enough predictions. But ultimately, they are nothing more than just that: guesses. It goes without saying that this method of connecting theory with experiment is deeply unsatisfying.

%\Out{Highlight again the distinction between pragmatic and realist worries.}
I should here clarify what exactly the pragmatic measurement problem is asking for. Importantly, it is not a metaphysical problem; I am not asking the sophomore to base their prediction on a metaphysical account of the measurement process. Rather, the pragmatic measurement problem is a methodological problem which applies equally well to anti-realist or pragmatic interpretations of quantum theory. Indeed, every scientific theory must provide us with a robust account of how predictions can and should be made from it. Namely, we need a satisfactory account of how one is allowed to \textit{model} both the system in question and the measurement process. The above critique thus highlights a devastating methodological failure of the sophomore's quantum theory; A theory-to-experiment linkage which relies so blatantly upon intuitive guessing simply cannot do the work we require of it.

%\Out{Clarify what is at risk in the pragmatic MP: evidential support and physical salience (cite Curiel).}
As I will now discuss, in comparison with the realist measurement problem, these pragmatic worries have far worse consequences if left unanswered. It is helpful to distinguish two senses in which our scientific theories are about reality. Firstly, they may have metaphysical aspirations of representing and/or describing reality. This is the domain of the realist measurement problem. But why should we believe that our scientific theories have any right to be ``about reality'' in the first place? Ultimately, our theories earn this right by a process of complex sustained bi-directional contact with experimental practice. That is, our scientific theories get their right to be meaningful from their (often messy \cite{Curiel}) connection to our systems of measurement devices and approximation techniques. This is the domain of the pragmatic measurement problem. 

To be clear, the subject of the pragmatic measurement problem is not the linkage between theory and reality which is mediated by successful reference. Instead, what is at risk here is the linkage between theory and reality which is mediated by real-life experimental practice. Without a clear understanding of this pragmatic connection between theory and experimental practice, quantum theory would be at risk of losing both its evidential support and its physical salience.

%\Out{Fortunately, these pragmatic worries are easy to solve: measurement framework (core) then measurement theory and observables (extended).}
Fortunately, however, as well as having worse consequences, the pragmatic measurement problem is also much easier to address than the realist's worries. To solve the core problem, all one needs to do is to develop a methodologically sound framework for modeling the relevant measurement processes in a sufficient level of dynamical detail~\cite{Curiel,LegitSin,BrownHarvey}.\footnote{There is ample room for discussion regarding the exact standards to which these models ought to conform; These standards ought to be high, but contextually reasonable. See Sec.~\ref{ChainsAndCuts} for further discussion.} As Sec.~\ref{ChainsAndCuts} will argue, it is pragmatically necessary to take a Heisenberg cut when modeling any real-life quantum measurement. That is, somewhere along the measurement chain it will be necessary to switch from a quantum model to a classical model. There are many ways of taking Heisenberg cuts available to us (see Sec.~\ref{SecTaxonomy}). 

Past this minimal solution to the pragmatic measurement problem, one can then strive to solve the extended problem by developing a unified measurement theory which is applicable to all (or nearly all) measurement processes. In Sec.~\ref{ChainsAndCuts} I will demonstrate in a non-relativistic context how one can begin from a case-by-case measurement framework based on measurement chains and then develop a wide-scoping measurement theory based upon PVMs and POVMs. Having such a measurement theory is not only highly convenient for modeling experiments but can also be theoretically fruitful: It can give us an empirically meaningful characterization of the theory's observables.

%\Out{Give priority to the pragmatic worries over realist ones; Spoils before ontology.} 
In light of the above discussion, it makes sense to address the pragmatic measurement problem before the realist one. As compelling as the realist's worries are, one might say: Let us first work on bringing home the spoils of quantum theory's experimental successes; We can then worry about providing a metaphysical account of the theory later, once we have better footing. 

%\Out{Anticipate a protest, the realist and pragmatic MPs don't have to be solved together.}
One might here protest that the realist and pragmatic measurement problems ought to be solved together. Indeed, this is a possibility: Developing a metaphysical account of the measurement process (e.g., Bohmian mechanics or a spontaneous collapse model) might show us how to model a great many different measurement processes. Importantly, however, it also might not; Having a satisfying metaphysical account of the measurement process does not automatically give us a tractable way of modeling quantum theory's key experimental successes. For instance, it may be the case that directly simulating the ontological development of an experiment is computationally infeasible. Alternatively, modeling these experiments may require us to go outside of the scope of whichever ontological account we have in mind (e.g., into QFT~\cite{WallaceBlueSkyTalk,WallaceBlueSkyPaper}). In either of these cases, we would still suffer the consequences of the pragmatic measurement problem. Thus, solving the realist measurement problem does not automatically address the pragmatic measurement problem. (Nor vice-versa.)

%\Out{Argue again that the pragmatic issues ought to be given priority}
As the above discussion has shown, the pragmatic and realist measurement problems are separate problems, with separate difficulties, consequences, and solution criteria. While one may hope to solve them simultaneously, this is far from compulsory. Indeed, given how contentious the ontology of quantum theory is, it makes sense to first address the pragmatic measurement problem in an ontology-neutral way. Even if one is committed to later give an ontology-laden solution, one can reasonably proceed this way thinking: At least in the meantime we will have a working understanding of quantum theory's measurement processes and its observables.

%\Out{In practice, this is exactly what has happened. The pragmatic measurement problem has been addressed while the realist one has not.}
In practice, this is exactly what has happened for non-relativistic quantum theory: While the realist measurement problem continues to be fiercely debated, the pragmatic measurement problem has long since been satisfactorily addressed (at least within the experimental purview of non-relativistic quantum theory).\footnote{The experimental purview of non-relativistic quantum theory might be notably smaller than one thinks~\cite{WallaceBlueSkyTalk,WallaceBlueSkyPaper}. See the quote from Wallace in the next subsection.} Said differently, while the ontology of non-relativistic quantum theory is still contentious, its experimental predictions are clear, as are the allowed methods for extracting these predictions from the theory. As I will discuss in Sec.~\ref{ChainsAndCuts}, the key notions here are measurement chains and Heisenberg cuts; In these terms, one can achieve an ontology-neutral solution to the pragmatic measurement problem, at least for non-relativistic quantum theory.

\subsection*{The Pragmatic Measurement Problem in QFT}
%\Out{Pivot to discussing the Prag. QFT MP} 
The main subject of this paper, however, is the pragmatic measurement problem in the context of quantum field theory (QFT). Namely, I consider \textit{The Pragmatic QFT Measurement Problem}:\footnote{As before, these pragmatic issues ought to be distinguished from the much-discussed realist/ontological issues within QFT.~\cite{BARRETT2014168,barrett_2005,Barrett2002,1681666,Halvorson:2006wj,OAOQFTChap10,OAOQFTChap11,Redhead1995,Malament1996-MALIDO}} How should one model measurement processes involving quantum fields? How must our measurement framework for QFT differ from our usual non-relativistic measurement framework? Can we model QFT-involved measurements using our usual measurement theory based on PVMs and POVMs? How must our characterization of the observables of QFT differ from the way we characterize the observables of non-relativistic quantum theory?

%\Out{Transfer lessons learned above into the QFT context. Advocate for spoils-before-ontology approach}
Much of the above discussion of the pragmatic vs realist measurement problems carries over unchanged into QFT. Namely, in comparison with the realist QFT measurement problem, the pragmatic QFT measurement problem has worse consequences if neglected; Quantum field theory would then be at risk of losing both its evidential support and its physical salience. Fortunately, as before, it is also much easier to address. In fact, the difficulty gap between the realist and pragmatic measurement problems arguably widens for QFT since certain ontological issues become notably more difficult in QFT.\footnote{Namely, certain strategies adopted by hidden variable and collapse approaches fail in relativistic contexts \cite{BARRETT2014168,barrett_2005}.} Hence, we have extra reason to seek out an ontology-neutral approach to the pragmatic QFT measurement problem. As I recommended above: Let us first work on bringing home the spoils of quantum field theory's experimental successes; We can then worry about the ontology of the theory later, once we have better footing. 

%\Out{Spell out consequences: troublesome/severe/catastrophic}
This spoils-before-ontology approach raises the following question: For quantum theory generally, where were these metaphorical spoils won? While establishing a detailed answer to this question is not essential for the main philosophical points made in this paper, it will help us to determine what is at stake. Namely, if quantum field theory is required in order to model some/many/most of quantum theory's key experimental successes, then failing to address the pragmatic QFT measurement problem is troublesome/severe/catastrophic. 

%\Out{Give Wallace quote about limited scope of NRQM}
%\begin{comment}
One might feel that the stakes here are significantly lower than in the non-relativistic context: Can't one adequately model almost all of quantum theory's key experimental successes without QFT? Wallace has recently argued for the following perhaps surprising claim:
\begin{quote}
For a quantum experiment to be modellable entirely within NRQM \dots not only the system being measured, but the apparatus doing the measurement, would have to be within the scope of NRQM. Such systems plausibly exist \dots But experiments like this comprise only quite a small fraction of the experiments performed within ‘non-relativistic’ quantum mechanics.~\cite[p.~21]{WallaceBlueSkyPaper}
\end{quote}
Importantly, Wallace arrives at this conclusion for fairly basic conceptual reasons, not a demand for hyper-accuracy. Ultimately, this would mean that if we cannot establish an adequate pragmatic link between QFT and experimental practice then not only quantum field theory but nearly the whole of quantum theory is at risk of losing its evidential support and physical salience.
%\end{comment}

%\Out{Q: But what novel issues arise in QFT? A: Essentially nothing changes.}
Given that the route home for some/many/most of quantum theory's experimental spoils runs through quantum field theory, our next question becomes: What novel issues arise when one attempts to model measurement processes involving quantum fields? Much of the physics literature on this topic stresses how our canonical modeling techniques (i.e., projective measurements) fail when naively implemented in QFT~\cite{Redhead1995,Malament1996-MALIDO,pologomez2021detectorbased,Jubb2022,BorstenJubbKells,fewster1,fewster2,fewster3,Anastopoulos2022,Sorkin,TaleOfTwo,Ruep2021,JoseMariaEdu,Dowker,Dowker2,borsten,alvaro,Adam,papageorgiou2023eliminating}. A central example of this are Sorkin's impossible measurements~\cite{Sorkin} which I will discuss in Sec.~\ref{ImpossibleMeasurements}. Roughly, in QFT one can identify projective operators which supported only over a localized region of spacetime. Intuitively, these ought to correspond to localized projective measurements. However, implementing these projective measurements straight-forwardly leads to faster-than-light signaling.

The key lesson to be drawn from Sorkin's impossible measurements is that having solved the pragmatic measurement problem for non-relativistic quantum theory does not automatically solve it for QFT. We do need to re-think how quantum measurements are to be modeled in QFT; In particular, the story of modeling measurement in QFT simply cannot end with a projective measurement theory as it did in the non-relativistic context. But does this then mean that we need to radically rethink the way we model measurement processes in QFT? Are our current experimental practices somehow misguided? I will answer no to both questions: \textit{Aside from some technical complications, moving into a quantum field theoretic context changes essentially nothing regarding how we can and should model quantum measurements.}

Since I am saying that ultimately nothing much changes as we move into QFT, I must start by discussing how the pragmatic measurement problem has already been solved for non-relativistic quantum theory. As I will discuss in Sec.~\ref{ChainsAndCuts}, our ability to casually invoke projective measurements in non-relativistic contexts is the end of a long story which begins with a discussion of measurement chains and Heisenberg cuts. We can use these notions to produce a measurement framework for non-relativistic quantum theory which is capable of modeling its key experimental successes, at least on a case-by-case basis. Past this, one can then strive for a wide-scoping measurement theory capable of modeling all (or nearly all) possible measurement processes. For non-relativistic quantum theory, this leads us to our usual projective measurement theory.

With this non-relativistic story established, I will then argue in Sec.~\ref{GenChainsAndCuts} that the pragmatic QFT measurement problem ought to be approached in exactly the same way. As I noted above, the story of modeling measurement in QFT cannot end with a projective measurement theory as it did before. As I will argue, however, the story nonetheless ought to start in the same way and then proceed in the same direction as it did before. In particular, we ought to first use measurement chains to build up a case-by-case measurement framework for QFT. Analogously to the non-relativistic case, it is pragmatically necessary that we switch from a QFT model to a non-QFT model at some point along the measurement chain. (I will call this ``taking a QFT-cut''.) Equipped with such a patchwork of measurement frameworks, we will be able to secure the spoils of QFT's key empirical successes, at least on a case-by-case basis. In my view, our current experimental practices are up to this task. (Although new high-precision experiments may bring this into question, see Sec.~\ref{SecIIIA}.)

Just as in the non-relativistic case, however, having such a case-by-case measurement framework does not entirely solve the pragmatic measurement problem. As I just mentioned, a case-by-case measurement framework might only temporarily cover QFT's experimental successes. New experimental successes may come along which are outside of the scope of our current best practices. Moreover, such a measurement framework does not allow us to characterize QFT's measurement processes in general. Nor does it allow us to talk in an empirically meaningful way about QFT's observables. To meet these ends, we must strive for both a unified measurement theory for QFT which is applicable to all (or nearly all) measurement processes.

More specifically, we need a way of making QFT-cuts which is near universally applicable in all measurement scenarios. It is only at this point that the story of measurement in QFT diverges substantially from the non-relativistic story. And it is at this point that significantly more theoretical work is needed. Finally, Sec.~\ref{StateOfTheArt} will review the state of the art in the physics literature regarding the modeling of QFT-involved measurement processes. What tools do physicists currently have available to them for making QFT-cuts? The primary two tools which I will discuss are the Fewster Verch (FV) framework~\cite{fewster1,fewster2,fewster3,Ruep2021,MeasurementSchemeFV}, and the Unruh-DeWitt detector model~\cite{pologomez2021detectorbased,Unruh1976,BLHu2007, Brown2013, Hotta2020, Zeromode,TaleOfTwo,Adam,Valentini1991, Reznik2003, Pozas-Kerstjens:2015,Menicucci, Terno2016, Cosmo, Henderson2018}.\footnote{Some additional comparison of these two approaches can be found in \cite{papageorgiou2023eliminating}.} A measurement theory for QFT based on Unruh-DeWitt detectors has recently been put forward~\cite{pologomez2021detectorbased}. As I will argue, this is (at least currently) the best approach available for achieving a wide-scoping measurement theory for QFT and for identifying its observables in an empirically meaningful way.

\section{Measurement Chains and Heisenberg Cuts}\label{ChainsAndCuts}
%\Out{Introduce the topics covered in this section: Chains and Cuts; Taxonomy; Solving the Core and Extended Problems; Rethinking Observables.}
This section will elaborate on my above claim that for non-relativistic quantum theory the pragmatic measurement problem has been satisfactorily solved. Namely, I will discuss how this problem can be addressed in an ontology-neutral way in terms of measurement chains and pragmatic Heisenberg cuts. Firstly, Sec.~\ref{SecFirstExamples} will introduce the notions of measurement chains and Heisenberg cuts via some example scenarios. Next, Sec.~\ref{SecTaxonomy} will introduce a helpful taxonomy regarding the different kinds of Heisenberg cuts which are available to us. Then, Sec.~\ref{SecCoreExt} will distinguish two versions of the pragmatic measurement problem and show how they can each be addressed using measurement chains and Heisenberg cuts. Finally, Sec.~\ref{SecObs} will make some progress towards identifying the observables of non-relativistic quantum theory (n.b., ``observables'' $\neq$ ``self-adjoint operators''). 

\subsection{Examples of Measurement Chains and Heisenberg Cuts}\label{SecFirstExamples}
%\Out{Introduce Measurement chains with an abstract example} 
In practice, we often model our experiments in terms of a measurement chain. Roughly, a measurement chain models an experiment as a sequence of interactions which carries the measured information from the systems being measured to some record-keeping device. To give an abstract example: System A interacts with system B which then interacts with system C which then ... which then interacts with system R, our record-keeping device. (More will be said momentarily about the freedom one has in starting and ending measurement chains.)

%\Out{Then give more concrete example: atomic experiment.}
For a more concrete example, we may be interested in a certain amplitude associated with an atom in a certain superposition. Our experiment may proceed as follows: An atom in a superposition emits a photon which is detected by a photo-multiplier which triggers a small current which turns on a transistor which ... which displays a number on a screen which the experimenter writes in her notepad.

%\Out{Measurement chains are a modeling tool, not a sequence of physical systems.}
To be clear, throughout this paper the term ``measurement chain'' does not refer to the linear sequence of \textit{physical systems/interactions} which carry the measured information, per se. Rather, here, the measurement chain is a formalization of these systems which the experimenter invents for the purposes of modeling her experiment. There may be multiple acceptable ways of parsing a given physical scenario into a formalized measurement chain. 

%\Out{There will often be ambiguity in where exactly we choose to start and stop the measurement chain.}
Indeed, given an experiment, it is not always clear where we ought to place either the start or the end of the measurement chain. Regarding initialization, we can always ask for the initial conditions of our initial condition. Regarding the late stages of the experiment, it is unclear where to stop: the computer screen, the experimenter, her notepad, etc. One promising way to proceed is to schematize the observer~\cite{Curiel}, thereby getting the laboratory inside the theory, so to speak. Such considerations would need to be built into whichever high but contextually reasonable standards we adopt for our modeling practices. The results of this paper do not depend sensitively on how this is done.%\footnote{There is also an interesting question regarding the middle of the measurement chain: Where along the measurement chain does the modeling-burden shift from the theorist to the experimenter? See Sec.~\ref{SecCoreExt} for further discussion.}

\begin{figure}
\includegraphics[width=0.5\textwidth]{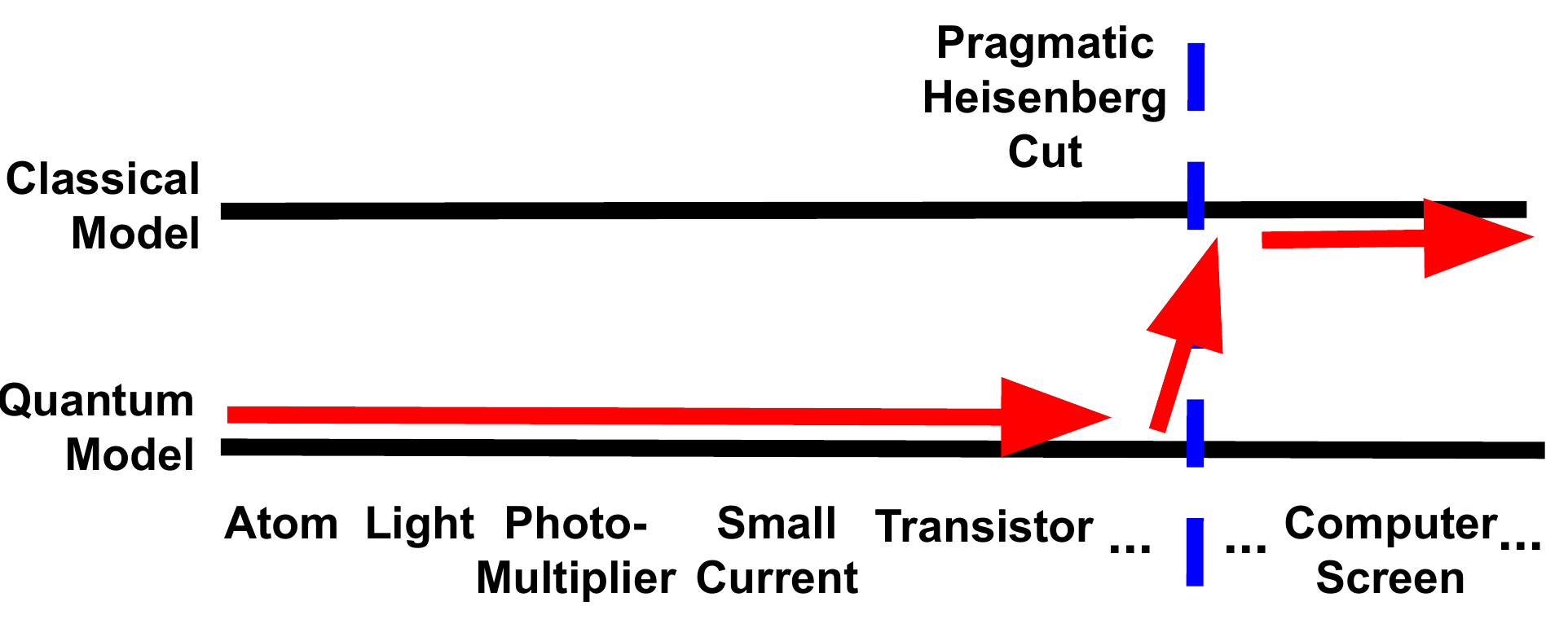}
\caption{The measurement chain of a simple atomic experiment. The black lines show two possible types of models: quantum or classical. The red arrow shows which part of the experiment we are modeling with which kind of model. The dashed blue line shows where we are taking the pragmatic Heisenberg cut. That is, where we switch from modeling the experiment in a quantum way to a classical way.}\label{FigHCut}
\end{figure}

\begin{figure*}[t]
\includegraphics[width=0.9\textwidth]{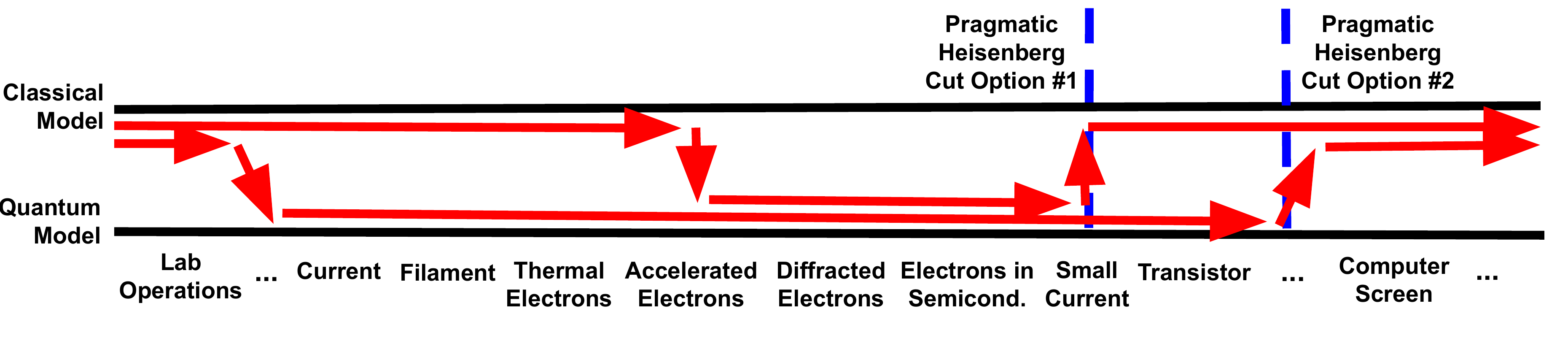}
\caption{One possible formalization of the double-slit experiment into a measurement chain is shown. The black lines show two possible types of models: quantum or classical. Each of the red arrows shows which parts of the experiment are being modeled with which kind of model theory. The bottom red line opts for a quantum model whenever possible, whereas the top red line opts for a classical model wherever possible. The dashed blue line shows where these two approaches to modeling this experiment place their respective pragmatic Heisenberg cuts.}\label{FigHCut2}
\end{figure*}

%\Out{Discuss first figure: What are the horizontal axes? What are the black and red lines?}
%\begin{comment}
The above-discussed example of a measurement chain is laid out horizontally in Fig. \ref{FigHCut}. It is important to note that while in this example, moving horizontally happens to move us into larger, more complex systems with more degrees of freedom, this is accidental. Horizontal movement in this diagram indicates only that we are moving from one system to another sequentially towards the end of the experiment. One can easily imagine experiments where advancing forward in the experiment temporarily moves the measured information into a smaller system. (Indeed, such a scenario is displayed in Fig.~\ref{FigHCut2}.) The two horizontal black lines in Fig. \ref{FigHCut} represent two types of model that we could have for each part of our experiment (here, either classical and quantum). The red arrows indicate how we are going to model each part of the experiment. 

%\end{comment}

%\Out{Note that the classical vs quantum distinction is ontology-neutral. Mention a Bohmian account.}
One thing which should be stressed here is that one can model one's experiment in terms of a measurement chain regardless of one's ontological preferences regarding non-relativistic quantum theory. For instance, for a Bohmian, a classical model would mean any model which does not include the wavefunction (i.e., classical physics) whereas quantum models do include the wavefunction and the guiding equation.

%\Out{Stress again that the location of the red arrow indicates a modeling choice, not anything ontological.} 
It is also important to note that the path that the red arrow takes through this diagram is, in large part, a free choice of the experimenter.\footnote{Towards the end of this subsection, the path of this red arrow in these figures will be related to the historical notion of Heisenberg cuts and to Bohr's doctrine of classical concepts.} The location of the red arrow does not mean that this or that system \textit{is} quantum/classical. All that this indicates is that, for the purposes of modeling this experiment, this particular experimenter has chosen to model this system as such.

%\Out{The red line is not completely divorced from reality, however. There will always be conceptual and pragmatic limitations.}
Importantly, however, it is not the case that any part of an experiment can be successfully modeled using any theory. In practice, there are always going to be some restrictions. Sometimes for the sake of accuracy it will be necessary to model a given system in a quantum way. Sometimes for computational or technological reasons it will not be feasible to model a given system in a quantum way (forcing us to model it classically). Sometimes it will be conceptually necessary to model a given system in a quantum way. (If one accepts Bohr's doctrine of classical concepts, then it is necessary for conceptual reasons to model the end of an experiment in a classical way.) For these and other reasons, the possible routes which the red arrows may make through these diagrams are limited.

%\Out{When these factors conspire against us, we simply cannot make predictions (except by guessing). Hence, we cannot gain empirical support.} 
With these restrictions in mind, it may occur that modeling some part of our measurement chain in a quantum way is both conceptually mandatory and technically infeasible. For instance, imagine a chemical reaction between two large biomolecules for which subtle quantum effects are unavoidably relevant. In this case, we simply cannot (yet) model this experiment satisfactorily. Hence, we cannot say what quantum theory predicts for this experiment (although we may have a sophomoric guess). Even if we do guess right, however, this experiment cannot in good faith be counted towards quantum theory's empirical support.

%\Out{Introduce the double-slit experiment measurement chain example and figure.}
%\begin{comment}
To better understand how these modeling restrictions work in practice, it is perhaps best to work through a familiar example. Consider a double-slit experiment conducted with electrons being modeled by the measurement chain shown in Fig. \ref{FigHCut2}. Note that two red arrows are shown. The bottom red line opts for a quantum model whenever possible, whereas the top red line opts for a classical model wherever possible.
%\end{comment}

%\Out{Discuss example 1/6}
The experiment begins with many lab operations. As discussed above, there is some freedom in picking where exactly the measurement chain starts. However, whatever one chooses, the preliminary lab operations can be described classically. Indeed, it is infeasible to model these lab operations with quantum theory. Recall that our purpose here is to provide an actual fully-modeled account of real-life experiments in order to extract statistical predictions from them. Thus both of the red arrows in Fig. \ref{FigHCut2} \textit{must} start on the top line.

%\Out{Discuss example 2/6}
These lab operations set up a current which travels through a filament in our cathode ray tube. This heats the filament which begins to thermally emit electrons. These electrons are then grabbed by an electric field and accelerated through a small aperture. All of these steps can be modeled classically without conceptual error or critical loss of accuracy. Hence the upper red arrow in Fig. \ref{FigHCut2} stays on the top row. All of these steps can be feasibly modeled quantumly. Hence the bottom red arrow in Fig. \ref{FigHCut2} jumps to the bottom row.

%\Out{Discuss example 3/6}
The next part of the experiment (the motion of these electrons through the double-slit apparatus) must be modeled with quantum theory. There are both conceptual issues and accuracy issues with modeling this part of the experiment classically. Hence both of the red arrows \textit{must} be on the bottom row here.\footnote{Note that there is nothing quantum per se about electrons moving through an aperture; Whether we can ignore quantum effects present at this point in the experiment depends sensitively on what's coming later. As I will discuss in Sec.~\ref{SecCoreExt}, this point is relevant for modeling experiments involving Wigner's friend.}

%\Out{Discuss example 4/6}
When the electrons reach the final screen, they enter into a semiconductor. There they are detected by causing a cascading avalanche of electric discharge. These electrons jumping over the semiconductor's band gap requires a quantum model. Hence both red arrows \textit{must} be on the bottom row here. 

%\Out{Discuss example 5/6}
However, once enough electrons are moving, we can describe them collectively as a small (but classical) current. This current activates a transistor. Some (but not all) transistors make use of quantum effects, but let's assume this one doesn't. All of these steps can be modeled classically without conceptual error or critical loss of accuracy. Hence the upper red arrow in Fig. \ref{FigHCut2} moves to the top row. All of these steps can feasibly be modeled quantumly. Hence the bottom red arrow in Fig. \ref{FigHCut2} moves along the bottom row.

%\Out{Discuss example 6/6}
The sequence of events which follows the activation of this transistor can all be described classically. Indeed, just as at the start of the experiment, it is infeasible to model the end of this experiment quantumly. As discussed above, there is some freedom in picking where exactly the measurement chain ends. However, whatever one chooses this part of the experiment can and must be described classically. Computer screens and humans and notepads are simply too large and complicated to model in a quantum way (at least for now and likely forever).

%\Out{Note that a pragmatic Heisenberg cut is pragmatically necessary when modeling any quantum measurement.}
%\begin{comment}
As this example hopefully makes clear, whenever we have a measurement chain, part of which requires a quantum model, we will have to at some point after this switch from modeling the measurement chain quantumly to non-quantumly (i.e., classically). In connection with the historical term, let us call wherever we happen to make this switch a \textit{Heisenberg cut}.\footnote{The view adopted here regarding Heisenberg cuts is compatible with how Heisenberg himself saw them. Before reading the following quote from Heisenberg it should be noted that for him the object–instrument divide and the quantum-classical divide coincide~\cite{SchlosshauerMaximilian2010WcDa}:
\begin{quote}
In this situation it follows automatically that, in a mathematical treatment of the process, a dividing line must be drawn between, on the one hand, the apparatus which we use as an aid in putting the question and thus, in a way, treat as part of ourselves, and on the other hand, the physical systems we wish to investigate. \dots The dividing line between the system to be observed and the measuring apparatus is immediately defined by the nature of the problem but it obviously signifies no discontinuity of the physical process. For this reason there must, within certain limits, exist complete freedom in choosing the position of the dividing line ~\cite[p.~3]{SchlosshauerMaximilian2010WcDa}.
\end{quote}
Note that for Heisenberg the location of the cut is not a physical discontinuity but is rather a free (albeit limited) choice made in the process of modeling.} When the pragmatic nature of this Heisenberg cut needs to be emphasized, I will describe it as a ``pragmatic Heisenberg cut''.
%\end{comment}

%\Out{Note also that the placement of the pragmatic Heisenberg cut reflects nothing fundamental/ontological}
This example should hopefully also make clear that there is nothing fundamental about the placement of the pragmatic Heisenberg cut. Indeed, one can believe the world to be quantum through-and-through and still make use of this cut for modeling purposes. Past the cut, we are no longer \textit{modeling} the measurement apparatus using quantum theory; This is very different from the measurement apparatus no longer \textit{being} quantum past the pragmatic Heisenberg cut.

%\Out{Entertain questions: But if the Heisenberg cut isn't fundamental, then why do we need it?}
At this point one may wonder: if the application of a pragmatic Heisenberg cut is a matter of non-fundamental pragmatic concern only, then do we really need it to make sense of the quantum measurement problem? One may ask: If we believe that the world is quantum through-and-through, then why would it be necessary to connect our quantum model of reality with a (known-to-be-incorrect) classical model of reality in order to model measurements within it? Can't we have a quantum-native understanding of quantum measurements?

%\Out{Re-iterate the difference between the pragmatic and realist problems. Making such a cut is pragmatically necessary.}
This line of questioning conflates the realist and pragmatic worries about quantum measurements which I have taken care to distinguish in Sec.~\ref{Introduction}: i.e., modeling versus understanding. If anyone wants to make such all-quantum-all-the-time demands on the realist side of the debate, they are more than welcome to. That is, one's ontological account of the measurement process might happen entirely within quantum theory. However, as the above discussion has hopefully shown, this attitude is not tenable on the pragmatic side. 

%\Out{There is nothing problematic about quantum theory depending on classical theory for its empirical support.}
It is no more problematic for quantum theory to depend on classical theory for its empirical support (and physical salience) than it is for general relativity to depend upon quantum theory (e.g., to model atomic clocks). Indeed, it is commonplace for scientific theories to depend on one another for metrological support; The biologist may rightfully outsource their metrological duties to an organic chemist when they are asked too many detailed questions about their measurement processes. What is important is that the scientific community collectively can give good models of its measurement processes. 

%\Out{Compare with mathematics: Full knowledge of a proof might be communal.}
The situation here is much like proof in mathematics, we do not require mathematicians to individually give all of the details of their proofs in terms of elementary logical operations. We do, however, demand that if we were to press the issue then they would collectively be able to give us such a long, detailed proof. Analogously, what we aspire to here is a computationally tractable connect-the-dots model-to-model account of real-life quantum experiments. This being possible is necessary in order to claim them as evidential support for quantum theory. My claim is that (at least currently and likely forever) a pragmatic Heisenberg cut is necessary for this. 

%\Out{The need for a Heisenberg cut might even be more than pragmatic. Reference Bohr's doctrine of classical concepts.}
It is perhaps possible (although I strongly doubt it) that we will one day be able to model the late parts of our experiments (including the experimenter) as quantum systems. However, even this possibility would not necessarily avoid the need for a Heisenberg cut. Suppose that one can somehow model an experiment up to and including the experimenter in a quantum way. It could still be the case that one can only parse the result of that experiment by means of taking some sort of classical approximation (i.e., taking a Heisenberg cut) on the experimenter right at the end~\cite{WallaceEmergentMultiverse}.\footnote{For a historical view of this kind, see Bohr's doctrine of classical concepts~\cite{SM2010}. It also should be noted that Bohmians may be able to avoid this last point: No last-minute classical approximation is needed on their theory since the classical result of the experiment (i.e., particle positions) is manifestly there in their description of the experiment's final state.} We cannot have a quantum-native understanding of measurement without a quantum-native understanding of the observer. Thus, in the absence of both tremendous computing capabilities and a quantum-native understanding of observers, taking a pragmatic Heisenberg cut is necessary for any satisfactory model of any quantum experiment.

%\Out{Finally, given the dangers involved our pragmatic Heisenberg cut out to be make explicitly and intentionally.}
In fact, not only is it pragmatically necessary to take a Heisenberg cut at some point, we ought to do so in an explicit and intentional way. Indeed, a mishandling of the pragmatic Heisenberg cut is one of the main dangers in trying to give a satisfactory model of quantum experiments. It is at the interface between our quantum and classical models that we need the most care both mathematically and conceptually. Handling this cut somewhere explicitly in the terms of either the dynamics or kinematics of our models is far superior to the sophomore's strategy of intuitively guessing. Indeed, as I will discuss in Sec.~\ref{SecCoreExt}, the success of the sophomore's hand-waving about PVMs/POVMs measurements is largely underwritten in terms of measurement chains and Heisenberg cuts. Before discussing this, however, it is worthwhile to provide a taxonomy of all the ways one might take a pragmatic Heisenberg cut.

\subsection{A Taxonomy of Heisenberg Cuts:\\ Vertical and Diagonal}\label{SecTaxonomy}
\begin{figure}
\includegraphics[width=0.45\textwidth]{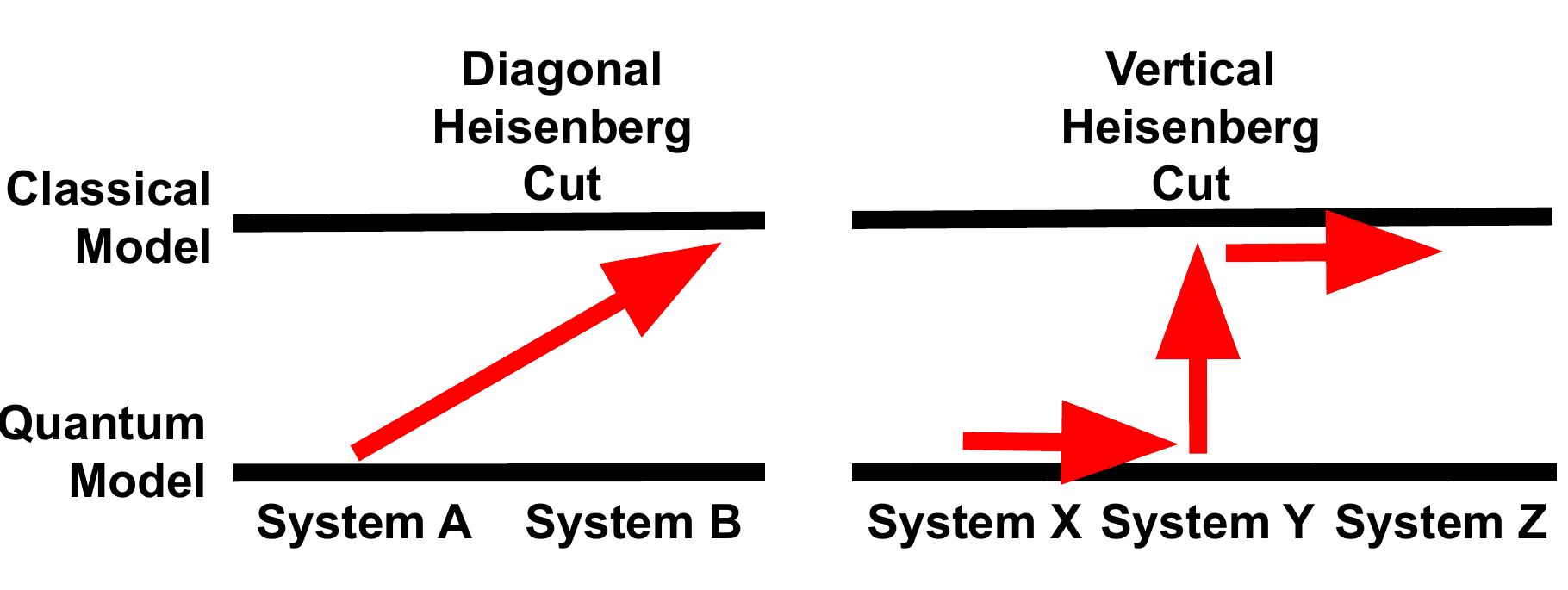}
\caption{The two possible ways of taking a pragmatic Heisenberg cut: diagonally (during an interaction) and vertically (in between interactions). On the left we have an example of a diagonal Heisenberg cut: system A is modeled quantumly and system B classically. Their interaction couples two systems modeled in different theories. On the right, we have an example of a vertical Heisenberg cut: The interaction between system X and Y is modeled within quantum theory, whereas the interaction between Y and Z is modeled classically. In between these interactions we apply some classical approximation scheme to Y while it is approximately isolated.}\label{FigHCut3}
\end{figure}

%\Out{Introduce the topic of this section: A Taxonomy of Heisenberg cuts: Vertical and Diagonal}
%\begin{comment}
As defined above, a pragmatic Heisenberg cut occurs wherever along the measurement chain we switch from modeling our experiment quantumly to classically. But how might we model our way across the quantum-classical divide? This subsection will introduce a useful taxonomy for classifying Heisenberg cuts. Given that a measurement chain is ultimately just a collection of interactions which are ordered in some way, there are only two ways to cross the divide: In between interactions, or during an interaction. Let us call these vertical and diagonal Heisenberg cuts respectively (for reasons which will become clear soon, see Fig.~\ref{FigHCut3}).
%\end{comment}

\subsubsection*{Vertical Heisenberg Cuts}
%\Out{Explain Vertical Cuts}
Regarding vertical Heisenberg cuts, consider a pair of interactions between three systems: system X (which we model quantumly) and system Y (which we can model either quantumly or classically) and system Z (which we model classically). We model the interaction between X and Y quantumly and the interaction between Y and Z classically. In between these two interactions (after X and before Z) we apply some classical approximation scheme to system Y in isolation. See the right side of Fig. \ref{FigHCut3} and notice that the red arrow moves vertically at system Y, hence the name ``vertical cut''.

%\Out{Give examples of vertical cuts}
%\begin{comment}
Some examples of vertical Heisenberg cuts of varying quality are:\footnote{We might also have vertical Heisenberg cuts which proceed in the reverse direction. That is, we may also have principled ways of mapping classical states onto quantum states. For example, the reverse of each of the above discussed examples are sometimes justified.}
\begin{enumerate}
 \item[1)] taking some sufficiently decohered quantum state, and using the Born rule to map it onto a probability distribution over classical states,
 \item[2)] taking a quantum state whose Wigner function~\cite{QMPhaseSpace} (i.e., the state's quasi-probability distribution in phase space) happens to be positive and reinterpreting it as a genuine probability distribution over a classical phase space,
 \item[3)] taking a minimum uncertainty quantum state and mapping it onto the definite classical state with matching expectation values.
 \item[4)] taking a Bohmian state (i.e., a wavefunction plus particle positions) and discarding the wavefunction for future calculations.
\end{enumerate}
among many other possibilities~\cite{Rosaler}.
%\end{comment}

%\Out{The validity of these approximations is fixed by experimental practice.}
In general, making such vertical cuts will be justified to differing degrees in different contexts. In order to address the pragmatic measurement problem, it is crucial that we understand when such classical approximations are and are not pragmatically justified. It is by-and-large experimental practice which grounds our knowledge of the regimes of applicability of such approximations~\cite{Curiel}.

%\Out{Note that the quantum-to-classical transition here happens at the level of kinematics.}
Vertical cuts push the quantum-to-classical transition onto the kinematics (as opposed to the dynamics). To see this, note that if one's measurement chain contains only vertical cuts then every system-to-system interaction is modeled as either classical-to-classical or quantum-to-quantum. Somewhere along the measurement chain, the state of some system must be able to be accurately (and feasibly) modeled in both ways.

\subsubsection*{Diagonal Heisenberg Cuts}
%\Out{Introduce diagonal cuts by pointing to the figure}
Regarding diagonal Heisenberg cuts, consider an interaction between system A (which we model quantumly) and system B (which we model classically). See the left side of Fig. \ref{FigHCut3} and notice that the red arrow moves diagonally between systems A and B, hence the name ``diagonal cut''. 

%\Out{Note that the quantum-to-classical transition here happens at the level of dynamics.}
Unlike with the vertical cuts discussed above, diagonal cuts push the quantum-to-classical transition onto the dynamics (as opposed to the kinematics). To see this, note that if one's measurement chain contains only diagonal cuts then every system is modeled as either classical or as quantum. Diagonal Heisenberg cuts occur wherever there is a dynamical quantum-to-classical interface.

%\Out{Give a simple example of a diagonal cut.}
%\begin{comment}
As a simple (and admittedly artificial) example consider the following pair of coupled differential equations:
\begin{align}
\nonumber
\partial_t\ket{\psi_\text{A}(t)} 
&= \left(\frac{\hat{p}_\text{A}^2}{2\,m_\text{A}} + U_\text{A}(\hat{x}_\text{A})+V(\hat{x}_\text{A}-y_\text{B}(t))\right)\ket{\psi_\text{A}(t)}\\
m_\text{B}\,\partial_t^2y_\text{B} 
&= -\partial_{y_\text{B}} U_\text{B}(y_\text{B})-\partial_{y_\text{B}}V(y_\text{B}-\langle\hat{x}_\text{A}(t)\rangle)
\end{align}
for some potential functions $U_\text{A}$, $U_\text{B}$, and $V$. Here we have a wavefunction, $\ket{\psi_\text{A}(t)}$, and a classical position, $y_\text{B}(t)$, each evolving under their free dynamics, $U_\text{A}$ and $U_\text{B}$, plus an interaction term, $V$. 
%\end{comment}

%\Out{Note that this simple example has back reactions and can be called semi-classical.}
Note that the dynamics of $\ket{\psi_\text{A}(t)}$ depends on $y_\text{B}(t)$. Note also that the dynamics of $y_\text{B}(t)$ depends on $\ket{\psi_\text{A}(t)}$ through its expectation value, \mbox{$\langle\hat{x}_\text{A}(t)\rangle=\bra{\psi_A(t)}\hat{x}_\text{A}\ket{\psi_\text{A}(t)}$}. These equations describe a two-way dynamical interface here between a quantum and a classical system (n.b., this model includes back-reactions). Such a quantum--to-classical coupling is typically called semi-classical.

%\Out{Talk about how one can go beyond a semi-classical treatment and invoke heuristic toy models. New physics is allowed so long as it is well-controlled and well-understood.}
%\begin{comment}
Beyond such semi-classical treatments, however, there has been a significant amount of research into hybrid theories which mix quantum and classical systems in more substantial ways. For an overview, see \cite{PhysRevA.86.042120} and references therein. To summarizing their findings:
\begin{quote}
Whereas it is certainly possible to construct hybrid systems, these constructions typically ask for the introduction of hybrid concepts absent in a straight classical-quantum product. These hybrid theories are not derivable from a straightforward purely quantum theory: They incorporate new physics. This explicitly warns us about the toy-model nature and heuristic character of the different frameworks analyzed above.~\cite[p.~3]{PhysRevA.86.042120} 
\end{quote}
Fortunately, however, for the purposes of modeling experiments heuristic toy-models which introduce new physics are allowed. Namely, they are allowed so long as the errors introduced by this ``new physics'' are small-enough, well-understood, and well-controlled. Roughly, we can use such toy models so long as the model-induced errors bars in the theoretical prediction are smaller than the experimental error bars.
%\end{comment}

\subsection{Addressing the Core and Extended Pragmatic Measurement Problems Non-relativistically}\label{SecCoreExt}
%\Out{Name drop the core and extended pragmatic measurement problems.}
Having introduced measurement chains and a taxonomy of Heisenberg cuts, we are now in a position to see how they can help us address the pragmatic measurement problem for non-relativistic quantum theory. Before this, however, allow me to distinguish between an easier and a harder version of this problem. I will call these the core and extended pragmatic measurement problems respectively. 

\subsubsection*{Introducing and Solving the Core Problem}
%\Out{Review what is at stake and discuss the minimal ``core'' solution}
At its core, what the pragmatic measurement problem threatens is our theory's evidential support and thereby its empirically supported connection with reality. As such any minimal solution to these worries must give us a \textit{measurement framework}: a satisfactory account of how to model the measurement processes of at least the theory's key experimental successes, potentially on a case-by-case basis. The task of developing such a measurement framework is the \textit{core pragmatic measurement problem}. Solving the core problem would restore empirical support to our theory's key experiments.

%\Out{Introduce road map metaphor}
It is easy to see how measurement chains and Heisenberg cuts can be used to solve the core pragmatic measurement problem for quantum theory: Analyzing any given quantum experiment in these terms gives us a road map to guide us in modeling its specific measurement processes. In particular, these road maps have the dangerous areas ahead clearly marked out (i.e., the quantum-classical divide). Fortunately, these maps also provide us with multiple possible routes for navigating around these dangers (i.e., we have a taxonomy of Heisenberg cuts). 

%\Out{Gesture at how they solve the core problem}
Using these road maps, we can go about giving satisfactory predictions for quantum experiments and gaining evidential support from them, at least on a case-by-case basis. This already gives us a working measurement framework for non-relativistic quantum theory. We are already in a much better position than relying on the sophomore's strategy of: 1) hoping that the measurement process in question can be modeled with a PVM, and 2) hoping that they have guessed the right PVM and applied it at the right time.

\subsubsection*{Introducing the Extended Problem}
%\Out{Solving the core problem doesn't allow us to talk about observables. This is the task of the extended problem.}
It should be noted, however, that solving the core problem does not allow us to say anything about measurement processes in general. For this, one would need a \textit{measurement theory}: a principled account regarding how to model all (or nearly all) of the theory's measurement processes in a holistic and wide-scoping way. The task of developing such a measurement theory is the \textit{extended pragmatic measurement problem}.

%\Out{The extended problem also allows for a division of labor between theory and experiment. Let's see how...}
As I will now discuss, solving the extended pragmatic measurement problem allows for the typical division of labor between theorists and experimenters. To see this first recall from above that in order to secure empirical support from an experiment it is necessary to model every part of the measurement chain in at least some dynamical detail. Let us briefly consider three cases of how this modeling burden might be split between the theorists and the experimenters.

%\Out{1) The theorist could do all of the work.}
In the first case, the modeling burden falls entirely on the theorist. They would then need to make full predictions of experimental outcomes: e.g., ``After a duration of one hour (as counted by this specific kind of clock) the gas will have this pressure (as measured by this specific kind of pressure gauge)''. This doesn't sound like the sort of thing theorists typically do. Typically, they share the modeling burden with the experimenter. 

%\Out{2) Or the modeling burden could be split. The theorist talks in terms of observables, relying on the fact that the experimenter can parse them} 
In the second case, the modeling burden is split between the theorist and the experimenter. In this case the theorist can talk directly in terms of observables making half-way predictions: e.g., ``After a duration of one hour, the gas will have this pressure. (\textit{Implicitly:} I trust that you, the experimenter, know what I mean and have robust techniques for measuring what I am calling `durations' and `pressures'.).'' Of course, while the theorist is here omitting many metrological details, they don't disappear; Instead, they must be accounted for by the experimenter. Namely, it then falls upon the experimenter to set up an appropriate (and sufficiently well-modeled) measurement apparatus.\footnote{One might want to include among the observables quantities which are only observable in principle or hypothetically. For instance, gravitational waves were hypothetically observable long before they were actually observed. Talk of hypothetical observables is meaningful insofar as we can reasonably expect that experimenters could, in principle, measure it.} Ultimately, the theorist's ability to talk so casually in terms of observables rests upon an (often under-discussed) mountain of experimental practice.

%\Out{3) Or the experimenter could do it all themselves. The theorist can speak casually in terms of observable, secure in the knowledge that ``someone, somewhere, knows how to measure something.''} 
A third possibility is that the experimenter takes on the full burden themselves, addressing the whole of their experiment from its initialization to its final outcome. In this scenario the theorists only play a background role working on the theories which the experimenter invokes in generating their models and making their predictions. Otherwise, the theorist is free to theorize. As in the previous case, the theorist here can adopt a habit of speaking casually in terms of observable, secure in the knowledge that ``someone, somewhere, knows how to measure something''.\footnote{This phrase is taken from a talk~\cite{FewsterRQITalk3} given by Chris Fewster. His work on the pragmatic QFT measurement problem will be discussed further in Sec.~\ref{StateOfTheArt}.}

%\Out{In sum: casual talk of observables requires a wide-scoping measurement theory. Let's next build a measurement theory and identify some observables.}
As these three scenarios demonstrate, the theorist's habit of casually talking in terms of observables depends upon the existence of a wide-scoping measurement theory for the theory in question. The remainder of this section will be spent developing a measurement theory for non-relativistic quantum theory. Following this, in Sec.~\ref{SecObs} I will then use this measurement theory to help us identify the observables of non-relativistic quantum theory (n.b., ``observables'' $\neq$ ``self-adjoint operators'').

%\Out{Before this, however, let us reflect on whether we can expect to find nice answers, a priori. I don't think we can.}
Before addressing the extended pragmatic measurement problem, however, it is worth briefly reflecting on the following two questions: Should we expect that our scientific theories generically have a unified wide-scoping account of their measurement processes (e.g., our PVM/POVM measurement theory)? And should we expect this measurement theory to give rise to a nice and tidy characterization of its observables (e.g., that they are subset of the POVMs)? I see no reason why we should expect either of these in general, (e.g., for QFT, or for quantum gravity); The fact that non-relativistic quantum theory has both of these features is remarkable to me. 

\subsubsection*{Solving the Extended Problem}
%\Out{Phrase the extended problem in terms of road maps. We need an always-available way of crossing the river}
What would it take to solve the extended pragmatic measurement problem for non-relativistic quantum theory? As I discussed in Sec.~\ref{SecFirstExamples}, when modeling experiments involving quantum systems one must in practice make a Heisenberg cut somewhere along the measurement chain. Continuing the road map metaphor introduced above, it is as if there is a long river (the quantum-classical divide) which we are required to cross somewhere. The above-discussed measurement framework gives us, for each experiment, various routes and crossing methods: We might ford the river here or we might swim across there. In order to upgrade this measurement framework into a measurement theory, we need to identify some way of crossing this river which is near-universally applicable: e.g., one can always take the ferry. 

%\Out{One way stands out: The decoherence way (plus the Born rule).}
In order to find a measurement theory for non-relativistic quantum theory, we need to find some standardized way of making Heisenberg cuts with near-universal applicability. We are tremendously lucky that this is possible. Indeed, in terms of wide applicability, one way of crossing the quantum-classical divide stands out from the rest:\footnote{As I mentioned in Sec.~\ref{SecTaxonomy}, there are many other ways of making pragmatic Heisenberg cuts. For instance, given a Bohmian state one might be justified in simply discarding the wavefunction from future calculations, keeping only the particle positions. Such a Bohmian Heisenberg cut might be justified under a wide range of conditions within non-relativistic quantum theory. However, its regime of applicability is limited to non-relativistic scenarios which do not cover a large portion of quantum theories empirical success~\cite{WallaceBlueSkyPaper,WallaceBlueSkyTalk}. Hence, Bohmian Heisenberg cuts are insufficient for my goal of recovering quantum theory's empirical support. More to the point, it is insufficient for my goal of learning lessons which are applicable to quantum field theory.} namely, by using decoherence theory and then the Born rule. As I will now discuss, by making such Heisenberg cuts one can justify (at least pragmatically) the sophomore's casual use of our usual PVM/POVM measurement theory.\footnote{It should be stressed that within my analysis the only thing special about crossing the quantum-classical divide in this decoherence way is its general applicability. If one could find another equally general way of crossing the quantum-classical divide, one can equally well use that to underwrite a different complementary measurement theory. This is an interesting possibility which may shed new light on justifications for the Born rule. Unfortunately, however, this falls outside of the scope of this paper.}

%\Out{When are we allowed to make such vertical cuts? Once enough decoherence has happened.}
Let us now restrict our attention to vertical Heisenberg cuts which are facilitated by decoherence theory and the Born rule. Where along the measurement chain are we justified in taking this specific kind of pragmatic Heisenberg cut? Luckily, to this we have a general answer: one can take such a pragmatic Heisenberg cut once enough decoherence has occurred that the possibility of spontaneous wide-scale recoherence (although not mathematically impossible) is practically inconceivable. That is, for modeling purposes, it does not matter where we put such a pragmatic Heisenberg cut so long as it is at a scale where quantum effects are (and will forever remain) irrelevant in practice.

%\Out{Explain ``and will forever remain''}
The above ``and will forever remain'' caveat is a critically important one. Indeed, the validity of such a Heisenberg cut will always depend on the context surrounding the measurement procedure under consideration. In particular, one cannot simply decide in the middle of modeling a quantum measurement to take such a Heisenberg cut without knowing beforehand what the rest of the measurement procedure will be like. 

%\Out{Introduce Wigner's friend-like puzzles.}
No matter how small quantum coherence effects appear to be in the middle of an experiment, there is always a possibility that the coherence effects are brought back to their full force. (Such a carefully orchestrated large-scale recoherence is, in fact, exactly what quantum computers are designed to do.) Moreover, even if within one experiment the quantum coherence effects never again become relevant, they may once again become relevant in other future measurements involving correlated systems. The consideration of measurements made by observers who themselves live inside of a giant quantum computer capable of wide-scale (observers included) recoherence, leads to interesting Wigner's friend-like puzzles.

%\Out{Putting these caveats aside, we have a wide-scoping measurement theory.}
As this caveat shows, the strategy of applying decoherence theory and then the Born rule is not applicable in all conceivable measurement scenarios. With these caveats noted, however, once enough decoherence has occurred one can take such a vertical Heisenberg cut by applying the Born rule. This method of analyzing measurement processes is near-universal in scope and so hence gives us, as desired, a measurement theory for non-relativistic quantum theory.

%\Out{Overview how a generic measurement can be analyzed in these terms.}
Let us next work out what measurement theory this is specifically. One can imagine modeling a generic measurement process along the following lines: The first steps of the measurement process transfer quantum information about the to-be-measured system into the measurement apparatus. This part of the measurement process can be modeled unitarily. Then decoherence happens, diagonalizing the state of the apparatus in whatever basis is dynamically picked out by the decoherence process. (See Sec. 1 of~\cite{sep-qm-decoherence}). This part of the measurement process can be modeled as projective. One then takes a vertical Heisenberg cut by applying the Born rule and models the rest of the experiment classically.

%\Out{This justifies us modeling measurements as POVMs. (Introduce these mathematically.)}
Combining all of these steps together (using Naimark's dilation theorem) one finds that the total effect can be modeled as a Positive Operator-Valued Measure (POVM). For notational simplicity, let us assume that there are only a finite possible number of outcomes (indexed by $\alpha$ in some alphabet, $\alpha\in A$). A POVM is then a collection of operators, $\{\hat{E}_\alpha\}$, with $0\leq \hat{E}_\alpha\leq \hat\openone$ and $\sum_{\alpha\in A} \hat{E}_\alpha=\hat\openone$. When such a measurement is applied to a given quantum state, $\hat\rho$, Born's rule says that the outcome labeled $\alpha$ occurs with probability $p_\alpha=\text{Tr}(\hat{E}_\alpha\hat\rho)$.

%\Out{Distinguish POVMs from PVMs and note that PVMs are thermodynamically impossible}
One notable fact about POVMs is that they are non-ideal in the following sense: Repeating the same POVM measurement twice back-to-back might yield different results each time. There is, however, a special subset of the POVMs (namely, the Projection-Valued Measures or PVMs) for which repeated measurement yields a fixed result. Such ideal measurements occur when the POVM operators, $\{\hat{E}_\alpha\}$, are a collection of orthogonal projectors, $\{\hat\pi_\alpha\}$. While such a PVM treatment of measurement is theoretically convenient, it ought to be thought of as an idealization, which is strictly speaking impossible. Indeed, ideal projective measurements have infinite resource costs and violate the third law of thermodynamics~\cite{Guryanova2020idealprojective}.

%\Out{Answer the sophomore's questions. Give POVM/PVM criteria. The specific POVM must be picked out by studying the dynamics.} 
We thus have some rough answers to the sophomore's questions: One is justified in modeling a quantum measurement as a POVM when a sufficient amount of decoherence has occurred (and Wigner's friend isn't around). One is moreover justified in using a PVM when one's measurement apparatus is sufficiently ideal (even if this is, strictly speaking, impossible). Let us call these the POVM and PVM criteria respectively. The sophomore is next asked: Supposing these criteria are satisfied, which PVM/POVM is one allowed to use, and how do we know this? Unsurprisingly, the answers to these questions follow from investigating the dynamical details of the particular measurement apparatus in question.

%\Out{Stress three surprising (and contingent) aspects of the above story. These underwrite our casual use of PVMs/POVMs.}
Three surprising aspects of the above story ought to be stressed. Firstly, it is surprising that such a uniform treatment of measurement processes arises from the mathematics of quantum theory (after assuming that the PVM/POVM criteria are satisfied); Secondly, it is surprising that such criteria exist which allow for such a broad and uniform analysis of measurement processes. Thirdly, it is a surprising contingent fact about the world that the PVM/POVM criteria are so very often satisfied in real-life experimental scenarios. It is these remarkable facts which underwrites (at least pragmatically) the sophomore's casual use of our usual PVM/POVM measurement theory.

%\Out{Summarize our present results.}
Our present results can be summarized as follows: Under minor assumptions, one is justified in modeling nearly any quantum measurement process as a POVM with the specific POVM being determined by detailed dynamical considerations. Indeed, it is the design of the measurement apparatus and the nature of its interaction with the to-be-measured system and its environment which determines what is being measured.

%\Out{The sophomore doesn't need to guess anymore.}
If the sophomore takes this lesson to heart, then all of their modeling issues are solved. Recall from Sec.~\ref{Introduction} that the primary issue with the sophomore's approach is that they are guessing: Their choice of PVM/POVM had no grounding in experimental practice. If, however, the sophomore were to follow along with the above story, then they might justify both the form of their guess (a PVM/POVM) and the specific PVM/POVM which they have guessed; This would make them guesses no more.

\subsection{Observables in Non-Relativistic Quantum Theory}\label{SecObs}
%\Out{Let us next revise the sophomore's understanding of observables (currently, ``observables'' = ``self-adjoint operators'').}
Before applying these lessons in a QFT context, let us first (partially) revise the sophomore's understanding of the term ``observables''. (A full revision must wait until Sec.~\ref{ImpossibleMeasurements}.) Recall from Sec.~\ref{Introduction} that the sophomore has been taught the following tautologies: All measurements are of some observable and all observables are measurable. Further, the sophomore has been taught (incorrectly) to equate the term ``observables'' with the term ``self-adjoint operators''. Hence, the sophomore currently thinks of measurements primarily in terms of self-adjoint operators. 

%\Out{Give angle example to prioritize PVMs over self-adjoint operators.}
After a bit of reflection, however, the sophomore can be convinced that it is better to instead think of measurements directly in terms of PVMs. As I discussed in Sec.~\ref{SecCoreExt}, the set of projectors, $\{\hat{\pi}_\alpha\}$, which an idealized measurement device implements is (at least, approximately) fixed by the dynamical details of the measurement process itself. The same is not true for any real values, $q_\alpha$, which may be associated with the measurement outcomes, $\alpha\in A$. To see this, imagine a measurement of some angle which is ultimately displayed by the position of an indicator needle against some marked scale. This scale is marked in three ways: in both radians and degrees as well as with a uniform marking of $\text{sin}(\theta)\in[0,1]$. For this measurement, it is ambiguous what the values, $q_\alpha$, should be in this case, but it remains clear what the projectors, $\hat\pi_\alpha$, are.

%\Out{Break the obsession with real-valued measurements.}
Indeed, it is not hard to think of measurements whose outcomes, $\alpha\in A$, cannot be reasonably associated with any real values, $q_\alpha\in\mathbb{R}$. For instance, a measurement outcome might be indicated by some flashing lights which are labeled with letters, not numbers. Alternatively, the flashing lights could be labeled by complex numbers. In general, these labels $\alpha\in A$ could be structured (or unstructured) in absolutely any way one wishes. Namely, these labels may or may not facilitate the computation of meaningful expectation values. 

%\Out{Break the obsession with expectation values.}
This is an important lesson as some physicists mistakenly think of observables as maps from quantum states into real expectation values. Namely, $\hat{Q}:\rho\mapsto\langle\hat{Q}\rangle\coloneqq\text{Tr}(\hat{Q}\,\hat{\rho})$ with $\langle\hat{Q}\rangle\in\mathbb{R}$ being the expectation value of the measurement of $\hat{Q}$. This understanding of measurement is incorrect principally because it is too narrow; Not all measurements admit value labels $q_\alpha$ which facilitate meaningful expectation value.\footnote{Moreover, while one can in general extract a unique PVM from a self-adjoint operator $\hat{Q}$, one cannot do so in general for POVMs; multiple distinct sets of POVMs, $\{\hat{E}_\alpha\}$, might sum to the same $\hat{Q}=\sum_\alpha q_\alpha\hat{E}_\alpha$ and hence yield exactly the same expectation values. Thus, pinning down an expectation value $\langle\hat{Q}\rangle$ does not fully specify the measurement.  Indeed, if one cares about post-measurement state updates then one must even go beyond POVMs. Namely, one must additionally specify a set of operators, $\{\hat{M}_\alpha\}$, with $\hat{E}_\alpha=\hat{M}_\alpha^\dagger \hat{M}_\alpha$.} There is something, however, which is right about the thought that observables are maps from quantum states into real numbers. Namely, any set of POVMs $\{\hat{E}_\alpha\}$ can be thought of as a set of maps from quantum states into real numbers, specifically the interval $[0,1]\subset \mathbb{R}$. Namely, they can be thought to act on states as $\hat{E}_\alpha:\rho\mapsto p_\alpha\coloneqq\text{Tr}(\hat{E}_\alpha\,\hat{\rho})$ with $p_\alpha\in\mathbb{R}$ being the probability of outcome $\alpha$. Thus, there is nonetheless a sense in which real numbers have a special place in quantum measurements. This is not because the value labels of the measurement outcomes, $q_\alpha$, must be real but rather because measurement probabilities, $p_\alpha$, must be real.

%\Out{Let us leave the sophomore here (for now) with ``observables'' = ``POVMs''.}
Taking the above lessons into account, the sophomore might now revise their view to equate the term ``observables'' with POVMs instead of self-adjoint operators. This is a step in the right direction, but as I will discuss in Sec.~\ref{ImpossibleMeasurements}, this is still not quite right: Not all POVMs are measureable, some are dynamically forbidden on grounds of symmetry, thermodynamics, and/or relativity. Further discussion of these impossible measurements and how we ought to respond to them, however, is best delayed until after we have moved into a QFT context.

\section{Approaching The Pragmatic QFT Measurement Problem}\label{GenChainsAndCuts}
%\Out{Reintroduce the Pragmatic QFT Measurement Problem}
The previous section has reviewed how measurement chains and Heisenberg cuts can be leveraged in order to solve the pragmatic measurement problems (both core and extended) for non-relativistic quantum theory. The rest of this paper will be spent applying the lessons we have learned so far to the \textit{The Pragmatic QFT Measurement Problem}: How should one model measurement processes involving quantum fields? How must our measurement framework for QFT differ from our usual non-relativistic measurement framework? Can we model QFT-involved measurements using our usual measurement theory based on PVMs and POVMs? How must our characterization of the observables of QFT differ from the way we characterize the observables of non-relativistic quantum theory?

%\Out{Can't we just use the old solution? No because of Sorkin (see later in the section)}
One's first thought may be as follows: Given that we have just solved the pragmatic measurement problem for non-relativistic quantum theory, can't we just transfer this solution over to QFT? Unfortunately, we cannot. As I will discuss in Sec.~\ref{ImpossibleMeasurements} in QFT almost all localized projective measurements violate causality, allowing for faster-than-light signaling; These are Sorkin's impossible measurements. Thus, the story of measurement in QFT cannot end with a projective measurement theory as it did before. Fortunately, however, not much else in the non-relativistic story needs to change as we move into QFT. Namely, I will now argue that we ought to begin by using measurement chains and cuts to establish a case-by-case measurement framework for QFT. From here, we can then strive for a wide-scoping measurement theory capable of modeling all (or nearly all) measurement processes.

\subsection{Correcting A Sophomoric Approach to the Core Pragmatic QFT Measurement Problem}\label{SecIIIA}
%\Out{Return to a sophomoric approach to modeling QFT-experiments}
As before, it will be helpful to begin by discussing a sophomoric approach to modeling measurements which (hopefully nearly) everyone finds dissatisfying. How are sophomores taught to model measurements involving quantum fields, for instance, in high-energy particle physics experiments? A sophomore might be taught to model such experiments as follows: One is first given some input states (e.g., spin-states, kinetic energies, relative phases) of some inbound particles. From here, one can use the given dynamics to compute the scattering amplitudes which emerge from their collision. Specifically, one computes the amplitude which is outbound within some solid angle (namely, in the direction of the experiment's particle detector). Finally, one applies the given PVM/POVM measurement (of particle number, or phase, or quadrature, etc.) to determine the detector's response rate.

%\Out{The sophomore runs into the same issues here as in the non-relativistic context}
Hopefully, this approach to modeling measurements is just as unsatisfying as it was in the non-relativistic case. Indeed, it fails for exactly the same reasons that the sophomore's quantum theory does (see Sec.~\ref{Introduction}). While the relevant mathematical objects may be given to the sophomore on their problem sheets, they will have no good answer to the following questions: Under what conditions is it appropriate to model this particle detector as implementing a POVM on the field? If it is appropriate, then exactly which POVMs am I allowed to use and when? How can one go about determining approximately which observable this apparatus measures? As before, if the sophomore is to do better than guessing, they will need to model in some detail the relevant measurement process. But how exactly should one go about modeling measurements involving quantum fields? This is the Pragmatic QFT Measurement Problem.

%\Out{The sophomore might guess what a Geiger counter measures. Their guess is strictly speaking impossible, but it still might be justified.}
To make things concrete, let us suppose the sophomore is confronted with an experiment in which $\beta$-radiation is picked up by a Geiger counter. The sophomore might intuitively guess that the Geiger counter is doing something like a local particle number measurement. Indeed, this is effectively what happens. (This, despite the fact that there are no well-defined local number operators in QFT~\cite{Redhead1995}.) The sophomore is able to intuitively guess roughly what the Geiger counter measures in just the same way they might intuitively guess that in the double-slit experiment the detection screen implements a measurement in the position basis. As before, however, a theory-to-experiment linkage which relies so blatantly upon intuitive guessing simply cannot do the work we require of it. The sophomore must justify their guess.

%\Out{Growing wise, the sophomore might outsource. Someone, however, owes us a satisfactory model.}
If pressed on this question the sophomore might give the following answer: ``You are asking the wrong person. The details regarding which quantum measurement the particle detector implements (including its fidelity and error rates) can be found written on the box in which it was delivered. Moreover, this tomographic information itself was painstakingly gathered by the manufacturer as a part of their quality control measures.'' As I discussed in Sec.~\ref{SecFirstExamples}, such an outsourcing answer is allowed. We must then follow up with the manufacturer. At some point, however, somebody is going to have to give us a satisfactory model of this Geiger counter as it sits in some measurement chain.

%\Out{Modeling QFT-involved experiments will require a QFT-cut. Our QFT-cuts ought to be out in the open}
Thus, we are led as before to address the pragmatic measurement problem in terms of measurement chains. In Sec.~\ref{SecFirstExamples} I argued that in modeling non-relativistic quantum experiments it is pragmatically necessary to make a Heisenberg cut (i.e., to cross the quantum-classical divide) at some point along the measurement chain. By the exact same reasoning, it is necessary to take a QFT-cut (i.e., to cross the QFT-non-QFT divide) somewhere along the measurement chain when modeling for any QFT-involved experiment. Indeed, just as in the non-relativistic context, one ought to do so explicitly and intentionally. It is at the interface of our QFT and non-QFT models that we need the most care both mathematically and conceptually. Handling one's QFT-cut intentionally and explicitly is much better than the sophomoric strategy of guessing which measurement is being implemented.

%\Out{As before these QFT-cuts will give us a road map for solving the core problem.}
As in the non-relativistic case, one can use these notions of chains and cuts to solve the core pragmatic QFT measurement problem. As before, what is at stake in the core problem is the linkage between theory and reality which is mediated by experimental practice. Without this connection, quantum field theory would lack empirical support and physical salience. As before, any minimal solution to these worries must give us a measurement framework for QFT: i.e., a satisfactory account of how to model the measurement processes of at least the QFT's key experimental successes, potentially on a case-by-case basis. Continuing the road map metaphor introduced in Sec.~\ref{SecCoreExt}, a good understanding of measurement chains and QFT-cuts would give us a road map for modeling any given QFT-involved experiment. Namely, these road maps would have the dangerous areas ahead clearly marked out (i.e., the QFT-non-QFT divide) as well as multiple routes around them (i.e., various possible QFT-cuts).

Indeed, in Sec.~\ref{StateOfTheArt} I will discuss some of the tools which physicists have available for making QFT-cuts. In my (well-informed but non-expert) opinion, the tools which we currently have available are collectively of wide enough scope to give us a good working measurement framework for QFT's current experimental successes.\footnote{This is not to say that I am validating current experimental practice. Ultimately, I am not qualified to evaluate the modeling capacities of the current state of experimental practice/methodologies. I am qualified, however, to make the following methodological points: Satisfactory modeling of QFT-involved experiments requires that we take QFT-cuts explicitly and openly at some point along the measurement chain. Moreover, if we are to unify our current patchwork of modeling techniques into a unified measurement theory, then we will need to find some way of making QFT-cuts which is applicable to all (or nearly all) measurement processes.} I see nothing which would prevent the engineers at the LHC or experimenters working in quantum optics from giving satisfactory models of their QFT-involved measurement apparatuses on a case-by-case basis. Recalling the above-discussed map metaphor, I trust that they know the lay of the land, at least currently. 

In sum, in my (again, well-informed but non-expert) assessment, we currently have a working solution to the core of the Pragmatic QFT Measurement problem. Importantly, however, the truth of this claim is liable to change over time. Higher precision measurements may require not just higher precision modeling but also brand-new modeling techniques. For instance, experimentalists are currently pushing our time resolution in atomic measurement to below the light-crossing time of the atom. Taking the radius of a Silicon atom to be $r=0.21\text{ nm}$, its light crossing time is $t = 2\,r/c = 1.4$ atto-seconds. For comparison, a recent experiment timing quantum tunneling events has a precision of a few atto-seconds~\cite{Yu2022}.

Consider now some experiment which records with such high precision the exact time when a photon is absorbed by a Silicon-based photo-detector. It will soon be possible via precision timing to determine whether the observed photon was absorbed by the front-half or by the back-half of the Silicon atom which it first met in the detector. Clearly, a non-relativistic understanding of electrons in orbitals is insufficient to model this detection process. Instead, it will require a QFT-level understanding of atomic orbitals which we currently lack. Hence, in the near future, our current modeling practices may become insufficient. There is nothing too alarming about this, however. It is the normal course of science for precision experimentation and precision modeling to develop side-by-side.

This example does expose, however, some ways in which having a measurement framework which solves only the core pragmatic measurement problem is less than ideal. Indeed, as this example shows, the sufficiency of any given measurement framework to cover the theory’s current experimental successes may only be temporary. Additionally, a solution to just the core problem does not allow us to do certain theoretical work regarding measurement processes generally (recall that measurement frameworks work on a case-by-case basis and have a limited scope). In particular, a solution to just the core problem does not allow us to characterize or talk abstractly about the observables of our theory.

Thus, as in the non-relativistic case, we may want to go beyond merely having a (temporary) solution to the core pragmatic measurement problem. Namely, we have substantial motivation to solve the extended pragmatic measurement problem by developing a unified measurement theory for QFT. At this point one might ask: Can't we just apply the projective measurement theory which we developed in Sec.~\ref{ChainsAndCuts}? Can't we model the detection of a photon in the front-half of a silicon atom by some PVM/POVM localized there? And the detection on the back-half of the atom by a different PVM/POVM localized there? Unfortunately, we cannot. As I will discuss in the next subsection in QFT almost all localized projective measurements violate causality, allowing for faster-than-light signaling. These are Sorkin's impossible measurements.

\subsection{More Sophomoric Issues:\\ Impossible Measurements and Observables}\label{ImpossibleMeasurements}
%\Out{Return to the sophomore's current understanding of observables. They would be shocked that some POVMs are unobservable.}
%\begin{comment}
Let us return to where we left our sophomore in Sec.~\ref{SecObs}. The sophomore knows the following tautologies: All measurements are of some observable and all observables are measureable. Furthermore, as we left them the sophomore currently equates the term ``observables'' with the term ``POVMs''. Given this, the sophomore would be shocked to learn that some measurements are impossible (or, equivalently, that some observables are unobservable). A much-discussed example of impossible measurements in the context of QFT are Sorkin's impossible measurements. Before discussing this, however, it should be noted that non-relativistic quantum theory has its own impossible measurements. As I mentioned in Sec.~\ref{SecObs}, some POVMs (namely, the PVMs) are impossible to measure on thermodynamic grounds. They have infinite resource costs and violate the third law of thermodynamics~\cite{Guryanova2020idealprojective}. 

Alternatively, the measurement of certain POVMs might violate other laws of physics (e.g., conservation laws and/or super-selection rules). To see this, first note that under charge-conserving dynamics it may be impossible for states with different charges (e.g., $\ket{q}$ and $\ket{q'}$) to be put into superposition. Now consider the self-adjoint operator $\hat{Q}=\ket{q}\bra{q'}+\ket{q'}\bra{q}$. The eigenstates of $\hat{Q}$ are disallowed by such a super-selection rule for charge. Hence, state update under the PVM corresponding to $\hat{Q}$ is dynamically impossible.

\begin{figure}
\includegraphics[width=0.45\textwidth]{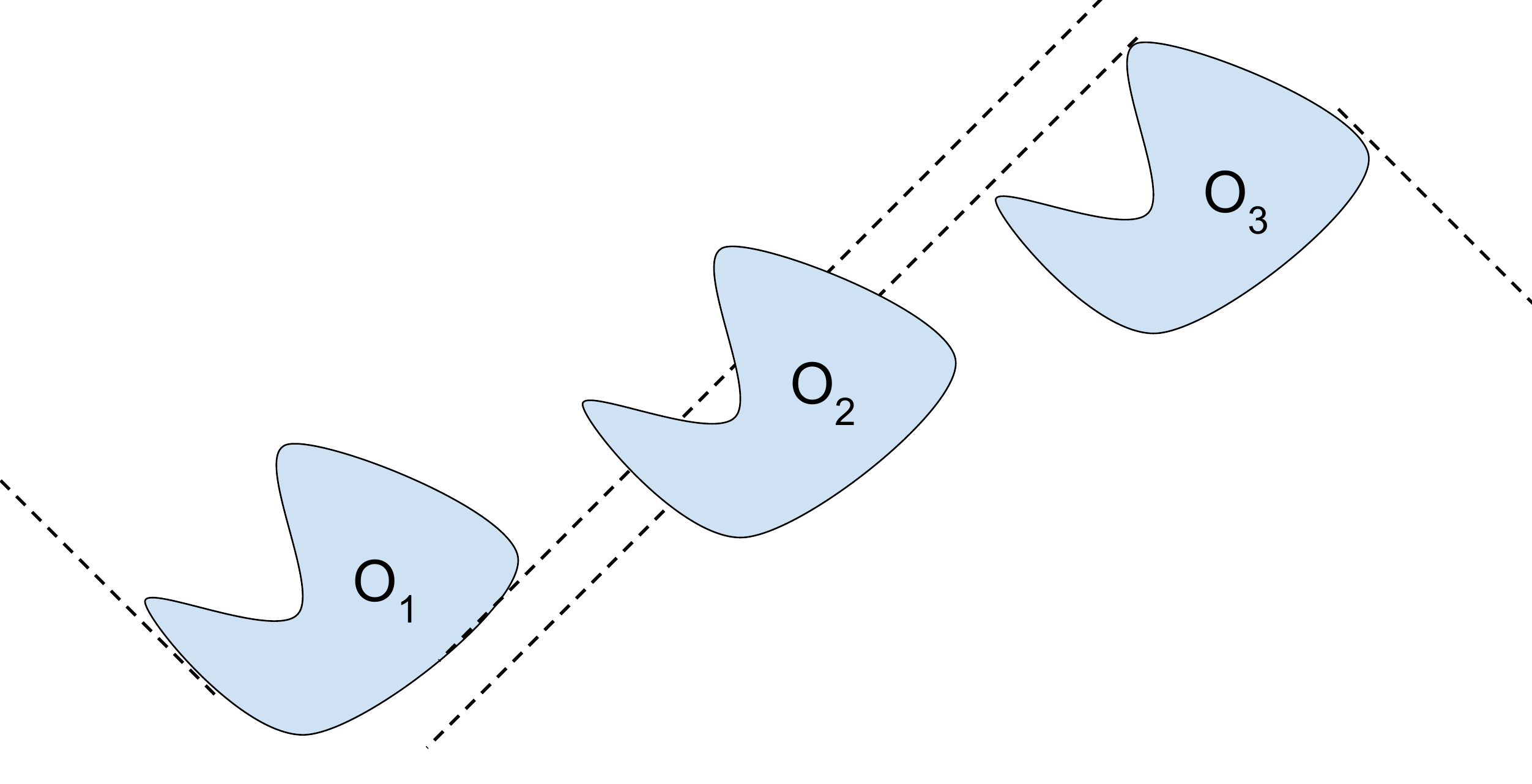}
\caption{The impossible measurement scenario considered in Sorkin's ~\cite{Sorkin}. Notice that regions $O_1$ and $O_3$ are space-like separated from one another. There is a mathematically well-defined local POVM in the algebra associated with $O_2$ which enables faster-than-light signaling from $O_1$ to $O_3$. Thus not all mathematically well-defined local operations are dynamically allowed.}\label{FigSorkin}
\end{figure}

%\Out{POVMs also might be impossible on grounds of conservation and super-selection.}
Another way in which some POVMs might be dynamically impossible is if they violate relativistic causality. Sorkin~\cite{Sorkin} was the first to notice that naively implementing PVM/POVM measurements in QFT results in faster-than-light signaling. See Fig. \ref{FigSorkin}. Roughly, there are some mathematically well-defined local PVMs/POVMs in region $O_2$ which will allow for signaling from region $O_1$ to region $O_3$. Importantly, however, not all localized POVMs yield faster-than-light signaling.
%\end{comment}

%\Out{Introduce the formal exact isolationist approach to observables}
An intuitive way to respond to such impossible measurements is to first formally categorize them, and then remove them from our official list of observables. Note that these impossible POVM measurements are well-defined in QFT and, moreover, the relativistic principles which they violate can also be formulated within QFT. Hence, the task of identifying exactly which POVMs are relativistically safe can, in principle, be conducted entirely within the formalism of QFT. Indeed, working entirely within QFT one can strive for an exact formal criterion which distinguishes the relativistically safe POVMs from the unsafe ones. We do, in fact, have such criteria (at least for real scalar QFT in a globally hyperbolic spacetime~\cite{Jubb2022,BorstenJubbKells}.) Let us call this the formal exact isolationist approach to identifying QFT's observables.

%\Out{This is wrong for several reasons: Firstly, there are other kinds of impossible measurements even non-relativistically (e.g., thermodynamic impossibility).}
One may be tempted to call the subset of POVMs which are relativistically-safe, the ``observables of QFT''. This would be wrong, however, for several reasons which I will now discuss in turn. Firstly, knowing that such measurements do not allow for faster-than-light signaling, does not guarantee that they are dynamically possible; They might violate other laws of physics (e.g, conservation laws, thermodynamics, etc.). Indeed, as I mentioned above, the fact that measuring some POVMs is dynamically impossible can already be seen in non-relativistic quantum theory. One could, in principle, identify every way in which POVMs might be dynamically impossible. This may, however, require us to look outside of QFT, so to speak, spoiling the isolationist aspect of the above-discussed formal exact isolationist approach to identifying QFT's observables.

%\Out{Patch up the formal exact isolationist approach. Are these dynamically-safe POVMs the observables of QFT?}
Setting this issue aside, suppose that we have identified every way in which POVMs might be dynamically impossible within QFT. Removing these from our consideration one would then be left with only the dynamically-safe POVMs. Presumably, one could derive some formal criteria which exactly identify these dynamically-safe POVMs. Conceivably, these formal criteria could be found through an isolated investigation of QFT (without connecting it to non-relativistic quantum theory or to classical theory). Thus, the formal exact approach seems to be alive and well (so far). Might the set of dynamically-safe POVMs arrived at in this way then be called the ``observables of QFT''?

%\Out{Exact formal criteria may be useful, but they cannot be physically salient.}
First, let me say that such exact formal criteria are definitely worth having in our tool belt. However, as tempting as it is to call these POVMs ``the observables of QFT'', I think that this formal exact isolationist approach is fundamentally wrong-headed. In my view, this desire for such a formal exact isolationist understanding of QFT's measurement processes is analogous to the dove's desire for empty space in the following Kantian metaphor:
\begin{quote}
The light dove cleaving in free flight the thin air, whose resistance it feels, might imagine that her movements would be far more free and rapid in air-less space.~\cite[p.~6]{KantCritiqueOfPureReason} 
\end{quote}
The humor here is that, while flying in a vacuum would be better in some sense (e.g, being resistance-free), it is exactly the messy complication of atmospheric resistance which makes flight possible in the first place. As I have discussed above, our theories get their empirical support and physical salience from their contact with experimental practice. This is ultimately a messy and approximate business~\cite{Curiel}. Any empirically meaningful notion of observables must come into a similarly messy contact with experimental practice.

%\Out{No purely formal characterization of observables can be empirically meaningful. Dog chases car.}
Indeed, as I discussed in Sec.~\ref{SecCoreExt}, the theorist's ability to talk so casually in terms of observables rests upon an (often under-discussed) mountain of experimental practice. Hence, a completely formal characterization of a theory's observables cannot be physically salient or empirically meaningful. Given such a formal characterization, we would then be (just like the sophomore) left to guess which observable our lab equipment measures. Chasing a theory's observables in this way is like a dog chasing a car, we would have no idea what to do with it once we caught it.\footnote{Indeed, the authors of~\cite{Jubb2022} recognize that ``it would be useful to construct an explicit dictionary between update maps and specific probe models''. A further connection would then need to be made between these probe models and real-life experimental practice.}

As I have discussed in Sec.~\ref{ChainsAndCuts}, for non-relativistic quantum theory, adequately linking theory to experiment requires that at least part of the measurement chain must be modeled outside of quantum theory (i.e., we must make a pragmatic Heisenberg cut somewhere). Similarly, in modeling QFT-involved measurements we will at some point need to invoke what I will call a \textit{QFT-cut}. That is, at some point along the measurement chain we must switch from using a QFT model to a non-QFT model.

%\Out{Nor can our approach be exact or isolationist. Crossing the quantum-classical divide is necessarily approximate.}
Applying this conclusion to quantum theory (including QFT) reveals two other issues with the formal exact isolationist approach. As I have discussed in Sec.~\ref{SecFirstExamples}, quantum theory (and hence QFT) depend upon classical theory for both their empirical support and physical salience. Hence, no isolationist approach to identifying observables is tractable. Moreover, taking either a Heisenberg cut or QFT-cut is unavoidably an approximate business. Therefore, we must (and, hence, are allowed to) pick out the observables of QFT in an inexact approximate way. This is what makes it okay for PVMs to sometimes be counted among our observables even though they are thermodynamically impossible. As I will discuss in Sec.~\ref{StateOfTheArt}, this is what makes it okay to sometimes model experiments using POVMs which allow for faster-than-light signaling (so long as these modeling errors are well-understood and well-controlled).

%\Out{Sum up: We must be informal, approximate, and make cross-theory cuts.}
In sum, as laudable as the formal exact isolationist approach is, its three attributes each come into conflict with a demand that our notion of ``observables'' is grounded in experimental practice. Instead, an empirically meaningful notion of observables must be informal, approximate, and arrived at through careful consideration of Heisenberg-like cuts. While the above discussion has been targeted at identifying the observables of QFT, the same conclusions also hold for identifying the observables of non-relativistic quantum theory. Thus, moving into a QFT context does not introduce any discontinuity in our how ought to approach observables and measurement processes.

In sum, if we want a unified measurement theory for QFT and an empirically meaningful characterization of its observables, then we need to solve the extended pragmatic measurement problem for QFT. This, in turn, requires us to work with QFT-cuts. In particular, it requires us to find a near universally applicable way of crossing the QFT-non-QFT divide. (Analogously to how decoherence theory and the Born rule provide us with a near universally applicable way of crossing the quantum-classical divide.) This raises the following question: Are any of the tools for making QFT-cuts which physicists currently have access to near-universally applicable in this way?

%\Out{Transition to next section. Name drop UDW detector.}
This concludes the philosophical portion of this paper. The next section will next review the state of the art in the physics literature as it applies to QFT-cuts. Those uninterested in these technical details can skip ahead to the conclusion. The purpose of this review is to give us a better handle on what types of QFT-cuts are available to us, and what their scopes are, both collectively and individually. As I will discuss, in my assessment, the current front-runner for getting us a wide-scoping measurement theory~\cite{pologomez2021detectorbased} for QFT is the Unruh-DeWitt detector model~\cite{Unruh1976,BLHu2007, Brown2013, Hotta2020, Zeromode,TaleOfTwo,Adam,Valentini1991, Reznik2003, Pozas-Kerstjens:2015,Menicucci, Terno2016, Cosmo, Henderson2018}. 

\section{The State of the Art}\label{StateOfTheArt}
%\Out{Ask a lot of questions to lead into the lit review}
Hopefully, the above discussion raises a great many questions for you: Do physicists have good tools for making QFT-cuts? What are the current possibilities and limitations for various kinds of QFT-cuts? Diagonal or vertical cuts? Crossing over into non-relativistic quantum theory or into classical physics? Are these tools collectively good enough to broadly cover all of quantum theory's QFT-involved experimental successes? Moreover, is any one of these tools of sufficient generality to allow us to induce a wide-scoping measurement theory for QFT from it? In order to answer these questions, I will need to review the current state of the art in the physics literature. Before this, however, some quick comments are needed regarding the term ``QFT-cut''.

\begin{figure}
\includegraphics[width=0.45\textwidth]{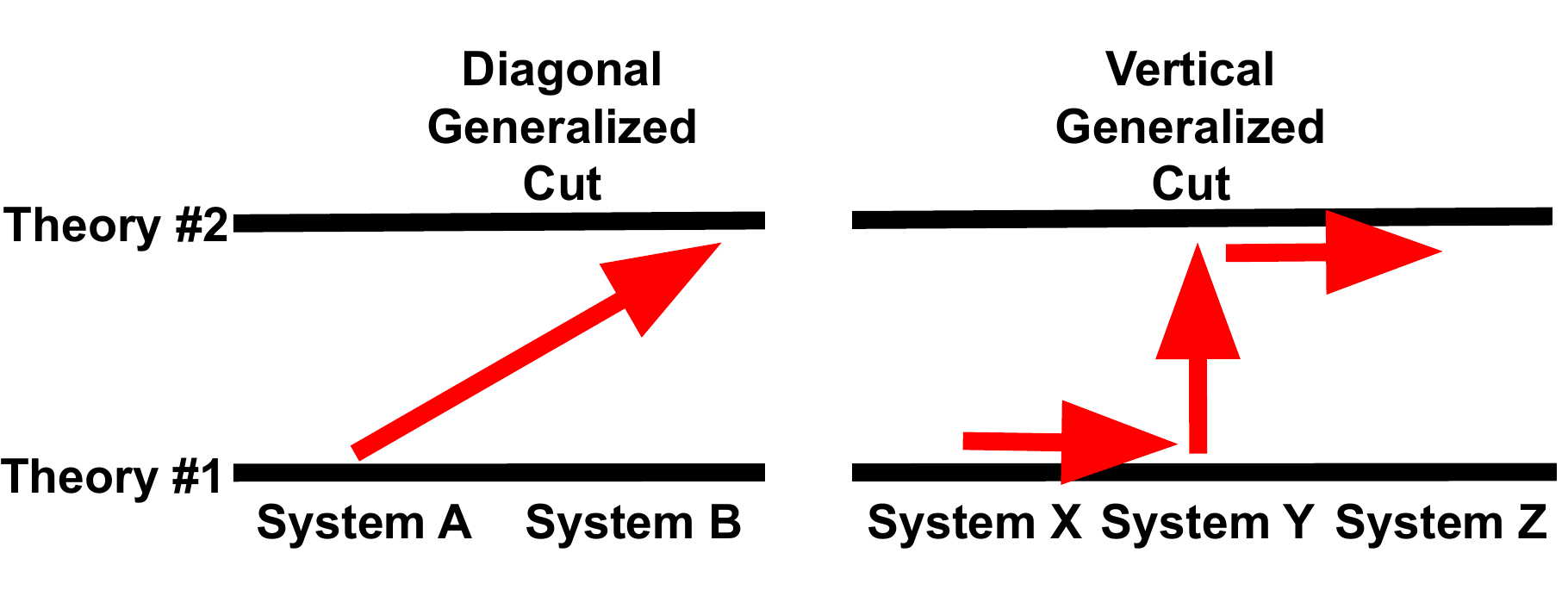}
\caption{The two possible ways of taking a generalized cut: diagonally (during an interaction) and vertically (in between interactions). On the left we have an example of a diagonal cut: system A is modeled in Theory 1 and system B in Theory 2. Their interaction couples two systems modeled in different theories. On the right, we have an example of a vertical cut: The interaction between system X and Y is modeled within Theory 1, whereas the interaction between Y and Z is modeled within Theory 2. In between these interactions we apply some approximation scheme to Y while it is isolated.}\label{FigGen}
\end{figure}

 %\Out{Comment on the name and suggest a variety of options}
%\begin{comment}
The notion of a QFT-cut has already been introduced in~\cite{TaleOfTwo} although it was there called a \textit{relativistic cut}. However, this name is apt to cause confusion because it is not \textit{relativity} per se which we must cut away from. To clarify this terminology, a QFT-cut occurs anywhere along the measurement chain where we switch from a QFT model to a non-QFT model. As I will discuss further below, there are a variety of ways in which one might go about making a QFT-cut. In addition to the vertical vs diagonal distinction introduced in Sec.~\ref{SecTaxonomy}, one can jump from QFT into a wide variety of different theories. See Fig~\ref{FigGen}.
%\end{comment}

%\Out{Mention field-cut}
A few other related cuts deserve mentioning and naming at this point. If one feels that the particularly troublesome part of QFT measurements is the fact that it describes things as a field, one might be interested in a field-cut. This is where we switch from modeling our measurement chain as a field (e.g., a relativistic quantum field, a non-relativistic quantum field, or a classical field) to not as a field (e.g., a qubit, a collection or classical point particles, or a nuclear spin degree of freedom).

%\Out{Mention Lorentz-cut}
Alternatively, one might feel that the troublesome part of QFT measurements is the fact that things are relativistic, i.e., that our models are set in a locally Lorentzian spacetime. In this case one might be interested in a relativistic cut. If that name is already taken, we might instead call this a Lorentz-cut where we switch from modeling our measurement chain in a locally Lorentzian spacetime to something else, e.g., a locally Galilean spacetime. For instance, we might move from relativistic QFT to non-relativistic QFT.

%\Out{Mention Type III algebra-cut}
Finally, one might feel that the troublesome part of QFT measurements is the fact that its algebraic structure is that of a Type III rather than a Type I von Neumann algebra~\cite{Witten,sep-qt-nvd}. In this case one might be interested in a Type III algebra-cut where we switch from modeling our measurement chain with a Type III algebraic structure to anything else, e.g. a Type I algebraic structure. For instance, we might move from relativistic QFT to non-relativistic quantum theory.

%\Out{If we start from a QFT, it is a QFT-cut}
One might be interested in all of the above, or many other subtle variations thereof. However, if we are beginning from a QFT then all of the above are examples of a QFT-cut (or in the terminology of \cite{TaleOfTwo}, a relativistic cut). As such, for the rest of this paper I will focus on QFT-cuts generally, where we switch from using a QFT model to using anything else. Of particular interest, however, are cases in which we switch from QFT to anything we know better how to model measurements of (e.g.: classical physics, special relativity, general relativity, or even non-relativistic quantum theory).

\begin{figure}
\includegraphics[width=0.45\textwidth]{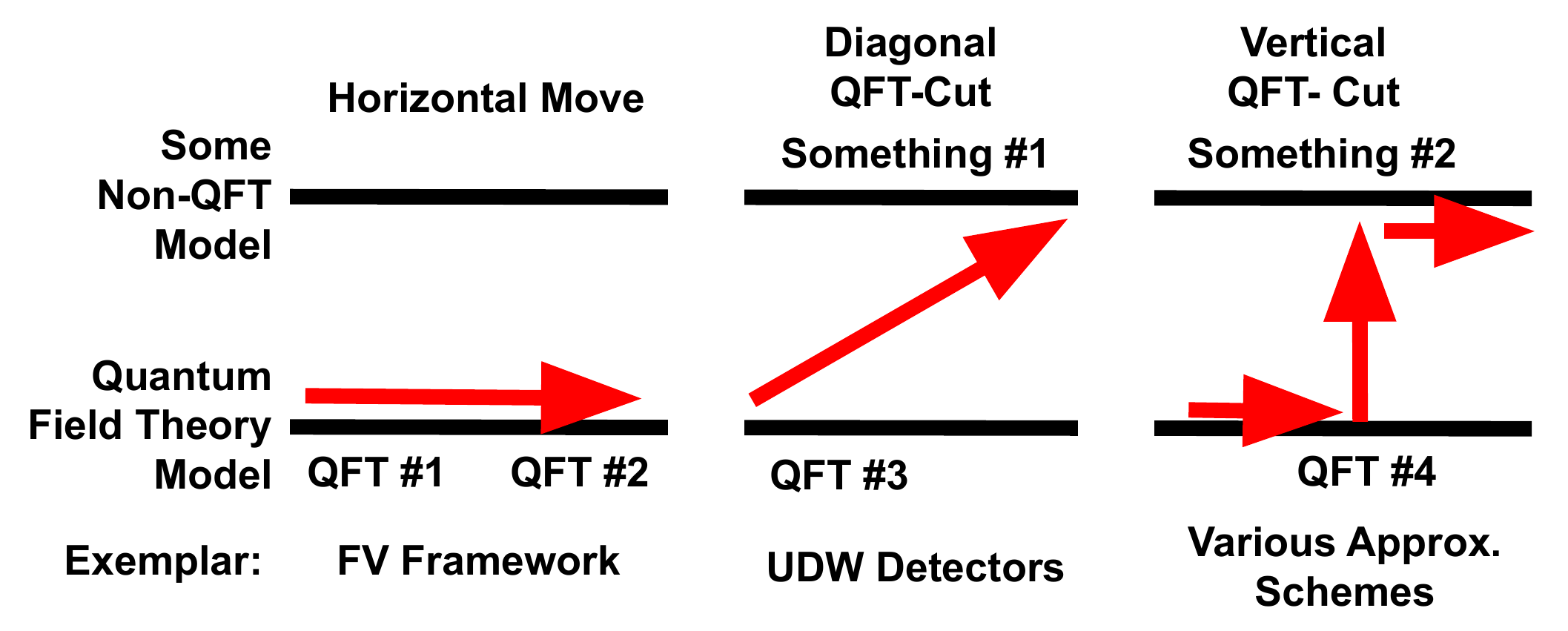}
\caption{This figure shows the three ways that one can approach and cross a QFT-cut: horizontal moves, diagonal cuts and vertical cuts. Exemplars of these three types of moves are the Fewster Verch (FV) framework~\cite{fewster1,fewster2,fewster3}, the Unruh-DeWitt (UDW) detector model~\cite{Unruh1976,BLHu2007, Brown2013, Hotta2020, Zeromode,TaleOfTwo,Adam,Valentini1991, Reznik2003, Pozas-Kerstjens:2015,Menicucci, Terno2016, Cosmo, Henderson2018,pologomez2021detectorbased}, and various approximation schemes~\cite{Rosaler,FlaminiaAchim} respectively.}\label{FigFVUDW}
\end{figure}

%\Out{Explain how this section is organized}
%\begin{comment}
Supposing that a theoretical or experimental physicist was interested in using an explicitly formalized QFT-cut in their modeling, what tools are currently available to them? What follows is an (incomplete) catalog of the various well-developed ways of approaching and crossing a QFT-cut in the physics literature. This catalog will be organized into three sections: horizontal moves, diagonal QFT-cuts, and vertical QFT-cuts. See Fig. \ref{FigFVUDW}. Following this I will briefly discuss the scope of experiments that these tools cover collectively. Additionally, I will briefly discuss whether any individual tool has the wide-scoping applicability needed to induce a measurement theory for QFT.
%\end{comment}

\subsection{Horizontal Moves}
%\Out{Point to the figure to explain what a horizontal move is}
Before discussing how one might make a QFT-cut, allow me to first talk about how to approach one horizontally. The general shape of a horizontal move is shown on the left side of Fig.~\ref{FigFVUDW}. Essentially, one QFT (QFT\#1) couples to another QFT (QFT\#2). Clearly, this is not a QFT-cut and so no collection of moves of this kind is the whole story.

%\Out{Since these moves do not cross the QFT-non-QFT divide they cannot be the full story.}
However, of course, such moves may still be helpful in advancing us along the measurement chain until we are in a better position to make a QFT-cut. There is no issue with using a horizontal move as \textit{part of} our measurement chain. The issue comes when one models an experiment involving QFT using only moves of this kind while neglecting to mention where exactly they take a QFT-cut. Or worse such an account might implicitly dismiss the need for a pragmatic QFT-cut altogether. %An example of this kind will be discussed later in this subsection.

%\Out{To start: What sort of QFTs can we describe in isolation?}
Before discussing what possibilities there are for describing interactions between QFTs, let's first talk about what sorts of isolated systems we know how to describe well with QFT. Neglecting momentarily their interaction, what types of systems QFT\#1 and QFT\#2 might go into the open slots in Fig.~\ref{FigFVUDW}? (Or QFT\#3 and QFT\#4 for that matter?) Firstly, it should be said that we know well how to describe a wide variety of free systems using Lagrangian QFT: free scalar fields with or without mass, free electrons, free neutrinos, free photons, free Higgs particles, free gravitons (in the linearized gravity regime), etc. We can even model systems like a free proton or neutron as a free massive spinor field (assuming, of course, we ignore their quarky internal structure). Bose-Einstein condensates and some other condensed matter systems can also be treated within QFT. We can also include any small perturbative interaction term between any of these and calculate their joint evolution within perturbation theory.

%\Out{Strongly interacting systems are more difficult} 
What is much more difficult to do within QFT is to describe strongly interacting systems, including bound states such as atoms, or bound quark systems, or atomic nuclei. In principle, one ought to be able to consider the electromagnetic field interacting with the electron field and the proton field (pretend such a thing exists). We then ought to be able to find bound state solutions to this strongly interacting QFT which correspond to the various energy states of a first-quantized Hydrogen atom. However, these bound states of QED are remarkably difficult to treat analytically. This is difficult for such systems in isolation, let alone interacting with an external field.

One appealing option is to simulate such strongly coupled QFTs via a lattice approximation (or, equivalently, via a hard UV cutoff). One can, for instance, model some QFT scenarios accurately as a lattice of coupled harmonic oscillators. Removing the UV degrees of freedom from our QFT in this way can make them tractable to simulate~\cite{PhysRevA.102.012619}. Careful work however is needed in relating the observables of the continuum QFT with the observables of the lattice QFT.   

%\Out{To summarize, even when stay within QFT-land we have limited mobility}
To summarize: even when we are just moving along the bottom line of Fig.~\ref{FigFVUDW} we have a rather limited mobility here currently. We have feasibility restrictions in terms of both what systems we can consider (i.e., QFT\#1 and QFT\#2) as well as how they might interact with each other.

%\Out{Discuss Lagrangian vs Algebraic QFT briefly}
Suppose that within computational feasibility, we have two QFTs in mind and an interaction between them. What mathematical formalisms do we have for modeling this interaction? As is typical in physics, one can begin from either a Lagrangian or from a Hamiltonian formulation. However, in the case of QFT some opt to put the theory on even more secure mathematical footing by formulating it in algebraic terms, namely in Algebraic QFT~\cite{sep-qt-nvd}. For a recent philosophical debate about the differences in these approaches see \cite{WallaceNaive,Wallace2011,Fraser2009,Fraser2011}. A significant trade-off between these approaches are their differences in mathematical rigor and in practical utility.

%\Out{Some turn to algebraic QFT} 
As discussed in Sec.\ref{GenChainsAndCuts}, there are many technical issues which arise when one tries to apply our projective measurement theory to QFT, see Fig.~\ref{FigSorkin}. As fraught as this area is with mathematical stumbling blocks, some have looked to Algebraic QFT in hopes of a more secure way to approach modeling (at least parts of) measurement processes of quantum fields. In particular, the Fewster Verch (FV) framework~\cite{fewster1,fewster2,fewster3} does this. Allow me to provide a brief overview.

\subsubsection*{Fewster Verch framework}
%\Out{Introduce the setup for FV}
Suppose that we can break at least a part of the measurement process down into a series of local interactions between QFTs. In particular, suppose that each of these interactions is localized in space and time, i.e. with one QFT acting as a local probe on another. The Fewster Verch (FV) framework~\cite{fewster1,fewster2,fewster3,Ruep2021,MeasurementSchemeFV} provides a model for such interactions within the mathematical rigor of Algebraic QFT.\footnote{Some additional discussion of the FV framework can be found in \cite{papageorgiou2023eliminating} where it is compared with the UDW approach discussed below.} By doing so, one can be assured to be completely respectful of the central `commandments' of relativity for at least part of the measurement chain. In particular, by describing this part of the measurement process entirely within Algebraic QFT, no causality violations of the kind shown in Fig.~\ref{FigSorkin} are possible; Algebraic QFT has the fundamental principles of relativity built right into it.

\begin{figure}
\includegraphics[width=0.45\textwidth]{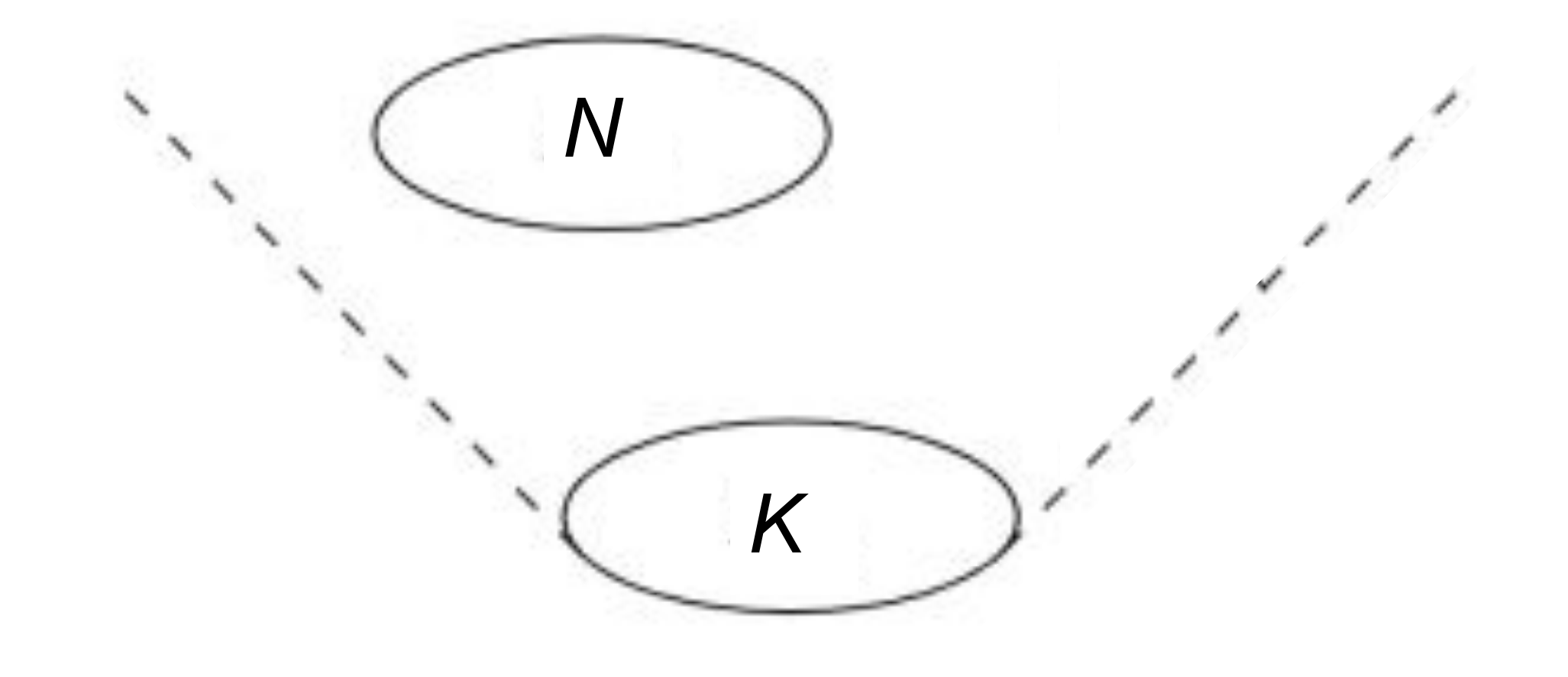}
\caption{The bounded spacetime regions considered in \cite{Ruep2021}. The two quantum fields interact only within the coupling region $K$. In the future of this interaction is the ``processing region'' $N$ where the probe field undergoes further measurement processes.}\label{FigFV}
\end{figure}

%\Out{Give example Lagrangian}
%\begin{comment}
To have something concrete in mind, let us consider a simple example (taken from~\cite{fewster1,Ruep2021}) which just so happens to have an equivalent representation in Lagrangian QFT. Consider a scenario where one massive Klein Gordon field (``the probe field''), $\hat\psi(t,x)$, acts as a local probe another (``the system field''), $\hat\phi(t,x)$. The joint Lagrangian for our simple example is,
\begin{align}\label{LagrangianFV}
\mathcal{L}
&=\underbrace{\frac{1}{2}(\nabla_\mu\hat\phi(t,x))(\nabla^\mu\hat\phi(t,x))
-\frac{m_1^2}{2}\hat\phi^2(t,x)}_{\mathcal{L}_\phi}\\
\nonumber
&+\underbrace{\frac{1}{2}(\nabla_\mu\hat\psi(t,x))(\nabla^\mu\hat\psi(t,x))
-\frac{m_2^2}{2}\hat\psi^2(t,x)}_{\mathcal{L}_\psi}\\
\nonumber
&-\underbrace{\lambda\,\rho(t,x)\,\hat\psi(t,x)\,\hat\phi(t,x)}_{\mathcal{L}_I}. 
\end{align}
The first and second terms are the free Lagrangians of the system field, $\hat\phi(t,x)$, and the probe field, $\hat\psi(t,x)$, respectively. The third term couples these two fields together. In the third term, $\lambda$ determines the strength of the interaction and $\rho(t,x)$ is a spacetime function which determines the interaction profile. That is, $\rho(t,x)$ determines where in space and time the two fields interact. For the purposes of modeling localized interactions, we can take $\rho(t,x)$ to be compactly supported in some spacetime region $K$, (i.e., $\rho(t,x)=0$ outside of $K$). See Fig. \ref{FigFV}. Here $N$ is some ``processing region'' in the future of $K$ where the probe field undergoes further measurement processes.
%\end{comment}

%\Out{The FV framework has a much wider scope than just this example}
It is important to note that the scope of interactions considered by the FV framework is much more general than this simple example. I have only specified the above interaction Lagrangian to have something concrete in mind for later comparison. In general, in the FV framework, one quantum field acts as a local probe upon another quantum field. The nature of these two fields is left completely open so long as they can both be formulated within Algebraic QFT. The nature of their interaction is left open, except that it all happens within a bounded spacetime region, $K$, see Fig.~\ref{FigFV}. The spacetime background for such an interaction is even left open, i.e., it could be curved.

%\Out{These sort of interactions cannot be the full story. Firstly, because they are computationally infeasible}
By using these sorts of local QFT-to-QFT interactions, one might be able to describe a part of a QFT-involved measurement chain. Of course, as discussed above, we cannot hope to provide a complete modeling of any real-life measurement process exclusively in terms of QFT-to-QFT interactions. This is for two reasons. Firstly, as discussed above, we currently have a rather limited technological and computational capacity for describing interacting bound states within Lagrangian QFT (let alone Algebraic QFT). Thus, the FV framework suffers here on grounds of computational feasibility, at least currently.

%\Out{Secondly, we still need to know how to measure the probe QFT}
Secondly, once one QFT acts as a probe on another, we are still left with the problem of how to model the measurement of the second QFT. Indeed, the FV framework does not claim to solve the quantum measurement problem (pragmatic or realist) but rather their interest is ``describing a link in the measurement chain, in a covariant spacetime context''~\cite{fewster1}. In particular, they ``take it for granted that the experimenter has some means of preparing, controlling and measuring the probe and sufficiently separating it from the QFT of interest''~\cite{fewster1} or put more simply that ``someone, somewhere, knows how to measure something''~\cite{FewsterRQITalk3}. 

%\Out{In sum, it may be helpful in solving the core problem.} 
In total therefore the FV framework is potentially useful (within its presently limited computational feasibility) for modeling parts of the QFT-involved measurement chains for real-life experiments. In combination with the yet-to-be-discussed diagonal and vertical cuts, it may be helpful in solving the core pragmatic measurement problem for QFT.

%\Out{But what about the extended problem}
But what about the extended pragmatic measurement problem? Our goal there is to give a unified wide-scoping account of measurements in QFT, i.e., to identify its observables. In this extended problem we care less about computational feasibility. One might therefore expect the FV framework (and horizontal moves generally) to be more useful in the extended problem. 

%\Out{Talk about their asymptotic measurement scheme for every observable of a quantum field.}
For instance, recent work claims to have used the FV framework to provide an ``Asymptotic measurement schemes for every observable of a quantum field theory''~\cite{MeasurementSchemeFV} in order to ``determine the set of system observables that can be measured by FV measurement schemes''. Concretely, their objective is ``to analyze how information about one physical structure (system) is transferred to another physical structure (probe) that is controlled by an external experimenter''~\cite{MaxTalk}. In particular their interest is in the case where both the system and field are QFTs and the information is transferred via a local interaction. This is exactly the sort of thing that the FV framework is good at: working out how information moves between quantum fields which interact with each other in localized regions.

%\Out{The limitation of their approach is assuming that we can characterize the probe field.}
The principal limitation in \cite{MeasurementSchemeFV} however is that (as is always the case with the FV framework) it explicitly assumes that the experimenter has full control over the probe field. In particular, it is assumed that they know how to extract classical information from the probe, i.e. ``someone, somewhere, knows how to measure something''~\cite{FewsterRQITalk3}. While this is potentially a step in the right direction, these results ultimately end up assuming that we know what the observables in the probe field are. Contrary to this methodology, I argue that the only empirically meaningful way to identify the observables within QFT is to connect them with observables outside of QFT by some measurement chain which includes a QFT-cut.

%\Out{Introduce some algebraic QFT terminology}
Allow me to briefly give the technical details of \cite{MeasurementSchemeFV} by first introducing some terminology. Within Algebraic QFT, each bounded region of spacetime $R$ is associated with an algebra, commonly called the ``algebra of observables''. However, this remains to be justified as what is an observable is exactly what is at question here. This algebra includes the field operator $\hat\phi(t,x)$ integrated against all smooth functions compactly supported over $R$. Additionally, the algebra includes products and sums of these smeared field operators. An FV measurement scheme for a field $\hat\phi(t,x)$ (``the system field'') specifies four things: a probe field $\hat\psi(t,x)$ labeled $\mathcal{P}$, an initial state for the probe field, $\rho_\mathcal{P}$ , a unitary interaction, $S$, between $\hat\phi(t,x)$ and $\hat\psi(t,x)$ localized in some region $K$, (e.g., Eq.~\eqref{LagrangianFV}) and finally an element of the probe algebra, $B$, associated with a processing region $N$ in the future of $K$, see Fig.~\ref{FigFV}. 

%\Out{Talk about induced observables}
%\begin{comment}
While the results of \cite{MeasurementSchemeFV} are proven in terms of Algebraic QFT, it here suffices to give their translation into the usual language of Hilbert spaces. A FV measurement scheme gives a way of indirectly addressing elements of the system algebra associated with the region $K$. Namely for every FV measurement scheme we have 
\begin{align}
\text{Tr}_{\mathcal{SP}}(\rho_\mathcal{S}\otimes\rho_{\mathcal{P}}\,S^\dagger \openone\otimes B\, S)
=\text{Tr}_{\mathcal{S}}(\rho_\mathcal{S}\,B_\text{ind})
\end{align}
for some induced $B_\text{ind}$ in the system algebra associated with $K$.
%\end{comment}

%\Out{Specify exactly which question they answer}
%\begin{comment}
Ultimately, the question addressed by \cite{MeasurementSchemeFV} is: Which elements of the system field's algebra at $K$ can be indirectly measured via an FV measurement scheme (assuming we can measure any element of the probe algebra at $N$)? Their answer is roughly that for every element $A$ of the system algebra at $K$ there exists a sequence of FV measurement schemes which indirectly measure it arbitrarily well in the limit. Namely, for every $A$ in the system algebra associated with $K$, we have~\cite{MaxTalk} 
\begin{align}
\text{Tr}_{\mathcal{SP}_\alpha}(\rho_\mathcal{S}\otimes\rho_{\mathcal{P}_\alpha}\,S_\alpha^\dagger \openone\otimes B_\alpha\, S_\alpha)
\to\text{Tr}_{\mathcal{S}}(\rho_\mathcal{S}\,A)
\end{align}
for some sequence of FV measurement schemes indexed by an integer $\alpha\in\mathbb{Z}$. Thus, given full control over the probe system in the region $N$ there is some sequence of probes, probe states and local interactions such that we can address any $A$ in the system's algebra arbitrarily well. 
%\end{comment}

%\Out{Sum up the usefulness of the FV framework}
%\begin{comment}
In summary, the FV framework is a great tool for working out how information moves between quantum fields which interact with each other in localized regions. However, to solve either the core or extended pragmatic measurement problems for QFT it alone is not enough. We need to at some point take a step outside of QFT via a QFT-cut.
%\end{comment}

\subsection{Vertical QFT-cuts}
%\Out{Introduce vertical cuts by pointing to the figure}
Let's next consider what vertical QFT-cuts are available to us currently. The general shape of a vertical QFT-cut is shown on the right side of Fig. \ref{FigFVUDW}. Essentially, we begin with something being modeled as a QFT (QFT\#4) and then take some sort of approximation on this to arrive the very same thing now modeled as something other than a QFT (Something\#2). Our freedoms in designing a vertical QFT-cut are: which theory we approximate into, what kind of QFT we begin from and relatedly what kind of non-QFT system we land on, as well as the details of our approximation scheme.

%\Out{Talk about taking the non-relativistic limit}
We have discussed already in the previous subsection the sorts of systems which we have a grip on how to model in QFT: free systems plus perturbative interactions and some condensed matter systems but not strongly interacting systems or bound states. Our options for QFT\#4 are fixed to be among these. As for which theory to cross into, the ``nearest'' theory to QFT would likely be non-relativistic QFT (i.e., QFT with $c\to\infty$, or rather QFT in a Galilean spacetime). For such an approximation to work our initial QFT\#4 must be massive, i.e., not light nor gravity. Massive fields may limit onto particles in a non-relativistic limit, but massless ones will not~\cite{Lamb1995,Rosaler}. For massive free states, taking such a limit gives us non-relativistic quantum particles. This alone is not enough unless we understand well how to model the measurement of non-relativistic QFTs. I am not aware of any research in this direction, but it could be a fruitful way forward.

%\Out{We could cut vertically into a hydrogen atom, but this is a bound QFT system}
The next nearest theory we could cut into is non-relativistic quantum theory. For instance, one might implement an effective-field-theory-style recovery of non-relativistic quantum theory and then make use of standard decoherence techniques. The question then is what we should take as Something\#2? An obvious experimentally relevant system would be a first-quantized Hydrogen atom. However, this would mean that QFT\#4 needs to be some second quantized description of the Hydrogen atom. As I have already discussed, describing such bound states in QFT is difficult. More research in this direction is warranted.

%\Out{Another intriguing option is to cut vertically into a UDW detector}
Another intriguing option for what non-relativistic quantum system to put for Something\#2 is an Unruh-DeWitt (UDW) detector~\cite{Unruh1976,BLHu2007, Brown2013, Hotta2020, Zeromode,TaleOfTwo,Adam,Valentini1991, Reznik2003, Pozas-Kerstjens:2015,Menicucci, Terno2016, Cosmo, Henderson2018,pologomez2021detectorbased}. These will be described in more detail in the next subsection, but roughly they are atom-like non-relativistic quantum systems which can be coupled to a quantum field in a way motivated by the light-matter interaction. In fact, recent work~\cite{FlaminiaAchim} has developed a ``second quantized UDW detector'', i.e., a QFT which reduces to a UDW detector in the non-relativistic limit. This an interesting avenue for future research, worthy of further development.

%\Out{Finally, we might cut vertically into classical physics.}
Finally, we can consider the possibility of approximating a QFT as a classical state of some sort. For instance, one might have a state of light described as a QFT and then switch to describing this as a classical electromagnetic field in Minkowski space. Vertical cuts of this kind seem to be experimentally relevant and deserve to be developed further. 

%\Out{In sum, vertical QFT cuts are promising.}
In summary, vertical QFT-cuts are a promising possibility deserving of further research. If technical limitations surrounding second quantized atoms can be overcome, then these could have a substantial scope including many experimentally relevant systems. 

\subsection{Diagonal QFT-cuts}\label{DiagonalUDW}
%\Out{Introduce diagonal cuts by pointing to figure}
Finally, let's next consider what diagonal QFT-cuts are available to us currently. The general shape of a diagonal QFT-cut is shown in the center of Fig. \ref{FigFVUDW}. Essentially one QFT (QFT\#3) couples directly to something which is not being modeled as a QFT (Something\#1). As is the case for any diagonal cut, the direct coupling between system's described in fundamentally different theories poses conceptual challenges. Here, there is an inherent risk that coupling to a non-QFT system will end up breaking one of the central `commandments' of relativity. 

%\Out{We have freedom indesigning out diagonal cuts}
However, on the bright side, since quantum field theory itself provides us with no prescription for how such systems ought to interact, we have a great deal of freedom in how one might model such an interaction. In particular, our freedoms in designing a diagonal QFT-cut are: what kind of QFT we begin from, which theory we cut into, what kind of non-QFT system we cut into, and the nature of the interaction between the systems.

%\Out{Talk about cutting into non-rel QFT}
We have discussed in the previous subsections the sorts of systems which we have a grip on how to model in QFT: free systems plus perturbative interactions and some condensed matter systems but not strongly interacting systems or bound states. Our options for QFT\#3 are fixed to be among these. As for which theory take a QFT-cut into, the ``nearest'' theory to QFT would likely be non-relativistic QFT (i.e., QFT with $c\to\infty$, or rather QFT in a Galilean spacetime). As far as I am aware, not much work has been put into the study of sensible dynamical couplings between QFT and non-relativistic QFT. One immediate concern is the possibility of uncontrolled faster-than-light signaling. We might have a relativistic field $\hat\phi_\text{Rel}(x)$ couples to $\hat\phi_\text{Non-Rel}(x)$ which can then send an instantaneous signal to $\hat\phi_\text{Non-Rel}(x+a)$ which is coupled to $\hat\phi_\text{Rel}(x+a)$. This seems problematic. More work may be needed in this direction.

%\Out{Mention cutting diagonally into non-rel QT, but delay this for later}
The next ``nearest'' theory we could try to take a QFT-cut into is non-relativistic quantum theory. Significant work has been done in this direction, which will be discussed momentarily.

%\Out{Mention cutting diagonally into quantum theory}
The last option which comes to mind is to couple our QFT to a classical system (e.g., something modeled in special relativity or general relativity). This doesn't seem terribly problematic, we could for instance have a classical Klein Gordon field interacting with the expectation value of a quantum Klein Gordon field, perhaps without back reaction. More work would need to be done motivating why such an interaction is an accurate reflection of some part of a real-life experiment, but there don't seem to be insurmountable technical difficulties here.

\subsubsection*{Unruh-DeWitt Detector Models}
%\Out{Return to diagonally cutting into non-rel QT. What could guide us?}
Let's return to the possibility of diagonal cuts into non-relativistic quantum theory. We have a lot of freedom here in designing this interaction. Of course, there are also certain things we want from this diagonal theory-to-theory coupling if it is going to be a productive part of an experimental prediction. So what ought to guide us in designing this interaction? As a first guide, we may rely on a desire to preserve the central `commandments' relativity (covariance, causality, and locality) as much as possible. Moreover, as a second guide we may rely on a desire to accurately model parts of real-life experiments. 

%\Out{We might cut into something atom-like}
For instance, we might take the Something\#1 system in Fig.~\ref{FigFVUDW} to be something atom-like and we might take QFT\#3 to be the electromagnetic field (or some scalar analog thereof). In this case, under such guidance, one is quickly led~\cite{Pablo,RichardEdu,Richard} to something very much like the Unruh-DeWitt (UDW) detector model first introduced in~\cite{Unruh1976}. 

%\Out{We might cut into a gravity wave detector}
Alternatively, we might take QFT\#3 to be a graviton field (in the linear gravity regime). In this case, one is quickly led to a certain variant of the Unruh-DeWitt detector model~\cite{GRQO,GravityDetector,GravityDetector2}. 

%\Out{We might cut into a Fermion detector}
One can also take Something\#1 to be a fermionic quantum system which interacts with QFT\#3 being a neutrino field. In this case one is led to another variant of the Unruh-DeWitt detector model~\cite{MatsasNeutrinos, perche2021antiparticle}. 

%\Out{These considerations generally yield UDW-like detector models}
The possibilities for which real-life interaction we might attempt to mimic here are very general. In each case, the resulting interaction model is within the family of Unruh-DeWitt-like models. Moreover, much of the above can be done in arbitrarily curved spacetime backgrounds as well~\cite{GRQO}.

%\Out{Give explicit Lagrangian (for comparison with FV)}
%\begin{comment}
Enough discussion of abstract possibilities, concretely what do these models look like? To have something concrete in mind, let us consider a simple example (taken from \cite{TaleOfTwo}). Consider a simple example in which a UDW detector $\hat{\mu}$ coupled to a massive Klein Gordon fields $\hat\phi(t,x)$ with joint Lagrangian,
\begin{align}\label{LagrangianUDW}
\mathcal{L}
&=\underbrace{\frac{1}{2}(\nabla_\mu\hat\phi(t,x))(\nabla^\mu\hat\phi(t,x))
-\frac{m_1^2}{2}\hat\phi^2(t,x)}_{\mathcal{L}_\phi}\\
\nonumber
&+\mathcal{L}_\text{UDW}-\underbrace{\lambda\,\rho(t,x)\,\hat\mu(\tau)\,\hat\phi(t,x)}_{\mathcal{L}_I}.
\end{align}
The first term is the free Lagrangian of the field. The second term $\mathcal{L}_\text{UDW}$ is the free Lagrangian of the non-relativistic probe system, the UDW detector. We have total freedom to pick the internal dynamics of the non-relativistic probe system. For instance, it could be a qubit, or a quantum harmonic oscillator, or a first quantized Hydrogen atom. 
%\end{comment}

%\Out{Explain what the interaction term is}
In the third term, $\lambda$ determines the strength with which the non-relativistic probe and field couple to each other. In the third term, $\rho(t,x)$ is a spacetime function determining the interaction profile. In this context, $\rho(t,x)$ is often called the probe's smearing function, and is taken to describe the size and shape of the probe through time (more will be said about this later). Just as in the FV framework, for the purposes of modeling local measurements, we can take $\rho(t,x)$ to be compactly supported in some spacetime region, $K$, (i.e., $\rho(t,x)=0$ outside of $K$). In the third term, $\hat{\mu}$ is the degree of freedom of the non-relativistic probe which couples to the quantum field. For instance, if the probe is a harmonic oscillator, $\hat{\mu}(\tau)$ might be its number operator $\hat{n}$ or one of its quadrature operators $\hat{q}$ or $\hat{p}$.

%\Out{Contrast with the FV approach}
The major difference between the FV and UDW approaches (Eq.~\eqref{LagrangianFV} and Eq.~\eqref{LagrangianUDW}) is just that in the first case the probe system is a quantum field, $\hat\psi(t,x)$, and in the second case it is a non-relativistic quantum system, $\hat{\mu}$. For a more in depth comparison of the UDW detector model and the FV framework, see~\cite{TaleOfTwo}.

%\Out{The UDW detector is not a pointer measurement but should rather be thought of as responding like an atom would}
It's important to note that the UDW detector is not designed as a Von Neumann ``pointer'' measurement device, i.e., one which translates a ``needle'' proportional to some targeted operator in the probed field. Rather, the UDW detector is designed to be atom-like. If one measures the UDW detector after its interaction with the field, one ought to interpret this roughly as one would if an atom had coupled to the field. For instance, if the UDW probe is initially in its ground state and then is later measured to be in an excited state, one might infer that it absorbed a photon from the quantum field.

%\Out{Discuss an entanglement harvesting application}
More complexly, one can use UDW detectors to model an entanglement harvesting experiment~\cite{Valentini1991, Reznik2003, Pozas-Kerstjens:2015,Menicucci, Terno2016, Cosmo, Henderson2018,Ruep2021}. Roughly, in such an experiment two initially uncorrelated probe systems interact locally with a quantum field in such a way that they do not have time to signal to each other. Despite this, these two probes become entangled because there was already entanglement present between the two space-like separated regions they interacted with. The benefit of such an experiment is that the entanglement in the field has been transferred into more accessible systems, both physically and mathematically. We cannot associate a Hilbert space to bounded regions in QFT and as such cannot straightforwardly compute the entanglement between these regions. The final entanglement of these probes is a witness to the initial entanglement in the field. 

%\Out{UDW detectors can be used to in curved spacetimes}
A great many theoretical investigations of this sort have been carried out using UDW detectors~\cite{Valentini1991, Reznik2003, Pozas-Kerstjens:2015,Menicucci, Terno2016, Cosmo, Henderson2018}. Such studies can even be done in curved spacetimes: one can study the entanglement structure around a black hole near the event horizon for instance. 

%\Out{In sum, UDW detectors are useful in a wide variety of situations}
Thus, in addition to providing a good model for many experimentally relevant systems, the UDW detector model covers a wide range of interesting hypothetical experiments, all while remaining computationally feasible.

%\Out{But how badly do they break relativity?}
How well do UDW-like detectors preserve the central `commandments' of relativity: covariance, causality, and locality? Do they for instance lead to uncontrolled faster-than-light signaling in the QFT? Before answering these questions, a distinction needs to be made between two modes of applications of the UDW model. It was mentioned above that we will often wish to localize the probe's smearing function $\rho(t,x)$ within some bounded spacetime region. For instance, when the UDW detector model is derived from the light matter interaction~\cite{Pablo,RichardEdu,Richard}, the smearing function $\rho(t,x)$ turns out to be determined by the overlaps of certain atomic orbitals. That is, $\rho(t,x)$ is a near-literal description of the shape of the atom in space and time. 

%\Out{UDW detectors could be either point-like or smeared}
When the size of the UDW is much smaller than all other relevant scales and the detailed shape of the detector doesn't matter much, we can also approximate the detector as being point-like. Consider a point-like detector traveling through spacetime on some time-like trajectory, $z(t)$. We can localize the interaction to this trajectory by taking \mbox{$\rho(t,x)=\chi(t)\delta(x-z(t))$}. Here $\chi(t)$ controls where along the trajectory the probe couples to the field (it may turn on and off) and the $\delta$ function localizes the interaction to the detector trajectory, $z(t)$. Let us call these the point-like detectors.

%\Out{There are some issues with the smeared detectors}
With this established, let us return to the question of how well do UDW-like detectors preserve the central `commandments' of relativity. In brief, they do so imperfectly, but with well understood and controllable issues~\cite{BrokenCovariance,JoseMariaEdu}. For smeared detectors (i.e., non-point-like detectors) there are some slight faster-than-light signaling issues. Essentially, the issue is that if some information is taken up by the left half of the detector it can ``immediately'' jump to the right half of the detector and then back into the field. Basically, signals can jump across the detector instantly. This breaks no-signaling and causes some issues with the relativity of simultaneity (this coupling does not treat all relativistically compatible time-orderings equally). 

%\Out{These issues are well-understood and well-controlled however.}
However, these issues are ultimately minor~\cite{BrokenCovariance,JoseMariaEdu}. The time ordering issues do not appear at the lowest orders of perturbation theory. If the light-matter interaction is weak enough, then the time-ordering issues are strongly suppressed. Moreover, the size of the no-signaling violations is set by the size of our detector. Recall $\rho(t,x)$ might have compact support. If the UDW detector has a width of $3\text{ nm}$ then signals can only arrive at most $10$ atto-seconds early. Ultimately if we care about such time-scales (of the order of the light crossing time for the atom) then we shouldn't even be allowed to talk about first-quantized atoms in the first place. Indeed, all of these commandment-breaking issues go away when we use point-like detectors.

\subsection{The Scope of these Tools}
%\Out{These tools can solve the core problem}
Having reviewed the state of the physics literature, for feasible ways of crossing the QFT-cut, what ultimately are the scope of these tools? In my assessment while each tool has its limitations and more development of each of them is needed, collectively these tools have a substantial scope. Thus, I believe that collectively these tools give us a good handle on a case-by-case measurement framework for QFT. That is, collectively they can give us a solution to the core pragmatic measurement problem for QFT. 

%\Out{But what about the extended problem?}
But what about the extended pragmatic measurement problem for QFT? Can these tools help us identify the observables of QFT? As discussed in Sec.~\ref{GenChainsAndCuts}, in order to get a wide-scoping unified measurement theory for QFT we would need some near-universally available way of making QFT-cuts. In particular, we would need for at least one of these tools to have a sufficiently wide range of applicability such that nearly all QFT-involving experiments can be modeled using it. Which of the above discussed tools has the widest scope?

%\Out{Of the tools reviewed, the UDW detector appears to have the widest scope}
As I have discussed above, at least currently the UDW detector model by far has the widest scope of applicability of any of the tools currently available. Thus, if one wants to develop a wide-scoping measurement theory for QFT and to identify its observables, this is currently the most promising way forward. Indeed, a recent paper claims to have used the UDW detector model to establish a detector-based measurement theory for quantum field theory~\cite{pologomez2021detectorbased}.

\section{Conclusion}\label{Conclusion}
%\Out{Briefly summarize the pragmatic vs realist measurement problems. What is at stake}
This paper began by distinguishing between the pragmatic and realist portions of the quantum measurement problem. Of these, I have argued that the pragmatic worries have worse consequences if left unanswered. If we lose the pragmatic connection between theory and experimental practice, then quantum theory is at risk of losing both its empirical support and its physical salience. 

%\Out{Briefly review how to solve the core and extended problems}
Fortunately, these pragmatic worries are not too hard to address. I next divided the pragmatic measurement problem into the core problem and the extended problem. The core pragmatic measurement problem calls for us to develop a case-by-case measurement framework for modeling quantum theory's key experimental successes. While solving the core problem would restore empirical support to quantum theory, it does not solve the pragmatic measurement problem entirely. The extended pragmatic measurement problem calls for us to develop a unified measurement theory capable of modeling all (or nearly all) possible measurement processes. Solving the extended problem is necessary in order to achieve an empirically meaningful characterization of our theory's observables and to permit talk of measurement processes in general.

%\Out{Review how these issues are solved in non-relativistic quantum theory}
In Sec.~\ref{ChainsAndCuts}, I discussed how both portions of the pragmatic measurement problem have been solved for non-relativistic quantum theory. Namely, thinking in terms of measurement chains gives us a road map for modeling each experiment's individual measurement processes. As I have argued, it is pragmatically necessary that we model our way across the quantum-classical divide at some point by invoking a Heisenberg cut. There are a wide variety of pragmatic Heisenberg cuts available to us. Having access to these various kinds of Heisenberg cuts allows us to develop a case-by-case measurement framework for modeling quantum theory's key experimental successes. This constitutes a solution to the core problem.

In order to solve the extended problem, we need to somehow unify this case-by-case measurement framework into a single measurement theory. In particular, we need to find some way of making pragmatic Heisenberg cuts which is applicable in all (or nearly all) measurement scenarios. As I have discussed, one way of achieving this goal is by appealing to decoherence theory and the Born rule. This justifies (at least pragmatically) our use of the canonical PVM/POVM measurement theory. Importantly, however, one cannot simply guess which projective measurement to use when modeling a given experiment. It is highly intuitive that the double-slit experiment ends with a measurement in the position basis once the electron hits the screen. As strong as this intuition may be, simply guessing the right projector is not a methodologically sound way of modeling measurements. One's choice of PVM/POVM must follow from a careful dynamical investigation of the system at hand. In particular, this analysis must include a pragmatic Heisenberg cut.

%\Out{My claim is that the same approach works in QFT}
But what changes when we move into a QFT context? A much-discussed complication which arises in QFT are Sorkin's impossible measurements~\cite{Sorkin}. In QFT almost all localized projective measurements violate causality, allowing for faster-than-light signaling. Thus, the story of measurement in QFT cannot end with a projective measurement theory as it did before. Fortunately, however, the beginning of the story and its overall structure can remain unchanged. As I argue in Sec.~\ref{GenChainsAndCuts}, we ought to first use measurement chains to build up a case-by-case measurement framework for QFT. This will require us to cross the QFT-non-QFT divide by using a pragmatic Heisenberg-like cut (what I call a QFT-cut).

We can then strive to unify this case-by-case measurement framework into a new wide-scoping measurement theory for QFT. In particular, we need to find some way of making pragmatic QFT cuts which is applicable in all (or nearly all) measurement scenarios. It is at this point that significantly more theoretical work is needed. It is only once we have such a unified measurement theory for QFT that we can talk about its measurement processes generally and achieve an empirically meaningful characterization of its observables.

%\Out{Contrast my approach with the formal exact isolationist approach}
My approach stands in strong contrast to what I have called the formal exact isolationist approach to identifying QFT's observables. This approach proceeds roughly as follows. ``Given that there exist impossible measurements within QFT (see Sorkin~\cite{Sorkin}), we ought identify them and get rid of them. Note that these problematic POVMs and the relativistic principles which they violate can both be formalized within QFT. Hence, it should be possible to achieve an exact formal characterization of them from entirely within QFT. Removing these impossible measurements from the set of all possible POVMs ought to yield a meaningful characterization of QFT's observables.'' 

As I have argued in Sec.~\ref{ImpossibleMeasurements}, useful as such a formal characterization might be, it cannot deliver us an empirically meaningful characterization of QFT's observables. Connecting QFT with experimental practice requires us to reach outside of QFT as we cross the QFT-non-QFT divide. This requires us to make some approximation, namely a QFT-cut. The ultimate justification for the validity of such an approximation is experimental practice~\cite{Curiel}. Hence, our understanding of QFT's observables must be informal, approximate, and arrived at through careful consideration of QFT-cuts.

%\Out{What tools are available? List some}
This is where the paper's philosophical argumentation ended. However, in light of this conclusion, it became important to understand what tools physicists have for making QFT-cuts. In Sec.~\ref{StateOfTheArt}, I have attempted to provide a (non-exhaustive) survey of several techniques which the physics community currently has of making various kinds of QFT-cuts. Indeed, physicists do have several good tools for approaching (the Fewster Verch (FV) framework~\cite{fewster1,fewster2,fewster3,Ruep2021}) and crossing the QFT-cut (UDW detector model~\cite{Unruh1976,BLHu2007, Brown2013, Hotta2020, Zeromode,TaleOfTwo,Adam,Valentini1991, Reznik2003, Pozas-Kerstjens:2015,Menicucci, Terno2016, Cosmo, Henderson2018}, and a handful of approximation schemes~\cite{Rosaler,FlaminiaAchim}.

%\Out{How good are these tools? Do they give us a wide-scoping measurement theory?}
But are their tools collectively good enough to secure evidential support for quantum theory? In my assessment, collectively these tools have a substantial scope. Thus, collectively these tools give us a good handle on a measurement framework for QFT solving its core pragmatic measurement problem. I see nothing which would prevent engineers and experimenters from satisfactorily modeling their QFT-involved measurement apparatus on a case-by-case basis.

But is any one of these tools on its own sufficient to give us wide-scoping unified measurement theory for QFT? Following our non-relativistic story discussed above, establishing a measurement theory for QFT would require that we find a way of making QFT-cuts which is near universally applicable across different measurement scenarios (i.e., like decoherence theory is). In my assessment, the UDW detector model has the widest scope of applicability of any of the tools currently available. Thus, if one wants to develop a wide-scoping measurement theory for QFT or to identify its observables, this appears to be the best way forward. Indeed, a recent paper claims to have used the UDW detector model to establish a detector-based measurement theory for quantum field theory~\cite{pologomez2021detectorbased}.

\section{Acknowledgements}
The authors thanks James Read, Ian Jubbs, Chris Fewster, Maximilian Ruep, Doreen Fraser, and the Barrio RQI (especially Eduardo Mart\'{i}n-Mart\'{i}nez, Jos\'{e} Polo-G\'{o}mez, Erickson Tjoa, Tales Rick Perch, and Bruno Torres) for their helpful feedback. I also received helpful feedback on this work from the Establishing the Philosophy of Supersymmetry Workshop and the Quantum Field Theory in Curved Spacetime workshop.

\section{Declarations}
The author did not receive support from any organization for the submitted work. The author has no competing interests to declare that are relevant to the content of this article.

\bibliographystyle{apsrev4-1}
\bibliography{references}

\begin{thebibliography}{80}%
\makeatletter
\providecommand \@ifxundefined [1]{%
 \@ifx{#1\undefined}
}%
\providecommand \@ifnum [1]{%
 \ifnum #1\expandafter \@firstoftwo
 \else \expandafter \@secondoftwo
 \fi
}%
\providecommand \@ifx [1]{%
 \ifx #1\expandafter \@firstoftwo
 \else \expandafter \@secondoftwo
 \fi
}%
\providecommand \natexlab [1]{#1}%
\providecommand \enquote  [1]{``#1''}%
\providecommand \bibnamefont  [1]{#1}%
\providecommand \bibfnamefont [1]{#1}%
\providecommand \citenamefont [1]{#1}%
\providecommand \href@noop [0]{\@secondoftwo}%
\providecommand \href [0]{\begingroup \@sanitize@url \@href}%
\providecommand \@href[1]{\@@startlink{#1}\@@href}%
\providecommand \@@href[1]{\endgroup#1\@@endlink}%
\providecommand \@sanitize@url [0]{\catcode `\\12\catcode `\$12\catcode
  `\&12\catcode `\#12\catcode `\^12\catcode `\_12\catcode `\%12\relax}%
\providecommand \@@startlink[1]{}%
\providecommand \@@endlink[0]{}%
\providecommand \url  [0]{\begingroup\@sanitize@url \@url }%
\providecommand \@url [1]{\endgroup\@href {#1}{\urlprefix }}%
\providecommand \urlprefix  [0]{URL }%
\providecommand \Eprint [0]{\href }%
\providecommand \doibase [0]{http://dx.doi.org/}%
\providecommand \selectlanguage [0]{\@gobble}%
\providecommand \bibinfo  [0]{\@secondoftwo}%
\providecommand \bibfield  [0]{\@secondoftwo}%
\providecommand \translation [1]{[#1]}%
\providecommand \BibitemOpen [0]{}%
\providecommand \bibitemStop [0]{}%
\providecommand \bibitemNoStop [0]{.\EOS\space}%
\providecommand \EOS [0]{\spacefactor3000\relax}%
\providecommand \BibitemShut  [1]{\csname bibitem#1\endcsname}%
\let\auto@bib@innerbib\@empty
%</preamble>
\bibitem [{\citenamefont {Maudlin}(1995)}]{Maudlin1995ThreeMP}%
  \BibitemOpen
  \bibfield  {author} {\bibinfo {author} {\bibfnamefont {T.}~\bibnamefont
  {Maudlin}},\ }\href@noop {} {\bibfield  {journal} {\bibinfo  {journal}
  {Topoi}\ }\textbf {\bibinfo {volume} {14}},\ \bibinfo {pages} {7} (\bibinfo
  {year} {1995})}\BibitemShut {NoStop}%
\bibitem [{\citenamefont {Muller}(2023)}]{muller2023measurement}%
  \BibitemOpen
  \bibfield  {author} {\bibinfo {author} {\bibfnamefont {F.~A.}\ \bibnamefont
  {Muller}},\ }\href@noop {} {\enquote {\bibinfo {title} {Six measurement
  problems of quantum mechanics},}\ } (\bibinfo {year} {2023}),\ \Eprint
  {http://arxiv.org/abs/2305.10206} {arXiv:2305.10206 [quant-ph]} \BibitemShut
  {NoStop}%
\bibitem [{\citenamefont {Myrvold}(2018)}]{sep-qt-issues}%
  \BibitemOpen
  \bibfield  {author} {\bibinfo {author} {\bibfnamefont {W.}~\bibnamefont
  {Myrvold}},\ }in\ \href {https://plato.stanford.edu/entries/qt-issues/}
  {\emph {\bibinfo {booktitle} {The {Stanford} Encyclopedia of Philosophy}}},\
  \bibinfo {editor} {edited by\ \bibinfo {editor} {\bibfnamefont {E.~N.}\
  \bibnamefont {Zalta}}}\ (\bibinfo  {publisher} {Metaphysics Research Lab,
  Stanford University},\ \bibinfo {year} {2018})\ \bibinfo {edition} {{F}all
  2018}\ ed.\BibitemShut {Stop}%
\bibitem [{\citenamefont {Bell}(1987)}]{1987Saui}%
  \BibitemOpen
  \bibfield  {author} {\bibinfo {author} {\bibfnamefont {J.}~\bibnamefont
  {Bell}},\ }\href@noop {} {\emph {\bibinfo {title} {Speakable and unspeakable
  in quantum mechanics : collected papers on quantum mechanics}}}\ (\bibinfo
  {publisher} {Cambridge University Press},\ \bibinfo {address} {Cambridge},\
  \bibinfo {year} {1987})\BibitemShut {NoStop}%
\bibitem [{\citenamefont {Wallace}(2020)}]{Wallace2020}%
  \BibitemOpen
  \bibfield  {author} {\bibinfo {author} {\bibfnamefont {D.}~\bibnamefont
  {Wallace}},\ }in\ \href {\doibase 10.1093/oso/9780198814979.003.0005} {\emph
  {\bibinfo {booktitle} {{Scientific Realism and the Quantum}}}}\ (\bibinfo
  {publisher} {Oxford University Press},\ \bibinfo {year} {2020})\ \Eprint
  {http://arxiv.org/abs/https://academic.oup.com/book/0/chapter/322307015/chapter-ag-pdf/44484139/book\_36983\_section\_322307015.ag.pdf}
  {https://academic.oup.com/book/0/chapter/322307015/chapter-ag-pdf/44484139/book\_36983\_section\_322307015.ag.pdf}
  \BibitemShut {NoStop}%
\bibitem [{\citenamefont {Dickson}(2007)}]{DICKSON2007275}%
  \BibitemOpen
  \bibfield  {author} {\bibinfo {author} {\bibfnamefont {M.}~\bibnamefont
  {Dickson}},\ }in\ \href {\doibase
  https://doi.org/10.1016/B978-044451560-5/50007-5} {\emph {\bibinfo
  {booktitle} {Philosophy of Physics}}},\ \bibinfo {series and number}
  {Handbook of the Philosophy of Science},\ \bibinfo {editor} {edited by\
  \bibinfo {editor} {\bibfnamefont {J.}~\bibnamefont {Butterfield}}\ and\
  \bibinfo {editor} {\bibfnamefont {J.}~\bibnamefont {Earman}}}\ (\bibinfo
  {publisher} {North-Holland},\ \bibinfo {address} {Amsterdam},\ \bibinfo
  {year} {2007})\ pp.\ \bibinfo {pages} {275--415}\BibitemShut {NoStop}%
\bibitem [{\citenamefont {Barrett}(2014)}]{BARRETT2014168}%
  \BibitemOpen
  \bibfield  {author} {\bibinfo {author} {\bibfnamefont {J.~A.}\ \bibnamefont
  {Barrett}},\ }\href {\doibase https://doi.org/10.1016/j.shpsb.2014.08.004}
  {\bibfield  {journal} {\bibinfo  {journal} {Studies in History and Philosophy
  of Science Part B: Studies in History and Philosophy of Modern Physics}\
  }\textbf {\bibinfo {volume} {48}},\ \bibinfo {pages} {168} (\bibinfo {year}
  {2014})},\ \bibinfo {note} {relativistic Causality}\BibitemShut {NoStop}%
\bibitem [{\citenamefont {Barrett}(2005)}]{barrett_2005}%
  \BibitemOpen
  \bibfield  {author} {\bibinfo {author} {\bibfnamefont {J.~A.}\ \bibnamefont
  {Barrett}},\ }\href {\doibase 10.1086/508948} {\bibfield  {journal} {\bibinfo
   {journal} {Philosophy of Science}\ }\textbf {\bibinfo {volume} {72}},\
  \bibinfo {pages} {802–813} (\bibinfo {year} {2005})}\BibitemShut {NoStop}%
\bibitem [{\citenamefont {Barrett}(2002)}]{Barrett2002}%
  \BibitemOpen
  \bibfield  {author} {\bibinfo {author} {\bibfnamefont {J.~A.}\ \bibnamefont
  {Barrett}},\ }in\ \href
  {http://ebookcentral.proquest.com/lib/oxford/detail.action?docID=1681666}
  {\emph {\bibinfo {booktitle} {Ontological Aspects Of Quantum Field
  Theory}}},\ \bibinfo {editor} {edited by\ \bibinfo {editor} {\bibfnamefont
  {M.}~\bibnamefont {Kuhlmann}}, \bibinfo {editor} {\bibfnamefont
  {H.}~\bibnamefont {Lyre}}, \bibinfo {editor} {\bibfnamefont {A.}~\bibnamefont
  {Wayne}}, \ and\ \bibinfo {editor} {\bibfnamefont {H.~T.}\ \bibnamefont
  {Leong}}}\ (\bibinfo  {publisher} {World Scientific Publishing Company},\
  \bibinfo {address} {Singapore, SINGAPORE},\ \bibinfo {year} {2002})\ \bibinfo
  {edition} {{W}inter 2021}\ ed.\BibitemShut {Stop}%
\bibitem [{\citenamefont {Kuhlmann}\ \emph {et~al.}(2002)\citenamefont
  {Kuhlmann}, \citenamefont {Lyre}, \citenamefont {Wayne},\ and\ \citenamefont
  {Leong}}]{1681666}%
  \BibitemOpen
  \bibfield  {author} {\bibinfo {author} {\bibfnamefont {M.}~\bibnamefont
  {Kuhlmann}}, \bibinfo {author} {\bibfnamefont {H.}~\bibnamefont {Lyre}},
  \bibinfo {author} {\bibfnamefont {A.}~\bibnamefont {Wayne}}, \ and\ \bibinfo
  {author} {\bibfnamefont {H.~T.}\ \bibnamefont {Leong}},\ }\href
  {http://ebookcentral.proquest.com/lib/oxford/detail.action?docID=1681666}
  {\emph {\bibinfo {title} {Ontological Aspects Of Quantum Field Theory}}}\
  (\bibinfo  {publisher} {World Scientific Publishing Company},\ \bibinfo
  {address} {Singapore, SINGAPORE},\ \bibinfo {year} {2002})\BibitemShut
  {NoStop}%
\bibitem [{\citenamefont {Halvorson}\ and\ \citenamefont
  {Muger}(2007)}]{Halvorson:2006wj}%
  \BibitemOpen
  \bibfield  {author} {\bibinfo {author} {\bibfnamefont {H.}~\bibnamefont
  {Halvorson}}\ and\ \bibinfo {author} {\bibfnamefont {M.}~\bibnamefont
  {Muger}},\ }\enquote {\bibinfo {title} {{Algebraic quantum field theory}},}\
  in\ \href {\doibase 10.1016/B978-044451560-5/50011-7} {\emph {\bibinfo
  {booktitle} {{Philosophy of physics}}}},\ \bibinfo {editor} {edited by\
  \bibinfo {editor} {\bibfnamefont {J.}~\bibnamefont {Butterfield}}\ and\
  \bibinfo {editor} {\bibfnamefont {J.}~\bibnamefont {Earman}}}\ (\bibinfo
  {publisher} {North-Holland},\ \bibinfo {year} {2007})\ pp.\ \bibinfo {pages}
  {731--864},\ \Eprint {http://arxiv.org/abs/math-ph/0602036}
  {arXiv:math-ph/0602036} \BibitemShut {NoStop}%
\bibitem [{\citenamefont {Halvorson}\ and\ \citenamefont
  {Clifton}(2002)}]{OAOQFTChap10}%
  \BibitemOpen
  \bibfield  {author} {\bibinfo {author} {\bibfnamefont {H.}~\bibnamefont
  {Halvorson}}\ and\ \bibinfo {author} {\bibfnamefont {R.}~\bibnamefont
  {Clifton}},\ }in\ \href
  {http://ebookcentral.proquest.com/lib/oxford/detail.action?docID=1681666}
  {\emph {\bibinfo {booktitle} {Ontological Aspects Of Quantum Field
  Theory}}},\ \bibinfo {editor} {edited by\ \bibinfo {editor} {\bibfnamefont
  {M.}~\bibnamefont {Kuhlmann}}, \bibinfo {editor} {\bibfnamefont
  {H.}~\bibnamefont {Lyre}}, \bibinfo {editor} {\bibfnamefont {A.}~\bibnamefont
  {Wayne}}, \ and\ \bibinfo {editor} {\bibfnamefont {H.~T.}\ \bibnamefont
  {Leong}}}\ (\bibinfo  {publisher} {World Scientific Publishing Company},\
  \bibinfo {address} {Singapore, SINGAPORE},\ \bibinfo {year} {2002})\ \bibinfo
  {edition} {{W}inter 2021}\ ed.\BibitemShut {Stop}%
\bibitem [{\citenamefont {Dieks}(2002)}]{OAOQFTChap11}%
  \BibitemOpen
  \bibfield  {author} {\bibinfo {author} {\bibfnamefont {D.}~\bibnamefont
  {Dieks}},\ }in\ \href
  {http://ebookcentral.proquest.com/lib/oxford/detail.action?docID=1681666}
  {\emph {\bibinfo {booktitle} {Ontological Aspects Of Quantum Field
  Theory}}},\ \bibinfo {editor} {edited by\ \bibinfo {editor} {\bibfnamefont
  {M.}~\bibnamefont {Kuhlmann}}, \bibinfo {editor} {\bibfnamefont
  {H.}~\bibnamefont {Lyre}}, \bibinfo {editor} {\bibfnamefont {A.}~\bibnamefont
  {Wayne}}, \ and\ \bibinfo {editor} {\bibfnamefont {H.~T.}\ \bibnamefont
  {Leong}}}\ (\bibinfo  {publisher} {World Scientific Publishing Company},\
  \bibinfo {address} {Singapore, SINGAPORE},\ \bibinfo {year} {2002})\ \bibinfo
  {edition} {{W}inter 2021}\ ed.\BibitemShut {Stop}%
\bibitem [{\citenamefont {Redhead}(1995)}]{Redhead1995}%
  \BibitemOpen
  \bibfield  {author} {\bibinfo {author} {\bibfnamefont {M.}~\bibnamefont
  {Redhead}},\ }\href {\doibase 10.1007/BF02054660} {\bibfield  {journal}
  {\bibinfo  {journal} {Foundations of Physics}\ }\textbf {\bibinfo {volume}
  {25}},\ \bibinfo {pages} {123} (\bibinfo {year} {1995})}\BibitemShut
  {NoStop}%
\bibitem [{\citenamefont {Malament}(1996)}]{Malament1996-MALIDO}%
  \BibitemOpen
  \bibfield  {author} {\bibinfo {author} {\bibfnamefont {D.}~\bibnamefont
  {Malament}},\ }in\ \href@noop {} {\emph {\bibinfo {booktitle} {Perspectives
  on Quantum Reality}}},\ \bibinfo {editor} {edited by\ \bibinfo {editor}
  {\bibfnamefont {R.}~\bibnamefont {Clifton}}}\ (\bibinfo  {publisher} {Kluwer
  Academic Publishers},\ \bibinfo {year} {1996})\BibitemShut {NoStop}%
\bibitem [{\citenamefont {Papageorgiou}\ and\ \citenamefont
  {Fraser}(2023)}]{papageorgiou2023eliminating}%
  \BibitemOpen
  \bibfield  {author} {\bibinfo {author} {\bibfnamefont {M.}~\bibnamefont
  {Papageorgiou}}\ and\ \bibinfo {author} {\bibfnamefont {D.}~\bibnamefont
  {Fraser}},\ }\href@noop {} {\enquote {\bibinfo {title} {Eliminating the
  "impossible": Recent progress on local measurement theory for quantum field
  theory},}\ } (\bibinfo {year} {2023}),\ \Eprint
  {http://arxiv.org/abs/2307.08524} {arXiv:2307.08524 [quant-ph]} \BibitemShut
  {NoStop}%
\bibitem [{\citenamefont {Curiel}(2020)}]{Curiel}%
  \BibitemOpen
  \bibfield  {author} {\bibinfo {author} {\bibfnamefont {E.}~\bibnamefont
  {Curiel}},\ }\href@noop {} {\enquote {\bibinfo {title} {Schematizing the
  observer and the epistemic content of theories},}\ } (\bibinfo {year}
  {2020}),\ \Eprint {http://arxiv.org/abs/1903.02182} {arXiv:1903.02182
  [physics.hist-ph]} \BibitemShut {NoStop}%
\bibitem [{\citenamefont {Giovanelli}(2014)}]{LegitSin}%
  \BibitemOpen
  \bibfield  {author} {\bibinfo {author} {\bibfnamefont {M.}~\bibnamefont
  {Giovanelli}},\ }\href {http://philsci-archive.pitt.edu/10922/} {\enquote
  {\bibinfo {title} {"but one must not legalize the mentioned sin".
  phenomenological vs. dynamical treatment of rods and clocks in einstein's
  thought},}\ } (\bibinfo {year} {2014})\BibitemShut {NoStop}%
\bibitem [{\citenamefont {Brown}(2005)}]{BrownHarvey}%
  \BibitemOpen
  \bibfield  {author} {\bibinfo {author} {\bibfnamefont {H.~R.}\ \bibnamefont
  {Brown}},\ }\href@noop {} {\emph {\bibinfo {title} {Physical Relativity}}}\
  (\bibinfo  {publisher} {Oxford University Press},\ \bibinfo {address}
  {Oxford},\ \bibinfo {year} {2005})\BibitemShut {NoStop}%
\bibitem [{\citenamefont {{D. Wallace}}(2021)}]{WallaceBlueSkyTalk}%
  \BibitemOpen
  \bibfield  {author} {\bibinfo {author} {\bibnamefont {{D. Wallace}}},\ }\href
  {https://youtu.be/tEwgNbfYn2E?t=2254} {\enquote {\bibinfo {title} {The sky is
  blue, and other reasons physics needs the everett interpretation},}\ }
  (\bibinfo {year} {2021}),\ \bibinfo {note} {[Oxford Philosophy of Physics
  Seminar, Michaelmas Term 2021, 4th Nov, Timestamp 37:34]}\BibitemShut
  {NoStop}%
\bibitem [{\citenamefont {Wallace}(2022)}]{WallaceBlueSkyPaper}%
  \BibitemOpen
  \bibfield  {author} {\bibinfo {author} {\bibfnamefont {D.}~\bibnamefont
  {Wallace}},\ }\href {\doibase 10.48550/ARXIV.2205.00568} {\enquote {\bibinfo
  {title} {The sky is blue, and other reasons quantum mechanics is not
  underdetermined by evidence},}\ } (\bibinfo {year} {2022})\BibitemShut
  {NoStop}%
\bibitem [{\citenamefont {Polo-Gómez}\ \emph {et~al.}(2021)\citenamefont
  {Polo-Gómez}, \citenamefont {Garay},\ and\ \citenamefont
  {Martín-Martínez}}]{pologomez2021detectorbased}%
  \BibitemOpen
  \bibfield  {author} {\bibinfo {author} {\bibfnamefont {J.}~\bibnamefont
  {Polo-Gómez}}, \bibinfo {author} {\bibfnamefont {L.~J.}\ \bibnamefont
  {Garay}}, \ and\ \bibinfo {author} {\bibfnamefont {E.}~\bibnamefont
  {Martín-Martínez}},\ }\href@noop {} {\enquote {\bibinfo {title} {A
  detector-based measurement theory for quantum field theory},}\ } (\bibinfo
  {year} {2021}),\ \Eprint {http://arxiv.org/abs/2108.02793} {arXiv:2108.02793
  [quant-ph]} \BibitemShut {NoStop}%
\bibitem [{\citenamefont {Jubb}(2022)}]{Jubb2022}%
  \BibitemOpen
  \bibfield  {author} {\bibinfo {author} {\bibfnamefont {I.}~\bibnamefont
  {Jubb}},\ }\href {\doibase 10.1103/PhysRevD.105.025003} {\bibfield  {journal}
  {\bibinfo  {journal} {Phys. Rev. D}\ }\textbf {\bibinfo {volume} {105}},\
  \bibinfo {pages} {025003} (\bibinfo {year} {2022})}\BibitemShut {NoStop}%
\bibitem [{\citenamefont {Borsten}\ \emph {et~al.}(2021)\citenamefont
  {Borsten}, \citenamefont {Jubb},\ and\ \citenamefont
  {Kells}}]{BorstenJubbKells}%
  \BibitemOpen
  \bibfield  {author} {\bibinfo {author} {\bibfnamefont {L.}~\bibnamefont
  {Borsten}}, \bibinfo {author} {\bibfnamefont {I.}~\bibnamefont {Jubb}}, \
  and\ \bibinfo {author} {\bibfnamefont {G.}~\bibnamefont {Kells}},\ }\href
  {\doibase 10.1103/PhysRevD.104.025012} {\bibfield  {journal} {\bibinfo
  {journal} {Phys. Rev. D}\ }\textbf {\bibinfo {volume} {104}},\ \bibinfo
  {pages} {025012} (\bibinfo {year} {2021})}\BibitemShut {NoStop}%
\bibitem [{\citenamefont {Fewster}\ and\ \citenamefont
  {Verch}(2020)}]{fewster1}%
  \BibitemOpen
  \bibfield  {author} {\bibinfo {author} {\bibfnamefont {C.~J.}\ \bibnamefont
  {Fewster}}\ and\ \bibinfo {author} {\bibfnamefont {R.}~\bibnamefont
  {Verch}},\ }\href {\doibase 10.1007/s00220-020-03800-6} {\bibfield  {journal}
  {\bibinfo  {journal} {Commun. Math. Phys.}\ }\textbf {\bibinfo {volume}
  {378}},\ \bibinfo {pages} {851–889} (\bibinfo {year} {2020})}\BibitemShut
  {NoStop}%
\bibitem [{\citenamefont {Fewster}(2019)}]{fewster2}%
  \BibitemOpen
  \bibfield  {author} {\bibinfo {author} {\bibfnamefont {C.~J.}\ \bibnamefont
  {Fewster}},\ }\href@noop {} {\enquote {\bibinfo {title} {A generally
  covariant measurement scheme for quantum field theory in curved
  spacetimes},}\ } (\bibinfo {year} {2019}),\ \Eprint
  {http://arxiv.org/abs/1904.06944} {arXiv:1904.06944 [gr-qc]} \BibitemShut
  {NoStop}%
\bibitem [{\citenamefont {Bostelmann}\ \emph {et~al.}(2021)\citenamefont
  {Bostelmann}, \citenamefont {Fewster},\ and\ \citenamefont
  {Ruep}}]{fewster3}%
  \BibitemOpen
  \bibfield  {author} {\bibinfo {author} {\bibfnamefont {H.}~\bibnamefont
  {Bostelmann}}, \bibinfo {author} {\bibfnamefont {C.~J.}\ \bibnamefont
  {Fewster}}, \ and\ \bibinfo {author} {\bibfnamefont {M.~H.}\ \bibnamefont
  {Ruep}},\ }\href {\doibase 10.1103/PhysRevD.103.025017} {\bibfield  {journal}
  {\bibinfo  {journal} {Phys. Rev. D}\ }\textbf {\bibinfo {volume} {103}},\
  \bibinfo {pages} {025017} (\bibinfo {year} {2021})}\BibitemShut {NoStop}%
\bibitem [{\citenamefont {Anastopoulos}\ and\ \citenamefont
  {Savvidou}(2022)}]{Anastopoulos2022}%
  \BibitemOpen
  \bibfield  {author} {\bibinfo {author} {\bibfnamefont {C.}~\bibnamefont
  {Anastopoulos}}\ and\ \bibinfo {author} {\bibfnamefont {N.}~\bibnamefont
  {Savvidou}},\ }\href {\doibase 10.3390/e24010004} {\bibfield  {journal}
  {\bibinfo  {journal} {Entropy}\ }\textbf {\bibinfo {volume} {24}} (\bibinfo
  {year} {2022}),\ 10.3390/e24010004}\BibitemShut {NoStop}%
\bibitem [{\citenamefont {Sorkin}(1993)}]{Sorkin}%
  \BibitemOpen
  \bibfield  {author} {\bibinfo {author} {\bibfnamefont {R.~D.}\ \bibnamefont
  {Sorkin}},\ }\href@noop {} {\enquote {\bibinfo {title} {Impossible
  measurements on quantum fields},}\ } (\bibinfo {year} {1993}),\ \Eprint
  {http://arxiv.org/abs/gr-qc/9302018} {arXiv:gr-qc/9302018 [gr-qc]}
  \BibitemShut {NoStop}%
\bibitem [{\citenamefont {Grimmer}\ \emph {et~al.}(2021)\citenamefont
  {Grimmer}, \citenamefont {Torres},\ and\ \citenamefont
  {Mart\'{\i}n-Mart\'{\i}nez}}]{TaleOfTwo}%
  \BibitemOpen
  \bibfield  {author} {\bibinfo {author} {\bibfnamefont {D.}~\bibnamefont
  {Grimmer}}, \bibinfo {author} {\bibfnamefont {B.~d. S.~L.}\ \bibnamefont
  {Torres}}, \ and\ \bibinfo {author} {\bibfnamefont {E.}~\bibnamefont
  {Mart\'{\i}n-Mart\'{\i}nez}},\ }\href {\doibase 10.1103/PhysRevD.104.085014}
  {\bibfield  {journal} {\bibinfo  {journal} {Phys. Rev. D}\ }\textbf {\bibinfo
  {volume} {104}},\ \bibinfo {pages} {085014} (\bibinfo {year}
  {2021})}\BibitemShut {NoStop}%
\bibitem [{\citenamefont {Ruep}(2021)}]{Ruep2021}%
  \BibitemOpen
  \bibfield  {author} {\bibinfo {author} {\bibfnamefont {M.~H.}\ \bibnamefont
  {Ruep}},\ }\href {\doibase 10.1088/1361-6382/ac1b08} {\bibfield  {journal}
  {\bibinfo  {journal} {Classical and Quantum Gravity}\ }\textbf {\bibinfo
  {volume} {38}},\ \bibinfo {pages} {195029} (\bibinfo {year}
  {2021})}\BibitemShut {NoStop}%
\bibitem [{\citenamefont {de~Ram\'on}\ \emph {et~al.}(2021)\citenamefont
  {de~Ram\'on}, \citenamefont {Papageorgiou},\ and\ \citenamefont
  {Mart\'{\i}n-Mart\'{\i}nez}}]{JoseMariaEdu}%
  \BibitemOpen
  \bibfield  {author} {\bibinfo {author} {\bibfnamefont {J.}~\bibnamefont
  {de~Ram\'on}}, \bibinfo {author} {\bibfnamefont {M.}~\bibnamefont
  {Papageorgiou}}, \ and\ \bibinfo {author} {\bibfnamefont {E.}~\bibnamefont
  {Mart\'{\i}n-Mart\'{\i}nez}},\ }\href {\doibase 10.1103/PhysRevD.103.085002}
  {\bibfield  {journal} {\bibinfo  {journal} {Phys. Rev. D}\ }\textbf {\bibinfo
  {volume} {103}},\ \bibinfo {pages} {085002} (\bibinfo {year}
  {2021})}\BibitemShut {NoStop}%
\bibitem [{\citenamefont {Dowker}(2011)}]{Dowker}%
  \BibitemOpen
  \bibfield  {author} {\bibinfo {author} {\bibfnamefont {F.}~\bibnamefont
  {Dowker}},\ }\href@noop {} {\enquote {\bibinfo {title} {Useless qubits in
  "relativistic quantum information"},}\ } (\bibinfo {year} {2011}),\ \Eprint
  {http://arxiv.org/abs/1111.2308} {arXiv:1111.2308 [quant-ph]} \BibitemShut
  {NoStop}%
\bibitem [{\citenamefont {Benincasa}\ \emph {et~al.}(2014)\citenamefont
  {Benincasa}, \citenamefont {Borsten}, \citenamefont {Buck},\ and\
  \citenamefont {Dowker}}]{Dowker2}%
  \BibitemOpen
  \bibfield  {author} {\bibinfo {author} {\bibfnamefont {D.~M.~T.}\
  \bibnamefont {Benincasa}}, \bibinfo {author} {\bibfnamefont {L.}~\bibnamefont
  {Borsten}}, \bibinfo {author} {\bibfnamefont {M.}~\bibnamefont {Buck}}, \
  and\ \bibinfo {author} {\bibfnamefont {F.}~\bibnamefont {Dowker}},\ }\href
  {\doibase 10.1088/0264-9381/31/7/075007} {\bibfield  {journal} {\bibinfo
  {journal} {Class. Quantum Gravity}\ }\textbf {\bibinfo {volume} {31}},\
  \bibinfo {pages} {075007} (\bibinfo {year} {2014})}\BibitemShut {NoStop}%
\bibitem [{\citenamefont {Borsten}\ \emph {et~al.}(2019)\citenamefont
  {Borsten}, \citenamefont {Jubb},\ and\ \citenamefont {Kells}}]{borsten}%
  \BibitemOpen
  \bibfield  {author} {\bibinfo {author} {\bibfnamefont {L.}~\bibnamefont
  {Borsten}}, \bibinfo {author} {\bibfnamefont {I.}~\bibnamefont {Jubb}}, \
  and\ \bibinfo {author} {\bibfnamefont {G.}~\bibnamefont {Kells}},\
  }\href@noop {} {\enquote {\bibinfo {title} {Impossible measurements
  revisited},}\ } (\bibinfo {year} {2019}),\ \Eprint
  {http://arxiv.org/abs/1912.06141} {arXiv:1912.06141 [quant-ph]} \BibitemShut
  {NoStop}%
\bibitem [{\citenamefont {Ortega}\ \emph {et~al.}(2019)\citenamefont {Ortega},
  \citenamefont {McKay}, \citenamefont {Alhambra},\ and\ \citenamefont
  {\mbox{Mart\'{\i}n-Mart\'{\i}nez}}}]{alvaro}%
  \BibitemOpen
  \bibfield  {author} {\bibinfo {author} {\bibfnamefont {A.}~\bibnamefont
  {Ortega}}, \bibinfo {author} {\bibfnamefont {E.}~\bibnamefont {McKay}},
  \bibinfo {author} {\bibfnamefont {A.~M.}\ \bibnamefont {Alhambra}}, \ and\
  \bibinfo {author} {\bibfnamefont {E.}~\bibnamefont
  {\mbox{Mart\'{\i}n-Mart\'{\i}nez}}},\ }\href {\doibase
  10.1103/PhysRevLett.122.240604} {\bibfield  {journal} {\bibinfo  {journal}
  {Phys. Rev. Lett.}\ }\textbf {\bibinfo {volume} {122}},\ \bibinfo {pages}
  {240604} (\bibinfo {year} {2019})}\BibitemShut {NoStop}%
\bibitem [{\citenamefont {Teixid\'o-Bonfill}\ \emph {et~al.}(2020)\citenamefont
  {Teixid\'o-Bonfill}, \citenamefont {Ortega},\ and\ \citenamefont
  {Mart\'{\i}n-Mart\'{\i}nez}}]{Adam}%
  \BibitemOpen
  \bibfield  {author} {\bibinfo {author} {\bibfnamefont {A.}~\bibnamefont
  {Teixid\'o-Bonfill}}, \bibinfo {author} {\bibfnamefont {A.}~\bibnamefont
  {Ortega}}, \ and\ \bibinfo {author} {\bibfnamefont {E.}~\bibnamefont
  {Mart\'{\i}n-Mart\'{\i}nez}},\ }\href {\doibase 10.1103/PhysRevA.102.052219}
  {\bibfield  {journal} {\bibinfo  {journal} {Phys. Rev. A}\ }\textbf {\bibinfo
  {volume} {102}},\ \bibinfo {pages} {052219} (\bibinfo {year}
  {2020})}\BibitemShut {NoStop}%
\bibitem [{\citenamefont {Fewster}\ \emph {et~al.}(2022)\citenamefont
  {Fewster}, \citenamefont {Jubb},\ and\ \citenamefont
  {Ruep}}]{MeasurementSchemeFV}%
  \BibitemOpen
  \bibfield  {author} {\bibinfo {author} {\bibfnamefont {C.~J.}\ \bibnamefont
  {Fewster}}, \bibinfo {author} {\bibfnamefont {I.}~\bibnamefont {Jubb}}, \
  and\ \bibinfo {author} {\bibfnamefont {M.~H.}\ \bibnamefont {Ruep}},\ }\href
  {\doibase 10.48550/ARXIV.2203.09529} {\enquote {\bibinfo {title} {Asymptotic
  measurement schemes for every observable of a quantum field theory},}\ }
  (\bibinfo {year} {2022})\BibitemShut {NoStop}%
\bibitem [{\citenamefont {Unruh}(1976)}]{Unruh1976}%
  \BibitemOpen
  \bibfield  {author} {\bibinfo {author} {\bibfnamefont {W.~G.}\ \bibnamefont
  {Unruh}},\ }\href {\doibase 10.1103/PhysRevD.14.870} {\bibfield  {journal}
  {\bibinfo  {journal} {Phys. Rev. D}\ }\textbf {\bibinfo {volume} {14}},\
  \bibinfo {pages} {870} (\bibinfo {year} {1976})}\BibitemShut {NoStop}%
\bibitem [{\citenamefont {Lin}\ and\ \citenamefont {Hu}(2007)}]{BLHu2007}%
  \BibitemOpen
  \bibfield  {author} {\bibinfo {author} {\bibfnamefont {S.-Y.}\ \bibnamefont
  {Lin}}\ and\ \bibinfo {author} {\bibfnamefont {B.~L.}\ \bibnamefont {Hu}},\
  }\href {\doibase 10.1103/PhysRevD.76.064008} {\bibfield  {journal} {\bibinfo
  {journal} {Phys. Rev. D}\ }\textbf {\bibinfo {volume} {76}},\ \bibinfo
  {pages} {064008} (\bibinfo {year} {2007})}\BibitemShut {NoStop}%
\bibitem [{\citenamefont {Brown}\ \emph {et~al.}(2013)\citenamefont {Brown},
  \citenamefont {Mart\'{\i}n-Mart\'{\i}nez}, \citenamefont {Menicucci},\ and\
  \citenamefont {Mann}}]{Brown2013}%
  \BibitemOpen
  \bibfield  {author} {\bibinfo {author} {\bibfnamefont {E.~G.}\ \bibnamefont
  {Brown}}, \bibinfo {author} {\bibfnamefont {E.}~\bibnamefont
  {Mart\'{\i}n-Mart\'{\i}nez}}, \bibinfo {author} {\bibfnamefont {N.~C.}\
  \bibnamefont {Menicucci}}, \ and\ \bibinfo {author} {\bibfnamefont {R.~B.}\
  \bibnamefont {Mann}},\ }\href {\doibase 10.1103/PhysRevD.87.084062}
  {\bibfield  {journal} {\bibinfo  {journal} {Phys. Rev. D}\ }\textbf {\bibinfo
  {volume} {87}},\ \bibinfo {pages} {084062} (\bibinfo {year}
  {2013})}\BibitemShut {NoStop}%
\bibitem [{\citenamefont {Hotta}\ \emph {et~al.}(2020)\citenamefont {Hotta},
  \citenamefont {Kempf}, \citenamefont {Mart\'{\i}n-Mart\'{\i}nez},
  \citenamefont {Tomitsuka},\ and\ \citenamefont {Yamaguchi}}]{Hotta2020}%
  \BibitemOpen
  \bibfield  {author} {\bibinfo {author} {\bibfnamefont {M.}~\bibnamefont
  {Hotta}}, \bibinfo {author} {\bibfnamefont {A.}~\bibnamefont {Kempf}},
  \bibinfo {author} {\bibfnamefont {E.}~\bibnamefont
  {Mart\'{\i}n-Mart\'{\i}nez}}, \bibinfo {author} {\bibfnamefont
  {T.}~\bibnamefont {Tomitsuka}}, \ and\ \bibinfo {author} {\bibfnamefont
  {K.}~\bibnamefont {Yamaguchi}},\ }\href {\doibase
  10.1103/PhysRevD.101.085017} {\bibfield  {journal} {\bibinfo  {journal}
  {Phys. Rev. D}\ }\textbf {\bibinfo {volume} {101}},\ \bibinfo {pages}
  {085017} (\bibinfo {year} {2020})}\BibitemShut {NoStop}%
\bibitem [{\citenamefont {Tjoa}\ and\ \citenamefont
  {Mart\'{\i}n-Mart\'{\i}nez}(2020)}]{Zeromode}%
  \BibitemOpen
  \bibfield  {author} {\bibinfo {author} {\bibfnamefont {E.}~\bibnamefont
  {Tjoa}}\ and\ \bibinfo {author} {\bibfnamefont {E.}~\bibnamefont
  {Mart\'{\i}n-Mart\'{\i}nez}},\ }\href {\doibase 10.1103/PhysRevD.101.125020}
  {\bibfield  {journal} {\bibinfo  {journal} {Phys. Rev. D}\ }\textbf {\bibinfo
  {volume} {101}},\ \bibinfo {pages} {125020} (\bibinfo {year}
  {2020})}\BibitemShut {NoStop}%
\bibitem [{\citenamefont {Valentini}(1991)}]{Valentini1991}%
  \BibitemOpen
  \bibfield  {author} {\bibinfo {author} {\bibfnamefont {A.}~\bibnamefont
  {Valentini}},\ }\href {\doibase
  http://dx.doi.org/10.1016/0375-9601(91)90952-5} {\bibfield  {journal}
  {\bibinfo  {journal} {Phys. Lett. A}\ }\textbf {\bibinfo {volume} {153}},\
  \bibinfo {pages} {321 } (\bibinfo {year} {1991})}\BibitemShut {NoStop}%
\bibitem [{\citenamefont {Reznik}(2003)}]{Reznik2003}%
  \BibitemOpen
  \bibfield  {author} {\bibinfo {author} {\bibfnamefont {B.}~\bibnamefont
  {Reznik}},\ }\href {\doibase 10.1023/A:1022875910744} {\bibfield  {journal}
  {\bibinfo  {journal} {Found. Phys.}\ }\textbf {\bibinfo {volume} {33}},\
  \bibinfo {pages} {167} (\bibinfo {year} {2003})}\BibitemShut {NoStop}%
\bibitem [{\citenamefont {Pozas-Kerstjens}\ and\ \citenamefont
  {Mart\'{i}n-Mart\'{i}nez}(2015)}]{Pozas-Kerstjens:2015}%
  \BibitemOpen
  \bibfield  {author} {\bibinfo {author} {\bibfnamefont {A.}~\bibnamefont
  {Pozas-Kerstjens}}\ and\ \bibinfo {author} {\bibfnamefont {E.}~\bibnamefont
  {Mart\'{i}n-Mart\'{i}nez}},\ }\href {\doibase 10.1103/PhysRevD.92.064042}
  {\bibfield  {journal} {\bibinfo  {journal} {Phys. Rev. D}\ }\textbf {\bibinfo
  {volume} {92}},\ \bibinfo {pages} {064042} (\bibinfo {year}
  {2015})}\BibitemShut {NoStop}%
\bibitem [{\citenamefont {Steeg}\ and\ \citenamefont
  {Menicucci}(2009)}]{Menicucci}%
  \BibitemOpen
  \bibfield  {author} {\bibinfo {author} {\bibfnamefont {G.~V.}\ \bibnamefont
  {Steeg}}\ and\ \bibinfo {author} {\bibfnamefont {N.~C.}\ \bibnamefont
  {Menicucci}},\ }\href {\doibase 10.1103/PhysRevD.79.044027} {\bibfield
  {journal} {\bibinfo  {journal} {Phys. Rev. D}\ }\textbf {\bibinfo {volume}
  {79}},\ \bibinfo {pages} {044027} (\bibinfo {year} {2009})}\BibitemShut
  {NoStop}%
\bibitem [{\citenamefont {Mart\'{\i}n-Mart\'{\i}nez}\ \emph
  {et~al.}(2016)\citenamefont {Mart\'{\i}n-Mart\'{\i}nez}, \citenamefont
  {Smith},\ and\ \citenamefont {Terno}}]{Terno2016}%
  \BibitemOpen
  \bibfield  {author} {\bibinfo {author} {\bibfnamefont {E.}~\bibnamefont
  {Mart\'{\i}n-Mart\'{\i}nez}}, \bibinfo {author} {\bibfnamefont {A.~R.~H.}\
  \bibnamefont {Smith}}, \ and\ \bibinfo {author} {\bibfnamefont {D.~R.}\
  \bibnamefont {Terno}},\ }\href {\doibase 10.1103/PhysRevD.93.044001}
  {\bibfield  {journal} {\bibinfo  {journal} {Phys. Rev. D}\ }\textbf {\bibinfo
  {volume} {93}},\ \bibinfo {pages} {044001} (\bibinfo {year}
  {2016})}\BibitemShut {NoStop}%
\bibitem [{\citenamefont {Mart{\'{\i}}n-Mart{\'{\i}}nez}\ and\ \citenamefont
  {Menicucci}(2012)}]{Cosmo}%
  \BibitemOpen
  \bibfield  {author} {\bibinfo {author} {\bibfnamefont {E.}~\bibnamefont
  {Mart{\'{\i}}n-Mart{\'{\i}}nez}}\ and\ \bibinfo {author} {\bibfnamefont
  {N.~C.}\ \bibnamefont {Menicucci}},\ }\href {\doibase
  10.1088/0264-9381/29/22/224003} {\bibfield  {journal} {\bibinfo  {journal}
  {Class. Quantum Gravity}\ }\textbf {\bibinfo {volume} {29}},\ \bibinfo
  {pages} {224003} (\bibinfo {year} {2012})}\BibitemShut {NoStop}%
\bibitem [{\citenamefont {Henderson}\ \emph {et~al.}(2018)\citenamefont
  {Henderson}, \citenamefont {Hennigar}, \citenamefont {Mann}, \citenamefont
  {Smith},\ and\ \citenamefont {Zhang}}]{Henderson2018}%
  \BibitemOpen
  \bibfield  {author} {\bibinfo {author} {\bibfnamefont {L.~J.}\ \bibnamefont
  {Henderson}}, \bibinfo {author} {\bibfnamefont {R.~A.}\ \bibnamefont
  {Hennigar}}, \bibinfo {author} {\bibfnamefont {R.~B.}\ \bibnamefont {Mann}},
  \bibinfo {author} {\bibfnamefont {A.~R.~H.}\ \bibnamefont {Smith}}, \ and\
  \bibinfo {author} {\bibfnamefont {J.}~\bibnamefont {Zhang}},\ }\href
  {\doibase 10.1088/1361-6382/aae27e} {\bibfield  {journal} {\bibinfo
  {journal} {Class. Quantum Gravity}\ }\textbf {\bibinfo {volume} {35}},\
  \bibinfo {pages} {21LT02} (\bibinfo {year} {2018})}\BibitemShut {NoStop}%
\bibitem [{\citenamefont {Schlosshauer}\ and\ \citenamefont
  {Camilleri}(2010)}]{SchlosshauerMaximilian2010WcDa}%
  \BibitemOpen
  \bibfield  {author} {\bibinfo {author} {\bibfnamefont {M.}~\bibnamefont
  {Schlosshauer}}\ and\ \bibinfo {author} {\bibfnamefont {K.}~\bibnamefont
  {Camilleri}}\ }(\bibinfo {year} {2010})\BibitemShut {NoStop}%
\bibitem [{\citenamefont {Wallace}(2012)}]{WallaceEmergentMultiverse}%
  \BibitemOpen
  \bibfield  {author} {\bibinfo {author} {\bibfnamefont {D.}~\bibnamefont
  {Wallace}},\ }\href@noop {} {\emph {\bibinfo {title} {The emergent multiverse
  [electronic resource] : quantum theory according to the Everett
  interpretation}}},\ Oxford scholarship online\ (\bibinfo  {publisher} {Oxford
  University Press},\ \bibinfo {address} {Oxford},\ \bibinfo {year}
  {2012})\BibitemShut {NoStop}%
\bibitem [{\citenamefont {Schlosshauer}\ and\ \citenamefont
  {Camilleri}(2008)}]{SM2010}%
  \BibitemOpen
  \bibfield  {author} {\bibinfo {author} {\bibfnamefont {M.}~\bibnamefont
  {Schlosshauer}}\ and\ \bibinfo {author} {\bibfnamefont {K.}~\bibnamefont
  {Camilleri}},\ }\href {\doibase 10.48550/ARXIV.0804.1609} {\enquote {\bibinfo
  {title} {The quantum-to-classical transition: Bohr's doctrine of classical
  concepts, emergent classicality, and decoherence},}\ } (\bibinfo {year}
  {2008})\BibitemShut {NoStop}%
\bibitem [{\citenamefont {Zachos}\ \emph {et~al.}(2005)\citenamefont {Zachos},
  \citenamefont {Fairlie},\ and\ \citenamefont {Curtright}}]{QMPhaseSpace}%
  \BibitemOpen
  \bibfield  {author} {\bibinfo {author} {\bibfnamefont {C.~K.}\ \bibnamefont
  {Zachos}}, \bibinfo {author} {\bibfnamefont {D.~B.}\ \bibnamefont {Fairlie}},
  \ and\ \bibinfo {author} {\bibfnamefont {T.~L.}\ \bibnamefont {Curtright}},\
  }\href {\doibase 10.1142/5287} {\emph {\bibinfo {title} {Quantum Mechanics in
  Phase Space}}}\ (\bibinfo  {publisher} {WORLD SCIENTIFIC},\ \bibinfo {year}
  {2005})\ \Eprint
  {http://arxiv.org/abs/https://www.worldscientific.com/doi/pdf/10.1142/5287}
  {https://www.worldscientific.com/doi/pdf/10.1142/5287} \BibitemShut {NoStop}%
\bibitem [{\citenamefont {Rosaler}(2013)}]{Rosaler}%
  \BibitemOpen
  \bibfield  {author} {\bibinfo {author} {\bibfnamefont {J.}~\bibnamefont
  {Rosaler}},\ }\emph {\bibinfo {title} {Inter-theory relations in physics:
  case studies from quantum mechanics and quantum field theory}},\ \href@noop
  {} {Ph.D. thesis},\ \bibinfo  {school} {University of Oxford} (\bibinfo
  {year} {2013})\BibitemShut {NoStop}%
\bibitem [{\citenamefont {Barcel\'o}\ \emph {et~al.}(2012)\citenamefont
  {Barcel\'o}, \citenamefont {Carballo-Rubio}, \citenamefont {Garay},\ and\
  \citenamefont {G\'omez-Escalante}}]{PhysRevA.86.042120}%
  \BibitemOpen
  \bibfield  {author} {\bibinfo {author} {\bibfnamefont {C.}~\bibnamefont
  {Barcel\'o}}, \bibinfo {author} {\bibfnamefont {R.}~\bibnamefont
  {Carballo-Rubio}}, \bibinfo {author} {\bibfnamefont {L.~J.}\ \bibnamefont
  {Garay}}, \ and\ \bibinfo {author} {\bibfnamefont {R.}~\bibnamefont
  {G\'omez-Escalante}},\ }\href {\doibase 10.1103/PhysRevA.86.042120}
  {\bibfield  {journal} {\bibinfo  {journal} {Phys. Rev. A}\ }\textbf {\bibinfo
  {volume} {86}},\ \bibinfo {pages} {042120} (\bibinfo {year}
  {2012})}\BibitemShut {NoStop}%
\bibitem [{\citenamefont {{C. J. Fewster}}(2021)}]{FewsterRQITalk3}%
  \BibitemOpen
  \bibfield  {author} {\bibinfo {author} {\bibnamefont {{C. J. Fewster}}},\
  }\href {https://www.youtube.com/watch?v=ldBqBW0tCjc&t=585s} {\enquote
  {\bibinfo {title} {Local measurement of quantum fields in curved
  spacetimes},}\ } (\bibinfo {year} {2021}),\ \bibinfo {note} {[Relativistic
  {Q}uantum information-{O}nline 2020/21 - {W}aterloo {S}ession 03: {W}ednesday
  {F}ebruary 10th, timestamp 09:45]}\BibitemShut {NoStop}%
\bibitem [{\citenamefont {Bacciagaluppi}(2020)}]{sep-qm-decoherence}%
  \BibitemOpen
  \bibfield  {author} {\bibinfo {author} {\bibfnamefont {G.}~\bibnamefont
  {Bacciagaluppi}},\ }in\ \href@noop {} {\emph {\bibinfo {booktitle} {The
  {Stanford} Encyclopedia of Philosophy}}},\ \bibinfo {editor} {edited by\
  \bibinfo {editor} {\bibfnamefont {E.~N.}\ \bibnamefont {Zalta}}}\ (\bibinfo
  {publisher} {Metaphysics Research Lab, Stanford University},\ \bibinfo {year}
  {2020})\ \bibinfo {edition} {{F}all 2020}\ ed.\BibitemShut {Stop}%
\bibitem [{\citenamefont {Guryanova}\ \emph {et~al.}(2020)\citenamefont
  {Guryanova}, \citenamefont {Friis},\ and\ \citenamefont
  {Huber}}]{Guryanova2020idealprojective}%
  \BibitemOpen
  \bibfield  {author} {\bibinfo {author} {\bibfnamefont {Y.}~\bibnamefont
  {Guryanova}}, \bibinfo {author} {\bibfnamefont {N.}~\bibnamefont {Friis}}, \
  and\ \bibinfo {author} {\bibfnamefont {M.}~\bibnamefont {Huber}},\ }\href
  {\doibase 10.22331/q-2020-01-13-222} {\bibfield  {journal} {\bibinfo
  {journal} {{Quantum}}\ }\textbf {\bibinfo {volume} {4}},\ \bibinfo {pages}
  {222} (\bibinfo {year} {2020})}\BibitemShut {NoStop}%
\bibitem [{\citenamefont {Yu}\ \emph {et~al.}(2022)\citenamefont {Yu},
  \citenamefont {Liu}, \citenamefont {Li}, \citenamefont {Yan}, \citenamefont
  {Cao}, \citenamefont {Tan}, \citenamefont {Liang}, \citenamefont {Guo},
  \citenamefont {Cao}, \citenamefont {Lan}, \citenamefont {Zhang},
  \citenamefont {Zhou},\ and\ \citenamefont {Lu}}]{Yu2022}%
  \BibitemOpen
  \bibfield  {author} {\bibinfo {author} {\bibfnamefont {M.}~\bibnamefont
  {Yu}}, \bibinfo {author} {\bibfnamefont {K.}~\bibnamefont {Liu}}, \bibinfo
  {author} {\bibfnamefont {M.}~\bibnamefont {Li}}, \bibinfo {author}
  {\bibfnamefont {J.}~\bibnamefont {Yan}}, \bibinfo {author} {\bibfnamefont
  {C.}~\bibnamefont {Cao}}, \bibinfo {author} {\bibfnamefont {J.}~\bibnamefont
  {Tan}}, \bibinfo {author} {\bibfnamefont {J.}~\bibnamefont {Liang}}, \bibinfo
  {author} {\bibfnamefont {K.}~\bibnamefont {Guo}}, \bibinfo {author}
  {\bibfnamefont {W.}~\bibnamefont {Cao}}, \bibinfo {author} {\bibfnamefont
  {P.}~\bibnamefont {Lan}}, \bibinfo {author} {\bibfnamefont {Q.}~\bibnamefont
  {Zhang}}, \bibinfo {author} {\bibfnamefont {Y.}~\bibnamefont {Zhou}}, \ and\
  \bibinfo {author} {\bibfnamefont {P.}~\bibnamefont {Lu}},\ }\href {\doibase
  10.1038/s41377-022-00911-8} {\bibfield  {journal} {\bibinfo  {journal}
  {Light: Science {\&} Applications}\ }\textbf {\bibinfo {volume} {11}},\
  \bibinfo {pages} {215} (\bibinfo {year} {2022})}\BibitemShut {NoStop}%
\bibitem [{\citenamefont {Kant}(1893)}]{KantCritiqueOfPureReason}%
  \BibitemOpen
  \bibfield  {author} {\bibinfo {author} {\bibfnamefont {.-.}\ \bibnamefont
  {Kant}, \bibfnamefont {Immanuel}},\ }\href@noop {} {\emph {\bibinfo {title}
  {Critique of pure reason}}}\ (\bibinfo  {publisher} {G. Bell \& Sons, 1893},\
  \bibinfo {address} {England},\ \bibinfo {year} {1893})\BibitemShut {NoStop}%
\bibitem [{\citenamefont {Witten}(2018)}]{Witten}%
  \BibitemOpen
  \bibfield  {author} {\bibinfo {author} {\bibfnamefont {E.}~\bibnamefont
  {Witten}},\ }\href {\doibase 10.1103/RevModPhys.90.045003} {\bibfield
  {journal} {\bibinfo  {journal} {Rev. Mod. Phys.}\ }\textbf {\bibinfo {volume}
  {90}},\ \bibinfo {pages} {045003} (\bibinfo {year} {2018})}\BibitemShut
  {NoStop}%
\bibitem [{\citenamefont {Kronz}\ and\ \citenamefont
  {Lupher}(2021)}]{sep-qt-nvd}%
  \BibitemOpen
  \bibfield  {author} {\bibinfo {author} {\bibfnamefont {F.}~\bibnamefont
  {Kronz}}\ and\ \bibinfo {author} {\bibfnamefont {T.}~\bibnamefont {Lupher}},\
  }in\ \href@noop {} {\emph {\bibinfo {booktitle} {The {Stanford} Encyclopedia
  of Philosophy}}},\ \bibinfo {editor} {edited by\ \bibinfo {editor}
  {\bibfnamefont {E.~N.}\ \bibnamefont {Zalta}}}\ (\bibinfo  {publisher}
  {Metaphysics Research Lab, Stanford University},\ \bibinfo {year} {2021})\
  \bibinfo {edition} {{W}inter 2021}\ ed.\BibitemShut {Stop}%
\bibitem [{\citenamefont {Giacomini}\ and\ \citenamefont
  {Kempf}(2022)}]{FlaminiaAchim}%
  \BibitemOpen
  \bibfield  {author} {\bibinfo {author} {\bibfnamefont {F.}~\bibnamefont
  {Giacomini}}\ and\ \bibinfo {author} {\bibfnamefont {A.}~\bibnamefont
  {Kempf}},\ }\href@noop {} {\enquote {\bibinfo {title} {Second-quantized
  unruh-dewitt detectors and their quantum reference frame transformations},}\
  } (\bibinfo {year} {2022}),\ \Eprint {http://arxiv.org/abs/2201.03120}
  {arXiv:2201.03120 [quant-ph]} \BibitemShut {NoStop}%
\bibitem [{\citenamefont {Klco}\ and\ \citenamefont
  {Savage}(2020)}]{PhysRevA.102.012619}%
  \BibitemOpen
  \bibfield  {author} {\bibinfo {author} {\bibfnamefont {N.}~\bibnamefont
  {Klco}}\ and\ \bibinfo {author} {\bibfnamefont {M.~J.}\ \bibnamefont
  {Savage}},\ }\href {\doibase 10.1103/PhysRevA.102.012619} {\bibfield
  {journal} {\bibinfo  {journal} {Phys. Rev. A}\ }\textbf {\bibinfo {volume}
  {102}},\ \bibinfo {pages} {012619} (\bibinfo {year} {2020})}\BibitemShut
  {NoStop}%
\bibitem [{\citenamefont {Wallace}(2006)}]{WallaceNaive}%
  \BibitemOpen
  \bibfield  {author} {\bibinfo {author} {\bibfnamefont {D.}~\bibnamefont
  {Wallace}},\ }\href {http://www.jstor.org/stable/20118789} {\bibfield
  {journal} {\bibinfo  {journal} {Synthese}\ }\textbf {\bibinfo {volume}
  {151}},\ \bibinfo {pages} {33} (\bibinfo {year} {2006})}\BibitemShut
  {NoStop}%
\bibitem [{\citenamefont {Wallace}(2011)}]{Wallace2011}%
  \BibitemOpen
  \bibfield  {author} {\bibinfo {author} {\bibfnamefont {D.}~\bibnamefont
  {Wallace}},\ }\href {http://philsci-archive.pitt.edu/8890/} {\enquote
  {\bibinfo {title} {Taking particle physics seriously: a critique of the
  algebraic approach to quantum field theory},}\ } (\bibinfo {year}
  {2011})\BibitemShut {NoStop}%
\bibitem [{\citenamefont {Fraser}(2009)}]{Fraser2009}%
  \BibitemOpen
  \bibfield  {author} {\bibinfo {author} {\bibfnamefont {D.}~\bibnamefont
  {Fraser}},\ }\href {\doibase 10.1086/649999} {\bibfield  {journal} {\bibinfo
  {journal} {Philosophy of Science}\ }\textbf {\bibinfo {volume} {76}},\
  \bibinfo {pages} {536} (\bibinfo {year} {2009})}\BibitemShut {NoStop}%
\bibitem [{\citenamefont {Fraser}(2011)}]{Fraser2011}%
  \BibitemOpen
  \bibfield  {author} {\bibinfo {author} {\bibfnamefont {D.}~\bibnamefont
  {Fraser}},\ }\href {\doibase https://doi.org/10.1016/j.shpsb.2011.02.002}
  {\bibfield  {journal} {\bibinfo  {journal} {Studies in History and Philosophy
  of Science Part B: Studies in History and Philosophy of Modern Physics}\
  }\textbf {\bibinfo {volume} {42}},\ \bibinfo {pages} {126} (\bibinfo {year}
  {2011})},\ \bibinfo {note} {philosophy of Quantum Field Theory}\BibitemShut
  {NoStop}%
\bibitem [{\citenamefont {{M. Ruep}}(2022)}]{MaxTalk}%
  \BibitemOpen
  \bibfield  {author} {\bibinfo {author} {\bibnamefont {{M. Ruep}}},\ }\href
  {https://youtu.be/NDi3HRPwEw8} {\enquote {\bibinfo {title} {Observing
  observables -- causal measurement schemes for every observable of the linear
  real scalar field in curved spacetime},}\ } (\bibinfo {year} {2022}),\
  \bibinfo {note} {[Contribution to the Quantum Field Theory in Curved
  Spacetimes Workshop (23-27 May 2022)]}\BibitemShut {NoStop}%
\bibitem [{\citenamefont {Lamb}(1995)}]{Lamb1995}%
  \BibitemOpen
  \bibfield  {author} {\bibinfo {author} {\bibfnamefont {W.~E.}\ \bibnamefont
  {Lamb}},\ }\href {\doibase 10.1007/BF01135846} {\bibfield  {journal}
  {\bibinfo  {journal} {Applied Physics B}\ }\textbf {\bibinfo {volume} {60}},\
  \bibinfo {pages} {77} (\bibinfo {year} {1995})}\BibitemShut {NoStop}%
\bibitem [{\citenamefont {Mart\'{i}n-Mart\'{i}nez}\ and\ \citenamefont
  {Rodriguez-Lopez}(2018)}]{Pablo}%
  \BibitemOpen
  \bibfield  {author} {\bibinfo {author} {\bibfnamefont {E.}~\bibnamefont
  {Mart\'{i}n-Mart\'{i}nez}}\ and\ \bibinfo {author} {\bibfnamefont
  {P.}~\bibnamefont {Rodriguez-Lopez}},\ }\href {\doibase
  10.1103/PhysRevD.97.105026} {\bibfield  {journal} {\bibinfo  {journal} {Phys.
  Rev. D}\ }\textbf {\bibinfo {volume} {97}},\ \bibinfo {pages} {105026}
  (\bibinfo {year} {2018})}\BibitemShut {NoStop}%
\bibitem [{\citenamefont {Lopp}\ and\ \citenamefont
  {Mart\'{\i}n-Mart\'{\i}nez}(2021{\natexlab{a}})}]{RichardEdu}%
  \BibitemOpen
  \bibfield  {author} {\bibinfo {author} {\bibfnamefont {R.}~\bibnamefont
  {Lopp}}\ and\ \bibinfo {author} {\bibfnamefont {E.}~\bibnamefont
  {Mart\'{\i}n-Mart\'{\i}nez}},\ }\href {\doibase 10.1103/PhysRevA.103.013703}
  {\bibfield  {journal} {\bibinfo  {journal} {Phys. Rev. A}\ }\textbf {\bibinfo
  {volume} {103}},\ \bibinfo {pages} {013703} (\bibinfo {year}
  {2021}{\natexlab{a}})}\BibitemShut {NoStop}%
\bibitem [{\citenamefont {Lopp}\ and\ \citenamefont
  {Mart\'{\i}n-Mart\'{\i}nez}(2021{\natexlab{b}})}]{Richard}%
  \BibitemOpen
  \bibfield  {author} {\bibinfo {author} {\bibfnamefont {R.}~\bibnamefont
  {Lopp}}\ and\ \bibinfo {author} {\bibfnamefont {E.}~\bibnamefont
  {Mart\'{\i}n-Mart\'{\i}nez}},\ }\href {\doibase 10.1103/PhysRevA.103.013703}
  {\bibfield  {journal} {\bibinfo  {journal} {Phys. Rev. A}\ }\textbf {\bibinfo
  {volume} {103}},\ \bibinfo {pages} {013703} (\bibinfo {year}
  {2021}{\natexlab{b}})}\BibitemShut {NoStop}%
\bibitem [{\citenamefont {Mart\'{\i}n-Mart\'{\i}nez}\ \emph
  {et~al.}(2020)\citenamefont {Mart\'{\i}n-Mart\'{\i}nez}, \citenamefont
  {Perche},\ and\ \citenamefont {de~S.~L.~Torres}}]{GRQO}%
  \BibitemOpen
  \bibfield  {author} {\bibinfo {author} {\bibfnamefont {E.}~\bibnamefont
  {Mart\'{\i}n-Mart\'{\i}nez}}, \bibinfo {author} {\bibfnamefont {T.~R.}\
  \bibnamefont {Perche}}, \ and\ \bibinfo {author} {\bibfnamefont
  {B.}~\bibnamefont {de~S.~L.~Torres}},\ }\href {\doibase
  10.1103/PhysRevD.101.045017} {\bibfield  {journal} {\bibinfo  {journal}
  {Phys. Rev. D}\ }\textbf {\bibinfo {volume} {101}},\ \bibinfo {pages}
  {045017} (\bibinfo {year} {2020})}\BibitemShut {NoStop}%
\bibitem [{\citenamefont {Faure}\ \emph {et~al.}(2020)\citenamefont {Faure},
  \citenamefont {Perche},\ and\ \citenamefont {Torres}}]{GravityDetector}%
  \BibitemOpen
  \bibfield  {author} {\bibinfo {author} {\bibfnamefont {R.}~\bibnamefont
  {Faure}}, \bibinfo {author} {\bibfnamefont {T.~R.}\ \bibnamefont {Perche}}, \
  and\ \bibinfo {author} {\bibfnamefont {B.~d. S.~L.}\ \bibnamefont {Torres}},\
  }\href {\doibase 10.1103/PhysRevD.101.125018} {\bibfield  {journal} {\bibinfo
   {journal} {Phys. Rev. D}\ }\textbf {\bibinfo {volume} {101}},\ \bibinfo
  {pages} {125018} (\bibinfo {year} {2020})}\BibitemShut {NoStop}%
\bibitem [{\citenamefont {Pitelli}\ and\ \citenamefont
  {Perche}(2021)}]{GravityDetector2}%
  \BibitemOpen
  \bibfield  {author} {\bibinfo {author} {\bibfnamefont {J.~P.~M.}\
  \bibnamefont {Pitelli}}\ and\ \bibinfo {author} {\bibfnamefont {T.~R.}\
  \bibnamefont {Perche}},\ }\href {\doibase 10.1103/PhysRevD.104.065016}
  {\bibfield  {journal} {\bibinfo  {journal} {Phys. Rev. D}\ }\textbf {\bibinfo
  {volume} {104}},\ \bibinfo {pages} {065016} (\bibinfo {year}
  {2021})}\BibitemShut {NoStop}%
\bibitem [{\citenamefont {Torres}\ \emph {et~al.}(2020)\citenamefont {Torres},
  \citenamefont {Perche}, \citenamefont {Landulfo},\ and\ \citenamefont
  {Matsas}}]{MatsasNeutrinos}%
  \BibitemOpen
  \bibfield  {author} {\bibinfo {author} {\bibfnamefont {B.~d. S.~L.}\
  \bibnamefont {Torres}}, \bibinfo {author} {\bibfnamefont {T.~R.}\
  \bibnamefont {Perche}}, \bibinfo {author} {\bibfnamefont {A.~G.~S.}\
  \bibnamefont {Landulfo}}, \ and\ \bibinfo {author} {\bibfnamefont {G.~E.~A.}\
  \bibnamefont {Matsas}},\ }\href {\doibase 10.1103/PhysRevD.102.093003}
  {\bibfield  {journal} {\bibinfo  {journal} {Phys. Rev. D}\ }\textbf {\bibinfo
  {volume} {102}},\ \bibinfo {pages} {093003} (\bibinfo {year}
  {2020})}\BibitemShut {NoStop}%
\bibitem [{\citenamefont {Perche}\ and\ \citenamefont
  {Mart\'{\i}n-Mart\'{\i}nez}(2021)}]{perche2021antiparticle}%
  \BibitemOpen
  \bibfield  {author} {\bibinfo {author} {\bibfnamefont {T.~R.}\ \bibnamefont
  {Perche}}\ and\ \bibinfo {author} {\bibfnamefont {E.}~\bibnamefont
  {Mart\'{\i}n-Mart\'{\i}nez}},\ }\href {\doibase 10.1103/PhysRevD.104.105021}
  {\bibfield  {journal} {\bibinfo  {journal} {Phys. Rev. D}\ }\textbf {\bibinfo
  {volume} {104}},\ \bibinfo {pages} {105021} (\bibinfo {year}
  {2021})}\BibitemShut {NoStop}%
\bibitem [{\citenamefont {Mart\'{\i}n-Mart\'{\i}nez}\ \emph
  {et~al.}(2021)\citenamefont {Mart\'{\i}n-Mart\'{\i}nez}, \citenamefont
  {Perche},\ and\ \citenamefont {Torres}}]{BrokenCovariance}%
  \BibitemOpen
  \bibfield  {author} {\bibinfo {author} {\bibfnamefont {E.}~\bibnamefont
  {Mart\'{\i}n-Mart\'{\i}nez}}, \bibinfo {author} {\bibfnamefont {T.~R.}\
  \bibnamefont {Perche}}, \ and\ \bibinfo {author} {\bibfnamefont {B.~d.
  S.~L.}\ \bibnamefont {Torres}},\ }\href {\doibase
  10.1103/PhysRevD.103.025007} {\bibfield  {journal} {\bibinfo  {journal}
  {Phys. Rev. D}\ }\textbf {\bibinfo {volume} {103}},\ \bibinfo {pages}
  {025007} (\bibinfo {year} {2021})}\BibitemShut {NoStop}%
\end{thebibliography}%

\end{document}